\documentclass[prd,aps,twocolumn,showpacs,nofootinbib]{revtex4-1}

\usepackage{amsfonts,color,graphicx,epsfig,amsmath,multirow}
\usepackage[colorlinks=true,linkcolor=blue,citecolor=blue,urlcolor=blue]{hyperref}
\usepackage{mathtools}
\usepackage{slashed}
\usepackage{graphicx}

\begin{document}

\title{Reconsideration of the $B \to K^*$ Transition Form Factors within the QCD Light-Cone Sum Rules}

\author{Wei Cheng$^1$}
\author{Xing-Gang Wu$^1$}
\email{wuxg@cqu.edu.cn}
\author{Hai-Bing Fu$^2$}

\address{$^1$Department of Physics, Chongqing University, Chongqing 401331, P.R. China}
\address{$^2$ School of Science, Guizhou Minzu University, Guiyang 550025, P.R. China}
\date{\today}

\begin{abstract}

In this paper, we study the $B \to K^*$ transition form factors (TFFs) within the QCD light-cone sum rules (LCSR) approach. Two correlators, i.e. the usual one and the right-handed one, are adopted in the LCSR calculation. The resultant LCSRs for the $B \to K^*$ TFFs are arranged according to the twist structure of the $K^*$-meson light-cone distribution amplitudes (LCDAs), whose twist-2, twist-3 and twist-4 terms behave quite differently by using different correlators. We observe that the twist-4 LCDAs, though generally small, shall have sizable contributions to the TFFs $A_{1/2}$, $V$ and $T_1$, thus the twist-4 terms should be kept for a sound prediction. We also observe that even though different choices of the correlator lead to different LCSRs with different twist contributions, the large correlation coefficients for most of the TFFs indicate that the LCSRs for different correlators are close to each order, not only for their values at the large recoil point $q^2=0$ but also for their ascending trends in whole $q^2$-region. Such a high degree of correlation is confirmed by their application to the branching fraction of the semi-leptonic decay $B \to K^* \mu^+ \mu^-$. Thus, a proper choice of correlator may inversely provide a chance for probing uncertain LCDAs, i.e. the contributions from those LCDAs can be amplified to a certain degree via a proper choice of correlator, thus amplifying the sensitivity of the TFFs, and hence their related observables, to those LCDAs.

\end{abstract}


\maketitle

\section{Introduction}

The heavy-to-light $B$-meson decay provides an excellent platform for testing the CP-violation phenomena and for seeking new physics beyond the Standard Model (SM). The heavy-to-light transition form factors (TFFs) are key components in those studies, which however are non-trivial due to the fact that for practical values of the momentum transfer and the $b$-quark mass ($m_b$), the soft contributions are always numerically important and are often dominant.

The Shifman-Vainshtein-Zakharov (SVZ) sum rules~\cite{Shifman:1978bx, Shifman:1978by}, which provides an important step forward for studying those non-perturbative hadron phenomenology. It is a method of expanding the correlation function (correlator) into the QCD vacuum condensates with subsequent matching via dispersion relations. The vacuum condensates are non-perturbative but universal, whose contributions follow from the usual power-counting rules at the large $q^2$-region and the first several ones are enough to achieve the required accuracy. Many successful hadron properties have been achieved since its invention, and the SVZ sum rules becomes a useful tool for studying the hadron phenomenology.

Following its strategy, one has to deal with the two-point correlator for the heavy-to-light transition form factors (TFFs)~\cite{Nesterenko:1982gc, Ioffe:1982qb, Reinders:1984sr}, which however will meet specific problems such as the breaking of power-counting and the contamination of sum rules by ``non-diagonal" transitions~\cite{Braun:1997kw}, severely restricting the precisions and applicabilities of the SVZ sum rules.

To avoid the problems of the two-point SVZ sum rules, the QCD light-cone sum rules (LCSR) has later been suggested to deal with the heavy-to-light TFFs~\cite{Balitsky:1989ry, Chernyak:1990ag, Ball:1991bs, Belyaev:1993wp, Ball:1997rj, Ball:2004rg}. Its main idea is to make a partial resummation of the operator product expansion (OPE) to all orders and reorganize the OPE expansion in terms of the twists of relevant operators rather than their dimensions. The vacuum condensates of the SVZ sum rules are then substituted by the light-meson's light-cone distribution amplitudes (LCDAs) of increasing twists. The LCDA, which relates the matrix elements of the nonlocal light-ray operators sandwiched between the hadronic state and the vacuum, has a direct physical significance and provides the underlying links between the hadronic phenomena at small and large distances.

Generally, contributions from the LCDAs suffer from the power counting rules basing on the twists, i.e. the high-twist LCDAs are usually powered suppressed to the lower twist ones in large $Q^2$-region, and the first several LCDAs shall usually provide dominant contributions to the LCSR. Since its invention, the LCSR approach has been widely adopted for studying the $B\to$ light meson decays. In the paper, we shall concentrate our attention on its application to the $B \to K^*$ decays, which is helpful for studying the $K^*$-meson LCDAs.

How to ``design" a proper correlator is a tricky problem for the LCSR approach. By choosing a proper correlator, one can not only study the  properties of the hadrons but also simplify the theoretical uncertainties effectively. Usually, the correlator is constructed by using the currents with definite quantum numbers, such as those with definite $J^P$, where $J$ is the total angular momentum and $P$ is the parity of the bound state. Such a direct way of constructing the correlator is not the only choice adopted in the literature, e.g. the chiral correlator with a chiral current in between the matrix element has also been suggested so as to suppress part of the hazy contributions from the uncertain LCDAs~\cite{Huang:1998gp, Huang:2001xb, Wan:2002hz, Zuo:2006dk, Wu:2007vi, Wu:2009kq}.

The LCDAs of the $K^*$-meson have a much complex structure than that of the light pseudo-scalar mesons. It contains two leading-twist (or twist-2) LCDAs $\phi^\bot_{2;K^*}$ and $\phi_{2;K^*}^\|$, seven twist-3 LCDAs $\phi_{3;K^*}^\bot$, $\psi_{3;K^*}^\bot$, $\Phi_{3;K^*}^\|$, $\widetilde\Phi_{3;K^*}^\|$, $\phi_{3;K^*}^\|$, $\psi_{3;K^*}^\|$ and $\Phi_{3;K^*}^\bot$, and twelve twist-4 LCDAs $\phi_{4;K^*}^\bot$, $\psi_{4;K^*}^\bot$, $\Psi_{4;K^*}^\bot$, $\widetilde{\Psi} _{4;K^*}^\bot$, $\Phi _{4;K^*}^{\bot(1)}$, $\Phi _{4;K^*}^{\bot(2)}$, $\Phi _{4;K^*}^{\bot(3)}$, $\Phi _{4;K^*}^{\bot(4)}$, $\phi_{4;K^*}^\|$, $\psi_{4;K^*}^\|$, $\widetilde \Phi_{4;K^*}^\|$ and $ \widetilde \Psi_{4;K^*}^\|$~\cite{Ball:2004rg}. By taking the usual correlator, we shall show that at the twist-4 accuracy, the LCSRs of the $B\to K^*$ TFFs shall contain almost all of the mentioned LCDAs, where the twist-3 and twist-4 LCDAs shall have sizable contributions to the $B\to K^*$ TFFs. At the present, some of the high-twist LCDAs have been studied within the QCD sum rules~\cite{Ball:2007rt, Ball:2007zt}, which, however, are still of large errors.

As an attempt, we have suggested to use a chiral correlator with a right-handed chiral current to deal with the $B\to K^*$ TFFs such that to suppress the uncertainties from the high-twist LCDA contributions~\cite{Fu:2014uea}. The resultant LCSRs derived there show the contributions from most of the high-twist LCDAs are suppressed by $\delta^2\sim (m_K^*/m_b)^2\sim 0.03$, thus uncertainties from high-twist LCDAs themselves are effectively suppressed. In previous discussions~\cite{Fu:2014uea}, some of the terms that are proportional to high-twist LCDAs have been omitted due to the $\delta$-power counting rule. In the paper, as a sound prediction, we shall keep all of them in our present calculations; as will be shown later, those terms shall provide sizable contributions for certain TFFs.

It is interesting to show whether the LCSRs under different choices of the correlator are consistent with each other. In the paper, as a step forward, we shall compare the LCSRs for the TFFs under the usual correlator and the right-handed chiral correlator with the help of the correlation coefficient $\rho_{XY}$~\cite{Agashe:2014kda}.

The remaining parts of the paper are organized as follows. In Sec.II, we present the calculation technology for the $B \to K^*$ TFFs within the LCSR approach, where the results for the usual correlator are presented. The results for the right-handed chiral correlator are presented in the Appendix. In Sec.III, we make a comparative study on various LCSRs for the $B\to K^*$ TFFs, and their application for the branching fraction $d{\mathcal B}(B\to K^* \mu^+\mu^-)/dq^2$ is also presented. Sec.IV is reserved for a summary.

\section{The $B \to K^*$ TFFs within the LCSR approach}

The $B \to K^*$ TFFs, $V(q^2)$, $A_{0,1,2}(q^2)$ and $T_{1,2,3}(q^2)$, are related with the matrix elements $\langle K^* |\bar{s}\gamma^\mu b|B\rangle$, $\langle K^* |\bar{s}\gamma^\mu\gamma^5 b|B\rangle$, $\langle K^* |\bar{s}\sigma^{\mu\nu}q_\nu b|B\rangle$ and $\langle K^* |\bar{s}\sigma^{\mu\nu}\gamma^5 q_\nu b|B\rangle$ via the following way~\cite{Ball:2004rg},
\begin{widetext}
\begin{eqnarray}
 \langle K^*(p,\lambda )|&&\bar s{\gamma _\mu }(1 - {\gamma _5})b|B(p+q)\rangle = - ie_\mu ^{*(\lambda )}(m_B + m_{K^*} )A_1(q^2) + i(e^{*(\lambda )}\cdot q)\frac{(2p + q)_\mu}{m_B + m_{K^*}}A_2(q^2) \nonumber\\
&&+ iq_\mu (e^{*(\lambda )}\cdot q)\frac{2 m_{K^*} }{q^2}\big[A_3(q^2)- A_0(q^2)\big]  +\epsilon_{\mu\nu\alpha\beta}e^{*(\lambda )\nu} q^\alpha p^\beta \frac{2V(q^2)}{m_B + m_{K^*}}  \label{BKstar:matrix1}
\end{eqnarray}
and
\begin{eqnarray}
\langle K^*(p,\lambda )|&&\bar s\sigma_{\mu \nu }q^\nu (1 + \gamma_5)b|B(p + q)\rangle = 2i\epsilon_{\mu\nu\alpha\beta}e^{*(\lambda )\nu} q^\alpha p^\beta T_1(q^2) + e_\mu ^{*(\lambda )} (m_B^2 - m_{K^*}^2)T_2(q^2) \nonumber\\
&& - (2p + q)_\mu (e^{*(\lambda )} \cdot q){\widetilde T_3}(q^2)  + q_\mu (e^{*(\lambda )} \cdot q)T_3(q^2),\label{BKstar:matrix2}
\end{eqnarray}
\end{widetext}
where $p$ is the momentum of $K^*$-meson and $q = p_B - p$ is the momentum transfer, $e^{(\lambda)}$ stands for the $K^*$-meson polarization vector with $\lambda$ being its transverse ($\bot$) or longitudinal ($\|$) component, respectively. The following relations are helpful,
\begin{eqnarray}
&&T_3(q^2) = \frac{m_B^2 - m_{K^*}^2}{q^2} [\widetilde T_3(q^2) - T_2(q^2)], \label{T3}
\\
&&A_3(q^2) = \frac{m_B+m_{K^*}}{2 m_{K^*}} A_1(q^2) - \frac{m_B-m_{K^*}}{2m_{K^*}} A_2(q^2) \label{A3}
\end{eqnarray}
and $A_0(0)=A_3(0)$ and $T_1(0)=T_2(0)=\widetilde T_3(0)$.

To derive the LCSRs for the $B\to K^*$ TFFs, we introduce the following correlator
\begin{eqnarray}
\Pi_{\mu}^{\rm I,II}(p,q) = -i \int d^4 x e^{iq\cdot x}\langle K^*(p,\lambda)|T\{j_W^{\rm I,II}(x),j_B^\dag (0)\}|0\rangle ,\nonumber\\ \label{correlators}
\end{eqnarray}
where the currents $j_W^{\rm I}(x) = \bar s (x){\gamma _\mu }(1 - {\gamma _5})b(x)$ and $J_W^{\rm II} (x)= \bar s (x){\sigma _{\mu \nu }}{q^\nu }(1 + {\gamma _5})b(x)$. The current $j_B^\dag (x)$ is usually chosen as $i m_b \bar b(x) \gamma_5 q(x)$, which has the same quantum state as the pseudoscalar $B$-meson with $J^{P}=0^-$. For simplicity, we call its corresponding LCSR as LCSR-${\cal U}$. As mentioned in the Introduction, the current $j_B^\dag (x)$ can also be chosen as a chiral current, e.g. the right-handed chiral current $i m_b \bar b(x)(1+ \gamma_5)q(x)$. We call its corresponding LCSR as LCSR-${\cal R}$. The calculation technology for the LCSR are the same for both cases, and we take $j_B^\dag (x)=i m_b \bar b(x) \gamma_5 q(x)$ as an explicit example to show how to derive the LCSRs for the $B\to K^*$ TFFs up to twist-4 accuracy.

The correlator (\ref{correlators}) is analytic in the whole $q^2$-region. In the time-like region, one can insert a complete series of the intermediate hadronic states in the correlator and obtain its hadronic representation by isolating out the pole term of the lowest pseudoscalar $B$-meson. More explicitly, the correlator $\Pi_\mu^{\rm H(I)}$ can be written as
\begin{widetext}
\begin{eqnarray}
\Pi_\mu^{\rm H(I)}(p,q)
&=& \frac{\langle K^*|\bar s \gamma_\mu (1 - \gamma_5)b|B\rangle \langle B|\bar bi m_b\gamma_5 q_1|0\rangle}{m_B^2 - (p + q)^2} + \sum\limits_{\rm H} \frac{\langle K^*|\bar s\gamma_\mu (1 - \gamma_5)b|B^{\rm H}\rangle \langle B^{\rm H}|\bar b i m_b \gamma_5 q_1|0\rangle}{m_{B^{\rm H}}^2 - (p + q)^2}  \nonumber\\
&=&\Pi_1^{\rm H (I)} e_\mu^{*(\lambda)} + \Pi_2^{\rm H (I)} (e^{*(\lambda)}\cdot q) (2p+q)_\mu  + \Pi _3^{\rm H (I)} (e^{*(\lambda )}\cdot q)q_\mu  + i\Pi_4^{\rm H (I)} \epsilon_\mu^{\nu \alpha \beta } e_\nu^{*(\lambda )} q_\alpha p_\beta ,
\end{eqnarray}
\end{widetext}
where the matrix element $\langle B|\bar b i m_b\gamma_5 q_1|0\rangle=m_B^2 f_B$, where $f_B$ is the $B$-meson decay constant. By replacing the contributions from the high resonances and continuum states with the dispersion relations, the invariant amplitudes can be written as
\begin{eqnarray}
\Pi _i^{\rm H(I)} &=& \frac{m_B^2 f_B(m_B+ m_{K^*})}{m_B^2 - (p  +  q)^2} \tilde{A}_{i}(q^2) \nonumber\\
&& + \int_{s_0}^\infty  \frac{\rho_i^{\rm H(I)}}{s - (p  +  q)^2}ds + \cdots,
\end{eqnarray}
where $i=(1,\cdots,4)$, $s_0$ is the threshold parameter, and the ellipsis stands for the subtraction constant or the finite $q^2$-polynomial which has no contribution to the final sum rules. The reduced functions $\tilde{A}_i$ are
\begin{eqnarray}
&& \tilde{A}_{1}=A_{1},\; \tilde{A}_{2}=\frac{A_{2}}{(m_B+ m_{K^*})^2},\;\nonumber\\
&& \tilde{A}_{3}=\frac{2m_{K^*}}{q^2 (m_B+ m_{K^*})}[A_3(q^2) - A_0(q^2)],\;\nonumber\\
&& \tilde{A}_{4}=\frac{2V(q^2)}{(m_B+ m_{K^*})^2}.
\end{eqnarray}
The spectral densities $\rho^{\rm H(I)}_{i}(s)$ is estimated by using the ansatz of the quark-hadron duality~\cite{Shifman:1978by}, i.e. $\rho^{\rm H(I)}_{i}(s)= \rho^{\rm QCD(I)}_{i}(s)\theta (s-s_0)$.

In the space-like region, the correlator can be calculated by using the operator production expansion (OPE). With the help of the $b$-quark propagator~\cite{Huang:1998gp},
\begin{eqnarray}
\langle 0|{\rm T}&&\{ b(x)\bar b (0)\} | 0\rangle  = i\int {\frac{{{d^4}k}}{{{{(2\pi )}^4}}}} {e^{ - ik \cdot x}}\frac{{\not\! k + {m_b}}}{{m_b^2 - {k^2}}}\nonumber\\
&&- i{g_s}\int {\frac{{{d^4}k}}{{{{(2\pi )}^4}}}} {e^{ - ik \cdot x}}\int_0^1 dv \bigg[\frac{1}{2}\frac{{\not\! k + m_b}}{(m_b^2 - k^2)^2}\nonumber\\
&&\times  {G^{\mu \nu }}(v x){\sigma _{\mu \nu }}+ \frac{1}{{m_b^2 - {k^2}}}v{x_\mu }{G^{\mu \nu }}(v x){\gamma _\nu }\bigg]~, \label{bd}
\end{eqnarray}
The correlator can be expressed as
\begin{widetext}
\begin{eqnarray}
\Pi_\mu ^{\rm OPE(I)}(p,q) &=& \int \frac{d^4 x d^4 k}{(2\pi )^4} \frac{e^{i(q-k)\cdot x}}{m_b^2 - k^2} \bigg\{k^\nu\langle K^*(p,\lambda)|{\rm T} \{ \bar s(x)\gamma _\mu \gamma _\nu \gamma _5 q(0) \}|0\rangle  + {k^\nu }\langle {K^*}(p,\lambda )|{\rm T}\{ {{{\bar s}}(x){\gamma _\mu }{\gamma _\nu }{q}(0)} \}|0\rangle  \nonumber \\
&& + m_b \langle K^*(p,\lambda)|{\rm T}\{ \bar s(x) \gamma_\mu \gamma_5 q(0)\}|0\rangle - m_b\langle K^*(p,\lambda )|{\rm T}\{ \bar s(x) \gamma_\mu q(0) \}|0\rangle +\cdots \bigg\},
\end{eqnarray}
\end{widetext}
The nonlocal matrix elements in the right-hand-side of the above equation can be reexpressed by the LCDAs of various twists~\cite{Ball:2004rg, Ball:2007zt}. We present the relations for the nonlocal matrix elements to the LCDAs in the Appendix.

The correlator $\Pi_\mu^{\rm II}(p,q)$ can be treated via a similar way. With the help of the analytic property of the correlator in different $q^2$-region, the LCSRs for the $B \to K^*$ TFFs are ready to be derived.

As a further step, we apply the usual Borel transformation to the sum rules, which removes the subtraction terms in the dispersion relation and effectively suppresses the contributions from the unknown excited resonances and continuum states heavier than $K^*$ meson. After applying the Borel transformation, our final LCSRs read
\begin{widetext}
\begin{eqnarray}
A_1^{\mathcal U}(q^2) &=& \frac{m_{K^*} m_b}{f_B m_B^2(m_B + m_{K^*})}\int_0^1 du e^{\left( {m_{B}^2 - s(u)} \right) / M^2}  \bigg\{ \frac{ m_{K^*} f_{K^*}^\bot \cal C}{2u^2 m_{K^*} ^2}\Theta(c(u,s_0))\phi_{2;{K^*}}^\bot(u) +\frac{m_{K^*} f_{K^*}^\bot}{2u} \Theta(c(u,s_0))\nonumber\\
&&\times \psi_{3;{K^*}}^\|(u)+ \frac{ m_b f_{K^*}^\parallel}{u} \Theta \left(c \left( u,s_0 \right) \right) \phi_{3;K^*}^ \bot \left( u \right) - m_{K^*} f_{K^*}^\bot\bigg[ \frac{m_b^2{\cal C}}{8u^4M^4} \widetilde{\widetilde\Theta}(c(u,s_0)) + \frac{{\cal C}-2m_b^2}{8u^3M^2} \widetilde\Theta(c(u,s_0))\nonumber\\
&&- \frac{1}{8u^2}\Theta(c(u,s_0))\bigg]\phi_{4;{K^*}}^\bot(u) - \frac{m_b m_{K^*}^2 f_{K^*}^\parallel }{u^2 M^2}\widetilde \Theta \left( c\left( u,s_0 \right) \right)C_{K^*}(u) - m_{K^*} f_{K^*}^\bot\bigg[\frac{\cal C}{u^3M^2}\widetilde\Theta(c(u,s_0))-\frac{1}{u^2} \nonumber\\
&&\times \Theta(c(u,s_0))\bigg]I_L(u) - m_{K^*} f_{K^*}^\bot \bigg[\frac{2m_b^2}{2u^2M^2}\widetilde \Theta(c(u,s_0))+\frac{1}{2u} \Theta(c(u,s_0)) \bigg] H_3(u)\bigg\}+\int_0^1 dv \int_0^1 du \int_0^1 d {\mathcal D}\nonumber\\
&&\times e^{\left( {m_{B}^2 - s(u)} \right) / M^2}\frac{\widetilde\Theta(c(u,s_0))}{u^2 M^2} \frac{m_b m_{K^*}^2} {12 f_B m_B^2( m_B + m_{K^*})} \bigg\{f_{K^*}^\bot \bigg[\widetilde {\Psi} _{4;{K^*}}^\bot (\underline \alpha ) - 12\bigg(\Psi _{4;{K^*}}^\bot (\underline \alpha )-2v \Psi _{4;{K^*}}^\bot (\underline \alpha )\nonumber\\
&&+2\Phi _{4;{K^*}}^{\bot(1)} (\underline \alpha ) -2\Phi _{4;{K^*}}^{\bot(2)} (\underline \alpha )+4v \Phi _{4;{K^*}}^{\bot(2)} (\underline \alpha )\bigg)\bigg]\bigg(m_B^2-m_{K^*}^2+2 u m_{K^*}^2\bigg)+ 2 m_b m_{K^*} f_{K^*}^\parallel \bigg(\widetilde {\Phi} _{3;{K^*}}^\parallel (\underline \alpha)\nonumber\\
&& + 12 \Phi _{3;{K^*}}^\parallel (\underline \alpha  )\bigg)\bigg\}, \label{tra A1}
\end{eqnarray}
\begin{eqnarray}
A_2^{\mathcal U}(q^2) &=& \frac{m_{K^*} m_b\left( m_B + m_{K^*} \right)}{2f_B m_B^2}\int_0^1 du e^{\left( {m_{B}^2 - s(u)} \right) / M^2} \bigg\{ \frac{m_{K^*} f_{K^*}^\bot}{u m_{K^*} ^2}\Theta(c(u,s_0))\phi_{2;{K^*}}^\bot(u)- \frac{m_b f_{K^*}^\bot}{u M^2} \widetilde\Theta (c(u,s_0))  \nonumber \\
&&\times\psi_{3;{K^*}}^{\|}(u) - \frac{m_{K^*} f_{K^*}^\bot}{4}\bigg[\frac{m_b^2}{u^3M^4} \widetilde{\widetilde\Theta}(c(u,s_0)) + \frac{1}{u^2M^2} \widetilde\Theta(c(u,s_0))\bigg]\phi_{4;{K^*}}^\bot(u)+ \frac{2 m_b f_{K^*}^\parallel }{u^2 M^2}\widetilde \Theta \left( c\left( {u,{s_0}} \right) \right)
\nonumber \\
&&\times A_{K^*}(u) - \frac{m_{K^*}^2 m_b^3 f_{K^*}^\parallel}{2u^4 M^6}\widetilde {\widetilde{\widetilde \Theta }}\left( c\left( u,s_0 \right) \right)B_{K^*}(u) + \frac{2 m_b m_{K^*}^2 f_{K^*}^\parallel }{u^2 M^4}\widetilde {\widetilde \Theta }\left( c\left( u,s_0 \right)\right)C_{K^*}(u) + 2 m_{K^*} f_{K^*}^\bot \nonumber\\
&&\times \bigg[\frac{{\cal C} - 2m_b^2}{u^3 M^4}\widetilde{\widetilde\Theta}(c(u,s_0)) - \frac{1}{u^2M^2}\widetilde\Theta (c(u,s_0))\bigg]I_L(u)- \frac{m_{K^*} f_{K^*}^\bot}{u M^2}\widetilde \Theta (c(u,s_0))H_3(u)\bigg\}+\int_0^1 dv \int_0^1 du\nonumber\\
&&\times \int_0^1 d {\mathcal D} e^{\left( {m_{B}^2 - s(u)} \right) / M^2} \frac{ m_b m_{K^*}^2 f_{K^*}^\bot} {12 f_B m_B^2} \frac{ m_B + m_{K^*}}{u^2 M^2}\widetilde\Theta(c(u,s_0))\bigg[\widetilde {\Psi} _{4;{K^*}}^\bot (\underline \alpha ) + 12\bigg( 2v \Psi _{4;{K^*}}^\bot (\underline \alpha )\nonumber\\
&&-\Psi _{4;{K^*}}^\bot (\underline \alpha )+(4v-2) \Phi _{4;{K^*}}^{\bot(1)} (\underline \alpha ) + 2\Phi _{4;{K^*}}^{\bot(2)} (\underline \alpha ) \bigg)\bigg], \label{tra A2}
\end{eqnarray}
\begin{eqnarray}
A_3^{\mathcal U}(q^2) &-& A_0^{\mathcal U}(q^2) = \frac{m_b q^2}{2f_B m_B^2}\int_0^1 du e^{\left( {m_{B}^2 - s(u)} \right) / M^2} \bigg\{ - \frac{f_{K^*}^\bot}{2u m_{K^*}}\Theta (c(u,s_0))\phi_{2;K^*}^\bot(u) - \frac{(2-u)m_{K^*}f_{K^*}^\bot}{2u^2M^2} \widetilde \Theta (c(u,s_0)) \nonumber\\
&&\times\psi_{3;K^*}^\| (u) + \frac{m_{K^*}f_{K^*}^\bot}{8}\bigg[\frac{m_b^2}{u^3M^4} \widetilde{\widetilde\Theta}(c(u,s_0))+ \frac{1}{u^2M^2} \widetilde\Theta(c(u,s_0))\bigg]\phi_{4;K^*}^\bot(u) -\frac{m_b f_{K^*}^\|}{m_{K^*}u^2M^2}\widetilde \Theta (c(u,s_0))A_{K^*}(u)\nonumber\\
&& + \frac{m_{K^*} m_b^3 f_{K^*}^\|}{u^4 M^6}\widetilde {\widetilde {\widetilde \Theta }}(c(u,s_0))B_{K^*}(u)+ \frac{m_bm_{K^*}f_{K^*}^\|(2 - u)}{u^3 M^4}\widetilde {\widetilde \Theta }(c(u,s_0))C_{K^*}(u) + m_{K^*} f_{K^*}^\bot \bigg[(4 - 2u) \nonumber\\
&&\times \bigg(\frac{\cal C}{2u^4 M^4} \widetilde{\widetilde\Theta}(c(u,s_0)) - \frac{1}{u^3 M^2}\widetilde\Theta(c(u,s_0))\bigg)+ \bigg(\frac{2m_b^2}{u^3 M^4}\widetilde{\widetilde\Theta}(c(u,s_0)) + \frac{1}{u^2M^2} \widetilde\Theta(c(u,s_0))\bigg)\bigg]I_L(u) \nonumber\\
&& - \frac{m_{K^*}(2-u)f_{K^*}^\bot }{2u^2M^2}\widetilde\Theta(c(u,s_0))H_3(u)\bigg\}-\int_0^1 dv \int_0^1 du \int_0^1 d {\mathcal D}  e^{\left( {m_{B}^2 - s(u)} \right) / M^2}\frac{ m_b q^2 f_{K^*}^\bot} {24 f_B m_B^2} \frac{\widetilde\Theta(c(u,s_0))}{u^2 M^2}\nonumber\\
&& \times \bigg[12\bigg(2v \Psi _{4;{K^*}}^\bot (\underline \alpha )-\Psi _{4;{K^*}}^\bot (\underline \alpha ) +(4v-2) \Phi _{4;{K^*}}^{\bot(1)} (\underline \alpha )+ 2\Phi _{4;{K^*}}^{\bot(2)} (\underline \alpha )\bigg)+\widetilde {\Psi} _{4;{K^*}}^\bot (\underline \alpha )\bigg],   \label{tra A30}
\end{eqnarray}
\begin{eqnarray}
T_1^{\mathcal U}(q^2) &=& \frac {m_b m_{K^*} } {2m_B^2 f_B} \int_0^1 du e^{\left( {m_{B}^2 - s(u)} \right) / M^2} \bigg\{\frac{m_b m_{K^*} f_{K^*}^\bot}{u m_{K^*}^2}\Theta (c(u,s_0))\phi _{2;K^*}^\bot (u)+ f_{K^*}^\| \bigg[ \frac {\mathcal E} {4u^2M^2} \widetilde \Theta (c(u,s_0)) + \frac1{4u}  \nonumber\\
&&\times \Theta(c(u,s_0)) \bigg] \psi_{3;K^*}^\bot(u) - \frac{ m_{K^*} m_b^3 f_{K^*}^\bot }{4u^3M^4} \widetilde{\widetilde \Theta} (c(u,s_0)) \phi_{4;K^*}^\bot(u) + f_{K^*}^\| \Theta (c(u,s_0))\phi _{3;K^*}^ \bot (u) + \frac{f_{K^*}^\|}{u}\nonumber\\
&&\times \Theta (c(u,s_0))A_{K^*}(u) - \bigg[ \frac1{4u^2 M^2}\widetilde \Theta(c(u,s_0)) + \frac {m_b^2} {4u^3 M^4} \widetilde{ \widetilde \Theta}(c(u,s_0)) \bigg] m_{K^*}^2 f_{K^*}^\| B_{K^*}(u)-\frac{2m_b m_{K^*} f_{K^*}^\bot}{u^2M^2} \nonumber\\
&& \times \widetilde\Theta(c(u,s_0)) I_L(u) - \frac{m_b m_{K^*} f_{K^*}^\bot}{uM^2} \widetilde\Theta(c(u,s_0)) H_3(u) \bigg\}+\int_0^1 dv \int_0^1 du \int_0^1 d {\mathcal D} e^{\left( {m_{B}^2 - s(u)} \right) / M^2} \frac{m_{K^*}}{12 f_B m_B^2}\nonumber\\
&&\times \frac{\widetilde\Theta(c(u,s_0))}{u^2 M^2}\bigg\{ m_{K^*} m_b f_{K^*}^\bot \bigg[\widetilde {\Psi} _{4;{K^*}}^\bot (\underline \alpha )-12\bigg(\Psi _{4;{K^*}}^\bot (\underline \alpha )+2\Phi _{4;{K^*}}^{\bot(1)} (\underline \alpha ) - 2 \Phi _{4;{K^*}}^{\bot(2)} (\underline \alpha )\bigg)\bigg] + f_{K^*}^\parallel \bigg [u m_{K^*}^2\nonumber\\
&&\times \bigg(12(1-2v)\Phi_{3;{K^*}}^\parallel (\underline \alpha )+\widetilde {\Phi} _{3;{K^*}}^\parallel (\underline \alpha )\bigg) -2 v u\widetilde {\Phi} _{3;{K^*}}^\parallel (\underline \alpha ) + v \bigg(-12\Phi_{3;{K^*}}^\parallel (\underline \alpha )+ \widetilde {\Phi} _{3;{K^*}}^\parallel (\underline \alpha )\bigg)\bigg( m_B ^2- m_{K^*}^2\bigg) \bigg] \bigg\}, \nonumber\\ \label{tra T1}
\end{eqnarray}
\begin{eqnarray}
T_2^{\mathcal U}(q^2) &=& \frac {m_b m_{K^*} } {2m_B^2f_B} \int_0^1 du e^{\left( {m_{B}^2 - s(u)} \right) / M^2} \bigg\{ \frac{m_b m_{K^*}f_{K^*}^\bot(1 - {\mathcal H})}{u m_{K^*}^2}\Theta (c(u,s_0)) \phi_{2;K^*}^\bot(u)+ f_{K^*}^\|\bigg[1 + \frac {(2 - u)\cal H} {u} \bigg]
\nonumber\\
&& \times \Theta (c(u,s_0)) \phi _{3;K^*}^ \bot (u) +\bigg[ \frac{\mathcal F (1-\mathcal H) - 4u m_{K^*}^2 \mathcal H}{4u^2 M^2}\widetilde \Theta (c(u,s_0))+ \frac {1 -\mathcal H} {4u} \Theta (c(u,s_0)) \bigg] f_{K^*}^\| \psi _{3;K^*}^\bot (u)
\nonumber\\
&& - \frac{m_{K^*}m_b^3 f_{K^*}^\bot } {4u^3M^4}(1 - {\mathcal H})\widetilde{\widetilde\Theta}(c(u,s_0))\phi_{4;K^*}^\bot(u) + \frac{f_{K^*}^\|(1 -\cal H)}{u}\Theta (c(u,s_0)) A_{K^*}(u)-\bigg[ \frac{1 - \mathcal H}{4u^2M^2} \widetilde \Theta (c(u,s_0)) \nonumber\\
&& + \frac{(1 - \mathcal H)m_b^2} {4u^3 M^4} \widetilde {\widetilde \Theta }(c(u,s_0))\bigg]m_{K^*}^2 f_{K^*}^\| B_{K^*}(u)- \frac{2m_b m_{K^*}f_{K^*}^\bot (1 - {\mathcal H})}{u^2M^2} \widetilde\Theta(c(u,s_0))I_L(u) - \frac{m_b m_{K^*} f_{K^*}^\bot }{uM^2}\nonumber\\
&&\times \bigg[ 1+ \bigg( \frac{2}{u}-1 \bigg){\mathcal H} \bigg]\widetilde\Theta (c(u,s_0)) H_3(u) \bigg\}+\int_0^1 dv \int_0^1 du \int_0^1 d {\mathcal D} e^{\left( {m_{B}^2 - s(u)} \right) / M^2} \frac{m_{K^*}} {12 f_B m_B^2} \frac{\widetilde\Theta(c(u,s_0))}{u^2 M^2}\nonumber\\
&&\times \bigg\{ m_{K^*}m_b f_{K^*}^\bot \bigg[\widetilde {\Psi} _{4;{K^*}}^\bot (\underline \alpha )-12\bigg(\Psi _{4;{K^*}}^\bot (\underline \alpha )+2\Phi _{4;{K^*}}^{\bot(1)} (\underline \alpha ) - 2 \Phi _{4;{K^*}}^{\bot(2)} (\underline \alpha )\bigg)\bigg]+ \frac{f_{K^*}^\parallel} {2(m_B^2-m_{K^*}^2)} \frac{\widetilde\Theta(c(u,s_0))}{u^2 M^2}\nonumber\\
&&\times \bigg[ v \bigg(\widetilde {\Phi} _{3;{K^*}}^\parallel (\underline \alpha )-12\Phi_{3;{K^*}}^\parallel (\underline \alpha ) \bigg) \bigg( m_B^2 - m_{K^*}^2\bigg)^2 + u m_{K^*}^2\bigg(\widetilde {\Phi} _{3;{K^*}}^\parallel (\underline \alpha )+12(1-2v)\Phi_{3;{K^*}}^\parallel (\underline \alpha )\bigg)\bigg( m_B - m_{K^*}^2\bigg) \nonumber\\
&& + 2 q^2 m_{K^*}^2 \bigg(-2v \widetilde {\Phi} _{3;{K^*}}^\parallel (\underline \alpha )+\widetilde {\Phi} _{3;{K^*}}^\parallel (\underline \alpha )+12 \Phi _{3;{K^*}}^\parallel (\underline \alpha ) \bigg)\bigg]\bigg\}, \nonumber\\  \label{tra T2}
\end{eqnarray}
\begin{eqnarray}
V^{\mathcal U}(q^2) &=& \frac{m_b \left( m_B + m_{K^*} \right)}{2f_B m_B^2}\int_0^1 du e^{\left( {m_{B}^2 - s(u)} \right) / M^2} \bigg\{f_{K^*}^\bot \Theta(c(u,s_0))\phi_{2;{K^*}}^\bot(u)+\frac{m_{K^*} m_b f_{K^*}^\parallel}{2u^2 M^2}\widetilde \Theta \left( c\left( u,s_0 \right) \right) \nonumber\\
&&\times \psi _{3;K^*}^ \bot (u) - \bigg[\frac{m_b^2}{u^2M^4}\widetilde {\widetilde\Theta}(c(u,s_0))+\frac{1}{uM^2}\widetilde\Theta(c(u,s_0))\bigg]
\frac{ m_{K^*}^2 f_{K^*}^\bot}{4}\phi_{4;{K^*}}^\bot (u)\bigg\}+ \int_0^1 dv \int_0^1 du\nonumber\\
&& \times \int_0^1 d {\mathcal D} e^{\left( {m_{B}^2 - s(u)} \right) / M^2} \frac{ m_{K^*}^2 f_{K^*}^\bot} {6} \frac{\widetilde\Theta(c(u,s_0))}{u^2 M^2}\bigg[(2v-1)\widetilde {\Psi} _{4;{K^*}}^\bot (\underline \alpha ) + 12\bigg(\Psi _{4;{K^*}}^\bot (\underline \alpha )-2(v-1)\nonumber\\
&&\times (\Phi _{4;{K^*}}^{\bot(1)} (\underline \alpha )-\Phi _{4;{K^*}}^{\bot(2)} (\underline \alpha ))\bigg)\bigg], \label{tra V}
\end{eqnarray}
\begin{eqnarray}
T_3^{\mathcal U}(q^2) &=& \frac{m_b m_{K^*} } {2m_B^2 f_B} \int_0^1 du e^{\left( {m_{B}^2 - s(u)} \right) / M^2} \bigg\{\frac{m_b m_{K^*}f_{K^*}^\bot}{u m_{K^*}^2} \Theta(c(u,s_0))\phi_{2;K^*}^\bot(u) + f_{K^*}^\|\left(1 - \frac2{u}\right)\Theta(c(u,s_0))\phi_{3;K^*}^\bot (u)  \nonumber\\
&&+ f_{K^*}^\|\bigg[ \frac{\mathcal F + 4um_{K^*}^2}{4u^2 M^2} \widetilde \Theta (c(u,s_0)) + \frac1{4u} \Theta (c(u,s_0))\bigg] \psi _{3;K^*}^ \bot (u)- \frac{ m_{K^*} m_b^3 f_{K^*}^\bot}{4u^3M^4}\widetilde{\widetilde\Theta}(c(u,s_0))\phi_{4;K^*}^\bot(u) + 2f_{K^*}^\|
\nonumber\\
&&\times \bigg[ \frac{m_B^2 - m_{K^*}^2}{u^2 M^2}\widetilde\Theta(c(u,s_0)) + \frac1{2u} \Theta (c(u,s_0))\bigg] A_{K^*}(u) -\bigg[ \frac1{4u^2 M^2}\widetilde \Theta (c(u,s_0))+ \frac{2q^2 + m_b^2 + 2\mathcal Q}{4u^3M^4}\nonumber\\
&&\times \widetilde {\widetilde \Theta }(c(u,s_0))+ \frac{(m_{K^*}^2 + \mathcal Q)m_b^2}{2u^4 M^6}\widetilde {\widetilde {\widetilde \Theta }}(c(u,{s_0})) \bigg] m_{K^*}^2f_{K^*}^\| B_{K^*}(u)- m_b m_{K^*} f_{K^*}^\bot \bigg[\frac{2}{u^2M^2}\widetilde\Theta(c(u,s_0))\nonumber\\
&&+ \frac{4}{u^3 M^4}\widetilde{\widetilde \Theta}(c(u,s_0))(m_B^2 - m_{K^*}^2)\bigg] I_L(u) + m_b m_{K^*} f_{K^*}^\bot \bigg[ \frac{2}{u^2M^2} - \frac{1}{uM^2} \bigg] \widetilde\Theta(c(u,s_0))H_3(u) \bigg\}\nonumber\\
&& + \int_0^1 dv \int_0^1 du \int_0^1 d {\mathcal D} e^{\left( {m_{B}^2 - s(u)} \right) / M^2} \frac{m_{K^*}} {12 f_B m_B^2} \frac{\widetilde\Theta(c(u,s_0))}{u^2 M^2}\bigg\{ m_{K^*}m_b f_{K^*}^\bot \bigg[\widetilde {\Psi} _{4;{K^*}}^\bot (\underline \alpha )-12\bigg(\Psi _{4;{K^*}}^\bot (\underline \alpha )\nonumber\\
&&+2\Phi _{4;{K^*}}^{\bot(1)} (\underline \alpha )- 2 \Phi _{4;{K^*}}^{\bot(2)} (\underline \alpha )\bigg)\bigg]+f_{K^*}^\parallel \bigg[ m_{K^*}^2 \bigg( \Phi_{3;{K^*}}^\parallel (\underline \alpha )(4v-u-2) + 12 \Phi_{3;{K^*}}^\parallel (\underline \alpha )(2vu-u-2) \bigg)\nonumber\\
&& - v \bigg( m_B^2 - m_{K^*}^2 \bigg) \bigg(\widetilde {\Phi} _{3;{K^*}}^\parallel (\underline \alpha )-12\Phi_{3;{K^*}}^\parallel (\underline \alpha )\bigg)\bigg]\bigg\}, \label{tra T3}
\end{eqnarray}
\end{widetext}
where the superscript ${\cal U}$ indicates those LCSRs are for the usual correlator with $j_B^\dag (x)=i m_b \bar b(x)\gamma_5 q(x)$. The LCDAs are generally scale-dependent, and for convenience we have implicitly omitted the factorization scale $\mu$ in the LCDAs. $\int d{\cal D}=\int d\alpha_1 d\alpha_2 d\alpha_3 \delta(1 - \sum \limits_{i \rm = 1}^{\rm{3}} {\alpha_i})$. ${\mathcal H} = q^2/(m_B^2 - m_{K^*}^2)$, ${\mathcal E} = m_b^2 - u^2 m_{K^*}^2 + q^2$, ${\mathcal C}=m_b^2+u^2m_{K^*}^2-q^2$, $\mathcal Q = m_B^2 - m_{K^*}^2 - q^2$, $\mathcal F = m_b^2 - u^2 m_{K^*}^2 - q^2$, $c(\varrho,s_0) = \varrho s_0 - m_b^2 + \bar \varrho q^2 - \varrho \bar \varrho m_{K^*}^2$ and $s(\varrho)=[ m_b^2 - \bar \varrho(q^2 - \varrho m_{K^*}^2)] / \varrho $ ($\varrho = u$) with $\bar \varrho = 1 - \varrho$. $\Theta(c(u,s_0))$ is the usual step function. $\widetilde \Theta(c(u,s_0))$ and $\widetilde {\widetilde \Theta}(c(u,s_0))$ come from the surface terms $\delta(c({u_0},{s_0}))$ and $\Delta (c({u_0},{s_0}))$, whose explicit forms have been given in Ref.\cite{Fu:2014uea}.

The reduced functions $I_L(u)$, $H_3(u)$, $A_{K^*}(u)$, $B_{K^*}(u)$, and $C_{K^*}(u)$ are defined as
\begin{eqnarray}
I_L(u)=&& \int_0^u dv \int_0^v dw \big[\phi_{3;{K^*}}^\|(w) -\frac{1}{2} \phi_{2;{K^*}}^\bot(w)
\nonumber\\&&
-\frac{1}{2} \psi_{4;{K^*}}^\bot(w)\big],\\
H_3(u)=&& \int_0^u dv \left[\psi_{4;{K^*}}^{\bot}(v)-\phi_{2;{K^*}}^\bot(v)\right],
\\
A_{K^*}(u) =&&\int_0^u dv \left[ \phi _{2;K^*}^\| (v) - \phi _{3;K^*}^ \bot (v) \right],
\\
B_{K^*}(u) =&&\int_0^u dv \phi _{4;K^*}^\| (v),
\\
C_{K^*}(u) =&&\int_0^u dv \int_0^v {dw} \left[\psi _{4;K^*}^\|(w) + \phi _{2;K^*}^\|(w)\right.
\nonumber\\
&&\left.- 2 \phi_{3;K^*}^\bot(w) \right].
\end{eqnarray}

By using the same correlator and keeping only the first term of the $b$-quark propagator (\ref{bd}), Ref.\cite{Ball:1997rj} calculated the LCSRs for the $B\to \rho$ TFFs $A_1$, $A_2$ and $V$, and Ref.\cite{Aliev:1996hb} calculated the LCSRs for the $B\to K^*$ TFFs $A_1$, $A_2$, $A_3 - A_0$, $V$, $T_1$, $T_2$ and $T_3$. All those LCSRs are given up to twist-3 accuracy \footnote{In those two references the surface terms have not be taken into consideration, and because only the 2-particle terms have been kept in the matrix elements, the twist-3 LCDAs involving 3-particle contributions have also been missed in the LCSRs.}. As a cross-check, we find if keeping the terms up to the same twist-3 accuracy and transforming to the same definitions for the form factors, we return to the same expressions listed in Refs.\cite{Ball:1997rj, Aliev:1996hb}.

\begin{table}[tb]
\centering
\begin{tabular}{ c c  c  c  }
\hline
~~ ~~ & ~~twist-2~~  & ~~twist-3~~ & ~~twist-4~~  \\
\hline
\multirow{3}{*}{{\textbf 1} ,$\gamma_5,\sigma_{\mu\nu}$}  & \multirow{3}{*}{$\phi_{2;K^*}^\bot$}  & $\phi_{3;K^*}^\|$, $\psi_{3;K^*}^\|$  & $\phi_{4;K^*}^\bot$, $\psi_{4;K^*}^\bot$, $\Psi_{4;K^*}^\bot$ \\
 & & $\Phi_{3;K^*}^\bot$ & $\widetilde{\Psi}_{4;K^*}^\bot$, $\Phi_{4;K^*}^{\bot(1)}$, $\Phi_{4;K^*}^{\bot(2)}$ \\
 & &   & $\Phi_{4;K^*}^{\bot(3)}$, $\Phi_{4;K^*}^{\bot(4)}$ \\
\hline
\multirow{2}{*}{$\gamma_{\mu}$,~$\gamma_{\mu}\gamma_5$} & \multirow{2}{*}{$\phi_{2;K^*}^\|$} & $\phi_{3;K^*}^\bot$, $\psi_{3;K^*}^\bot$  & $\phi_{4;K^*}^\|$, $\psi_{4;K^*}^\|$  \\
 & & $\Phi_{3;K^*}^\|$, $\widetilde\Phi_{3;K^*}^\|$ & $\widetilde \Phi_{4;K^*}^\|$, $ \widetilde \Psi_{4;K^*}^\|$ \\
\hline
\end{tabular}
\caption{Following the idea of Ref.\cite{Ball:2004rg}, we rewrite the $K^*$-meson LCDAs with different $\Gamma$-structures in the non-perturbative hadronic matrix elements.} \label{DA_del}
\end{table}

Up to twist-4 accuracy, we present the required $K^*$-meson LCDAs in Table~\ref{DA_del}. All of those LCDAs are emerged in the LCSRs (\ref{tra A1}, \ref{tra A2}, \ref{tra A30}, \ref{tra T1}, \ref{tra T2}, \ref{tra V}, \ref{tra T3}). The accuracy of the LCSRs thus depend heavily on how well we know those LCDAs.

In general cases, the contributions from the twist-4 terms are numerically small, thus the uncertainties from the twist-4 LCDAs themselves are highly suppressed. We shall directly adopt the twist-4 LCDAs derived by applying the conformal expansion of the matrix element~\cite{Ball:2007zt} to do the numerical calculation.

The twist-3 contributions are generally suppressed by certain $\delta$-powers ($\delta=m_{K^*}/m_b\sim 0.17$) and $1/M^2$-powers to the leading twist-2 terms. For examples, the twist-3 contributions from the LCDAs $\phi_{3;K^*}^\bot$, $\psi_{3;K^*}^\bot$, $\Phi_{3;K^*}^\|$ and $\widetilde\Phi_{3;K^*}^\|$ are suppressed by $\delta^1$ and the twist-3 contributions from the LCDAs $\phi_{3;K^*}^\|$, $\psi_{3;K^*}^\|$ and $\Phi_{3;K^*}^\bot$ are suppressed by $\delta^2$. However, the twist-3 contributions are sizable and important in certain kinematic region, a special effect should be paid for a precise prediction.

On the one hand, one may use more accurate twist-2 LCDAs to predict the twist-3 contributions. This can be achieved by applying the relations among the twist-2 and twist-3 LCDAs. For example, under the Wandzura-Wilczek approximation~\cite{Wandzura:1977qf}, the 2-particle twist-3 LCDAs $\psi_{3;K^*}^\bot$, $\phi_{3;K^*}^\bot$, $\psi_{3;K^*}^\|$ and $\phi_{3;K^*}^\|$ can be related to the twist-2 LCDAs $\phi_{2;K^*}^\|$ and $\phi_{2;K^*}^\bot$ via the following relations~\cite{Ball:1997rj}
\begin{eqnarray}
\psi_{3;K^*}^\bot(u)&=&2\left[\bar u \int_0^u dv \frac {\phi_{2;K^*}^\|(v)}{\bar v}+ u \int_u^1 dv \frac {\phi_{2;K^*}^\|(v)}{v}\right],\nonumber\\
\phi_{3;K^*}^\bot(u)&=&\frac1{2}\left[\int_0^u dv \frac {\phi_{2;K^*}^\|(v)}{\bar v}+\int_u^1 dv \frac {\phi_{2;K^*}^\|(v)}{v}\right],\nonumber\\
\psi_{3;K^*}^\|(u)&=&2\left[\bar u \int_0^u dv \frac {\phi_{2;K^*}^\bot(v)}{\bar v}+ u \int_u^1 dv \frac {\phi_{2;K^*}^\bot(v)}{v}\right],\nonumber\\
\phi_{3;K^*}^\|(u)&=&(1-2\bar u)\left[\int_0^u dv \frac {\phi_{2;K^*}^\bot(v)}{\bar v}-\int_u^1 dv \frac {\phi_{2;K^*}^\bot(v)}{v}\right],\nonumber
\end{eqnarray}
where $\bar{u} =1-u$ and $\bar{v} =1-v$. The contributions from the remaining three 3-particle twist-3 LCDAs to the $B\to K^*$ TFFs are numerically small, thus as the same as the twist-4 LCDAs, we shall directly take them as the ones from Ref.\cite{Ball:2007zt}.

On the other hand, it has been suggested that by using the improved LCSR approach~\cite{Huang:1998gp, Huang:2001xb} and by taking a chiral correlator, the less certain high-twist contributions could be highly suppressed. Ref.\cite{Fu:2014uea} has shown that by taking a right-handed correlator with $j_B^\dag (x)=i m_b \bar b(x)(1+ \gamma_5)q(x)$, the twist-3 LCDAs, $\phi_{3;K^*}^\bot$, $\psi_{3;K^*}^\bot$, $\Phi_{3;K^*}^\|$, $\widetilde\Phi_{3;K^*}^\|$, and even the twist-2 LCDA $\phi_{2;K^*}^\|$ disappear in the LCSRs. Thus the uncertain twist-3 contributions can be highly suppressed.

Following the standard LCSR procedures, and by keeping all the terms that contribute to the LCSRs up to twist-4 accuracy, we recalculate the $B\to K^*$ TFFs for the right-handed chiral correlator. Our final LCSRs are presented in the Appendix.

The hadronic representation of the chiral correlator contains an extra resonance $J^P = 0^+$ in addition to the usual one with $J^P=0^-$, introducing extra uncertainty to the LCSR-${\cal R}$. The LCSR-${\cal R}$ eliminates the large uncertainties from the twist-2 and twist-3 structures which are at the $\delta^1$-order and we can also suppress its pollution by a proper choice of continuum threshold $s_0$, thus it is worthwhile to use a chiral correlator. Numerically, we confirm our previous observation that the final LCSRs have slight $s_0$ dependence~\cite{Fu:2014uea}, thus the uncertainties from $J^P = 0^+$ resonance are small.

\section{numerical analysis} \label{numerical analysis}

\subsection{Basic inputs} \label{basic input}

In doing the numerical calculation, we take the $K^*$-meson decay constants $f^\bot_{K^*}= 0.185(9) {\rm GeV}$ and $f^\|_{K^*}= 0.220(5) \rm GeV$~\cite{Ball:2007zt}, the $b$-quark mass $m_b = 4.80 \pm 0.05 \rm GeV$, the $K^*$-meson mass $m_{K^*} = 0.892 \rm GeV$, the $B$-meson mass $m_B =5.279 \rm GeV$~\cite{Agashe:2014kda}, and the $B$-meson decay constant $f_B =0.160 \pm 0.019 \rm GeV$~\cite{Fu:2014pba}. The factorization scale $\mu$ is set as typical momentum of the heavy $b$-quark, i.e. $\mu \simeq (m^2_B-m^2_b)^{1/2}\sim 2.2 {\rm GeV}$~\cite{Belyaev:1994zk, Ball:1998tj}, and we predict its error by taking $\Delta\mu=\pm 1.0 {\rm GeV}$.

The choices of twist-3 and twist-4 LCDAs have been explained in last subsection. As for the twist-2 LCDAs, we adopt the model, following the idea of Wu-Huang model for the pion LCDA~\cite{Wu:2010zc}, to do the calculation~\cite{Fu:2014uea}
\begin{widetext}
\begin{eqnarray}
\phi_{2;K^*}^{\lambda}(x) &=& \frac{A_{2;K^*}^\lambda \sqrt{3x \bar x} \rm Y}{8\pi^{3/2}\widetilde f_{K^*}^\lambda b_{2;K^*}^\lambda} [1 + B_{2;K^*}^\lambda C_1^{3/2}(\xi) +C_{2;K^*}^\lambda C_2^{3/2}(\xi) ]\exp \left[ - b_{2;K^*}^{\lambda 2}\frac{\bar x m_s^2 + x m_q^2 - \rm Y ^2} {x\bar x} \right]  \nonumber\\
&& \times\left[ \text{Erf}\left( b_{2;K^*}^\lambda \sqrt {\frac {\mu _0^2 + \rm Y^2} {x\bar x}}  \right) - \text{Erf}\left( b_{2;K^*}^\lambda \sqrt {\frac{\rm Y^2}{x\bar x}}  \right) \right], \nonumber\\ \label{phi2}
\end{eqnarray}
\end{widetext}
where $\lambda=\|$ or $\perp$, $\widetilde f_{K^*}^\perp=f_{K^*}^\perp/\sqrt{3}$ and $\widetilde f_{K^*}^\|=f_{K^*}^\|/\sqrt{5}$ are reduced decay constants, $\xi=2x-1$, $Y= \overline x m_s + x m_q$, the error function, ${\rm Erf}(x) = \frac2{\sqrt \pi}\int_0^x e^{ - t^2}dt$. The model cooperates the transverse momentum dependence with the longitudinal one under the Brodsky-Huang-Lepage prescription~\cite{BHL} and the Wigner-Melosh rotation~\cite{Huang:1994dy, Cao:1997hw, Huang:2004su}. Such a cooperation of transverse effect in the light meson wavefunction is helpful for an effective suppression of the end-point singularity for high-energy processes involving light mesons, cf. a review~\cite{Wu:2013lga}.

The model parameters $A_{2;K^*}^\lambda$, $B_{2;K^*}^\lambda$, $C_{2;K^*}^\lambda$ and $b_{2;K^*}^\lambda$ can be fixed by applying the criteria,
\begin{itemize}
\item The normalization condition of the twist-2 LCDA, i.e. $\int \phi_{2;K^*}^{\lambda}(x)dx=1$;
\item As shown by Ref.\cite{Fu:2014uea}, the average of the squared transverse momentum $\langle{\bf k}_\bot^2 \rangle_{K^*}^{1/2}$ could be determined from the light-cone wavefunction which is related to the LCDA by integrating out its transverse momentum dependence. To fix the parameter, we adopt $\langle{\bf k}_\bot^2 \rangle_{K^*}^{1/2} = 0.37(2) {\rm GeV}$~\cite{Wu:2010zc, Huang:2013yya}.
\item Generally, the twist-2 LCDA can be expanded as a Gegenbauer polynomial,
    \begin{equation}
     \phi_{2;K^*}^{\lambda}(x)=6x{\bar x}\left(1+\sum_{n=1,2,\cdots} a_n^{\lambda} C_n^{3/2}(\xi)\right),
    \end{equation}
    whose Gegenbauer moment $a_n^{\lambda}$ can be calculated via the following way due to the orthogonality of the Gegenbauer functions, i.e.
    \begin{eqnarray}
     a_n^\lambda= \dfrac{\int_0^1 dx \phi_{2;K^*}^{\lambda}(x) C_n^{3/2}(\xi)}{ \int_0^1 6x \bar x [C_n^{3/2}(\xi)]^2}.
    \end{eqnarray}
    Generally, the behavior of the twist-2 LCDA is dominated by its first several terms. We adopt the first two Gegenbauer moments derived from the QCD sum rules~\cite{Ball:2007zt} to fix the parameters, i.e. $a_1^\bot (1{\rm GeV}) = 0.04(3)$ and $a_2^\bot (1{\rm GeV}) = 0.10(8)$ for $\phi^{\perp}_{2;K^*}$, and $a_1^\| (1{\rm GeV}) = 0.03(2)$ and $a_2^\| (1{\rm GeV}) = 0.11(9)$ for $\phi^{\|}_{2;K^*}$.
\end{itemize}
This way, we get the LCDA at the scale of $1$ GeV, and its behavior at any other scale can be achieved via the renormalization group evolution~\cite{Ball:2006nr}.

\begin{table}[htb]
\centering
\begin{tabular}{cccccc}
\hline
~~$a_1^{\|}$~~ & ~~$a_2^{\|}$~~ & ~~$B_{2;K^*}^{\|}$~~ & ~~$C_{2;K^*}^{\|}$~~ & ~~$A_{2;K^*}^{\|}$~~ & ~~$b_{2;K^*}^{\|}$~~\\
\hline
0.03&0.11&-0.007&0.178&26.645&0.629\\
0.01&0.11&-0.029&0.180&26.777&0.630\\
0.05&0.11&-0.014&0.176&26.519&0.628\\
0.03&0.20&-0.008&0.275&24.256&0.599\\
0.03&0.02&-0.001&0.078&27.530&0.642\\
\hline
\end{tabular}
\caption{Parameters of the twist-2 LCDA $\phi_{2;K^*}^{\|}$ determined for $a_1^\| (1{\rm GeV}) = 0.03(2)$ and $a_2^\| (1{\rm GeV}) = 0.11(9)$. } \label{parameters}
\end{table}

\begin{figure}[htb]
\centering
\includegraphics[width=0.45\textwidth]{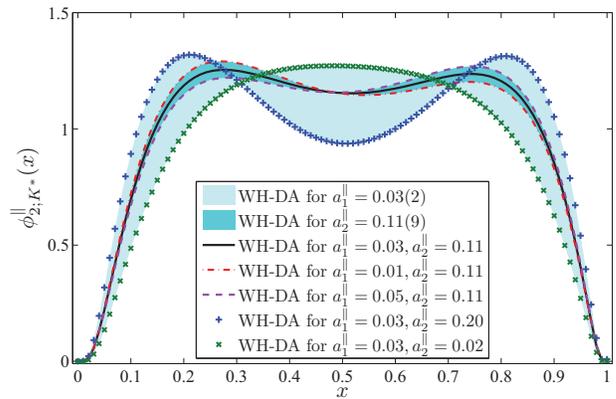}
\caption{The twist-2 LCDA $\phi_{2;K^*}^\|(x)$ at the scale $1$ GeV. } \label{phinew}
\end{figure}

The parameters of $\phi_{2;K^*}^{\perp}$ for $a_1^\bot (1{\rm GeV}) = 0.04(3)$ and $a_2^\bot (1{\rm GeV}) = 0.10(8)$ have been given in Ref.\cite{Fu:2014uea}. We present the parameters of $\phi_{2;K^*}^{\|}$ for $a_1^\| (1{\rm GeV}) = 0.03(2)$ and $a_2^\| (1{\rm GeV}) = 0.11(9)$ in Table~\ref{parameters}, and the corresponding LCDA behavior in Fig.\ref{phinew}.

\subsection{Criteria for the LCSRs}
\label{Criteria for the LCSRs}

The Borel parameter $M^2$ and the continuum threshold $s_0$ are determined by the criteria,
\begin{itemize}
\item The continuum contribution, which is the part of the dispersive integral from $s_0$ to $\infty$, should not be too large. We take it to be less than $50\%$ of the total LCSR,
    \begin{displaymath}
      \frac{\int_{s_0}^\propto ds \rho^{\rm tot}(s)e^{-s/M^2}} {\int_{m_b^2}^\propto ds\rho^{\rm tot}(s)e^{-s/M^2}} \le 50\%.
    \end{displaymath}
\item All high-twist LCDAs' contributions are less than $35\%$ of the total LCSR, qualitatively ensuring the usual power counting of twist contributions.
\item The derivatives of the TFFs with respect to $1/M^2$ give the LCSRs for $m_B$. We require all predicted $B$-meson masses to be full-filled with high accuracy, e.g. $\left| m_B^{\rm LCSR} - m_B^{\exp } \right| / m_B^{\exp} \le 0.1\% $.
\end{itemize}
The determined continuum threshold $s_0$ and the Borel parameter $M^2$ for various $B\to K^*$ TFFs at the large recoil point $q^2 =0$ are listed in Table~\ref{Borel}.

\begin{table}[tb]
\centering
\begin{tabular}{  c c c  c  c c c c c c }
\hline
      & ~~~$A_1$~~~      & ~~~$A_2$~~~        & ~~~$A_{3-0}$~~~     & ~~~$V$~~~        & ~~~$T_1$~~~      & ~~~$T_3$~~~  \\
\hline
$M^2_{\cal{R}}$ &   6.0      &    4.5         &     5.0       &   4.5        &  3.0        &    8.5  \\
$s_{0\cal{R}}$  &   36.0     &    33.0        &      30.0     &    32.0      &     33.0    &    34.0 \\
$M^2_{\cal{U}}$ &   25.0     &    24.0        &     25.0      &   4.5        &   32.0      &    28.0 \\
$s_{0\cal{U}}$  &   38.0     &    37.0        &      39.0     &    37.0      &     37.0    &    37.0 \\
\hline
\end{tabular}
\caption{The Borel parameter $M^2$ and the continuum threshold $s_0$ (in units: GeV$^2$) for the $B \to K^*\mu^+\mu^-$ TFFs at $q^2=0$. The subscripts $\cal{R}$ and $\cal{U}$ stand for the cases of the right-handed and the usual correlators, respectively.} \label{Borel}
\end{table}

\subsection{Properties of the LCSRs}

\begin{figure*}[htb]
\centering
\includegraphics[width=0.25\textwidth]{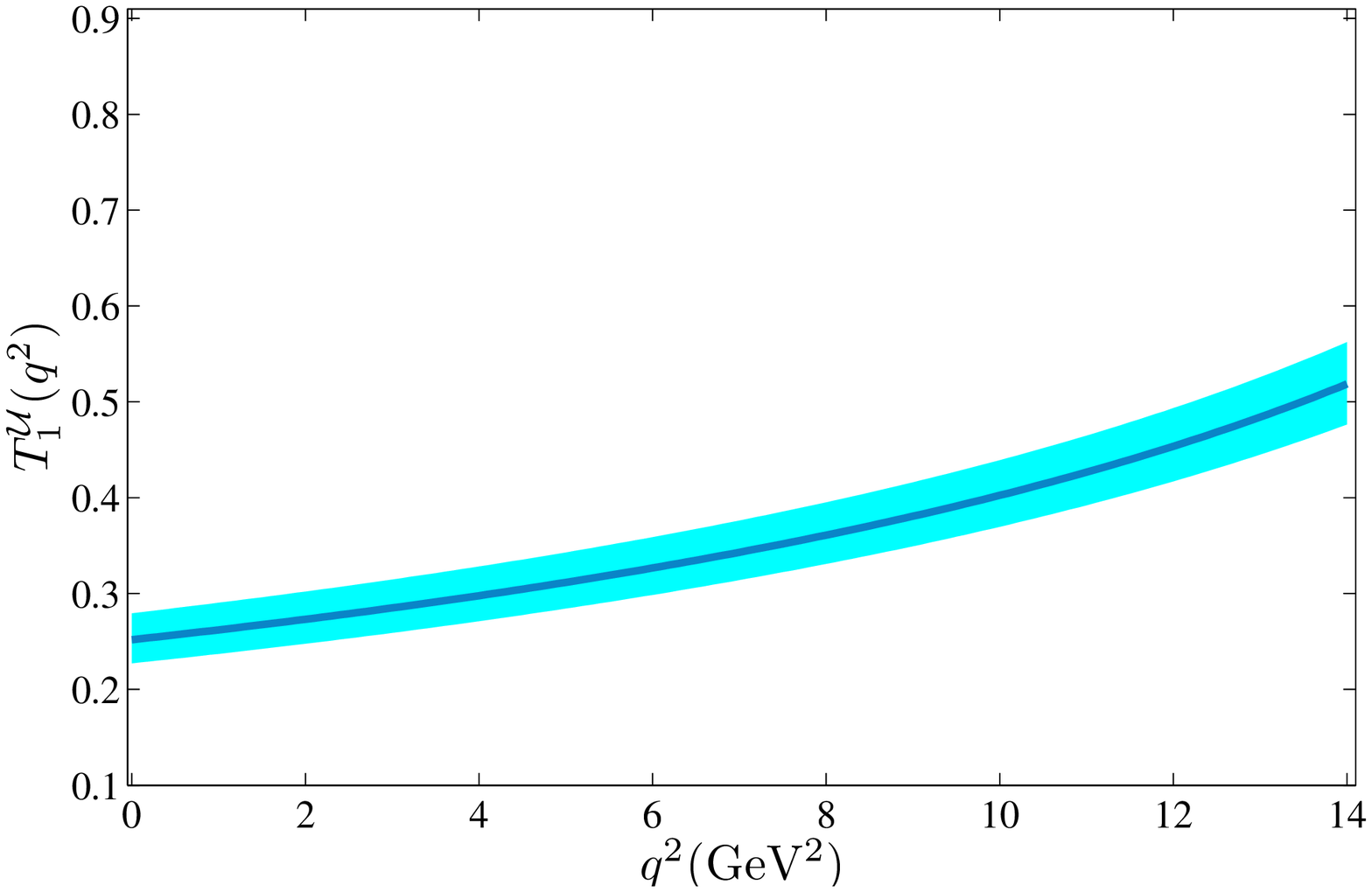}
\includegraphics[width=0.25\textwidth]{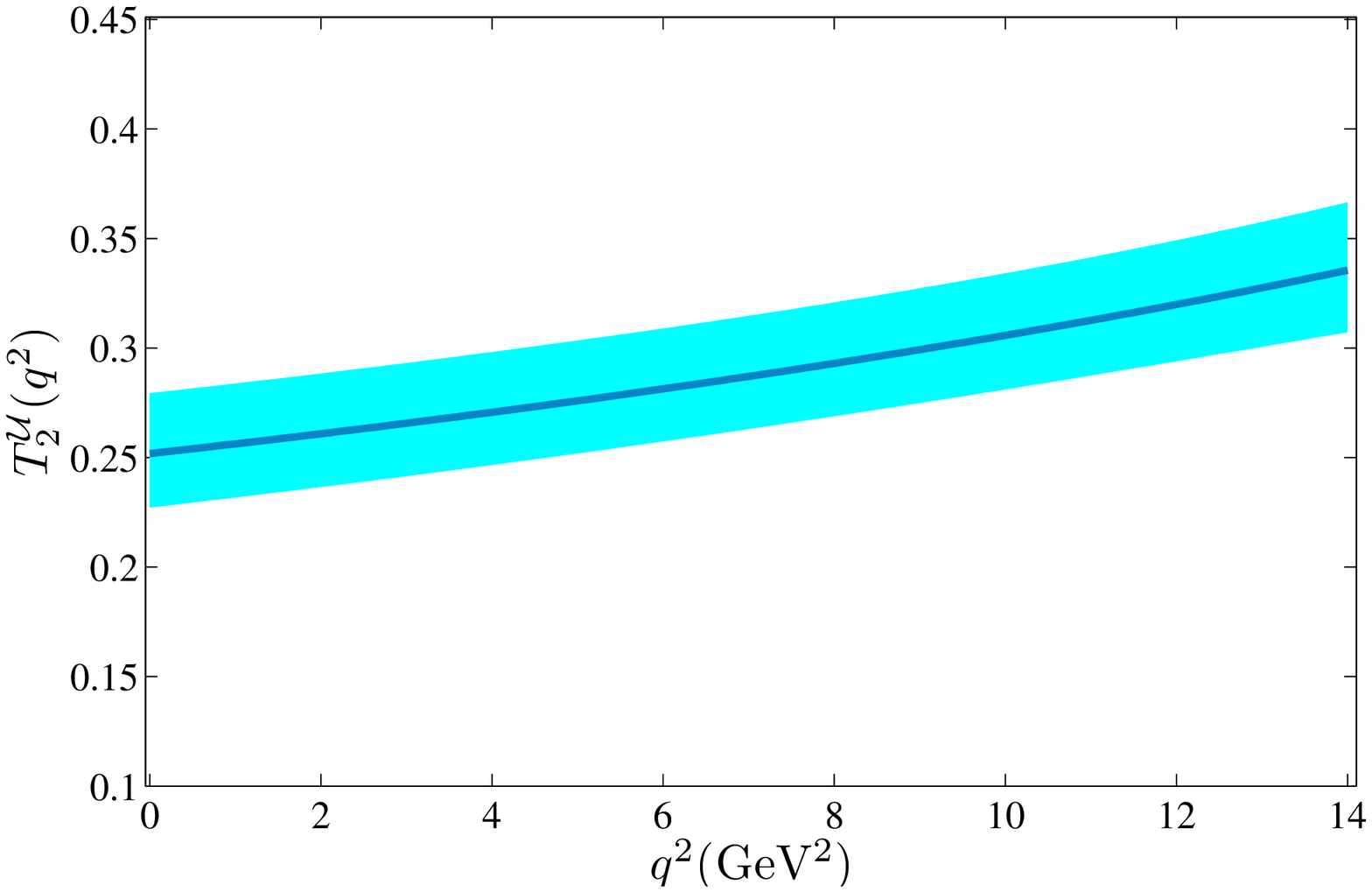}
\includegraphics[width=0.25\textwidth]{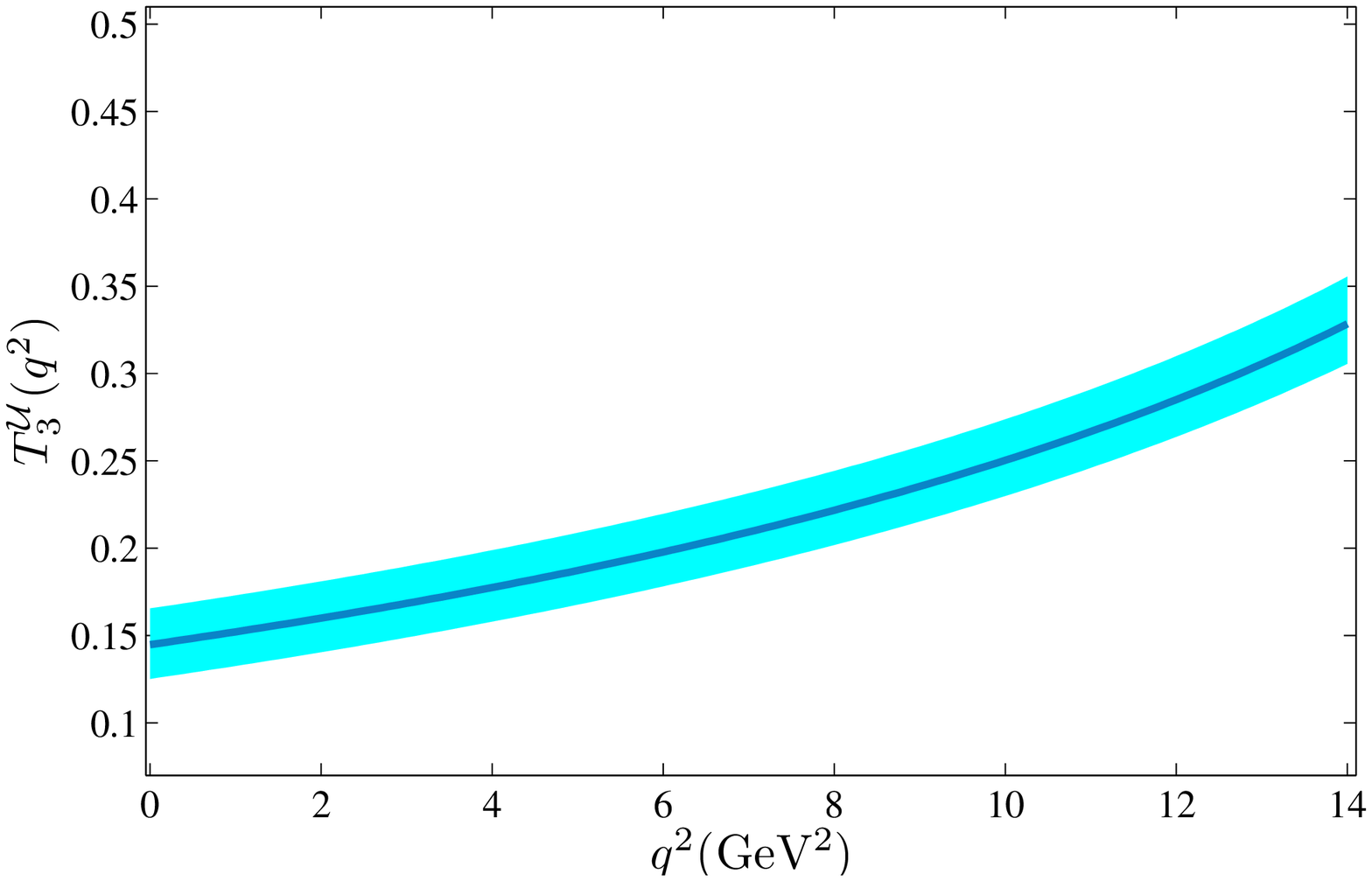}
\includegraphics[width=0.25\textwidth]{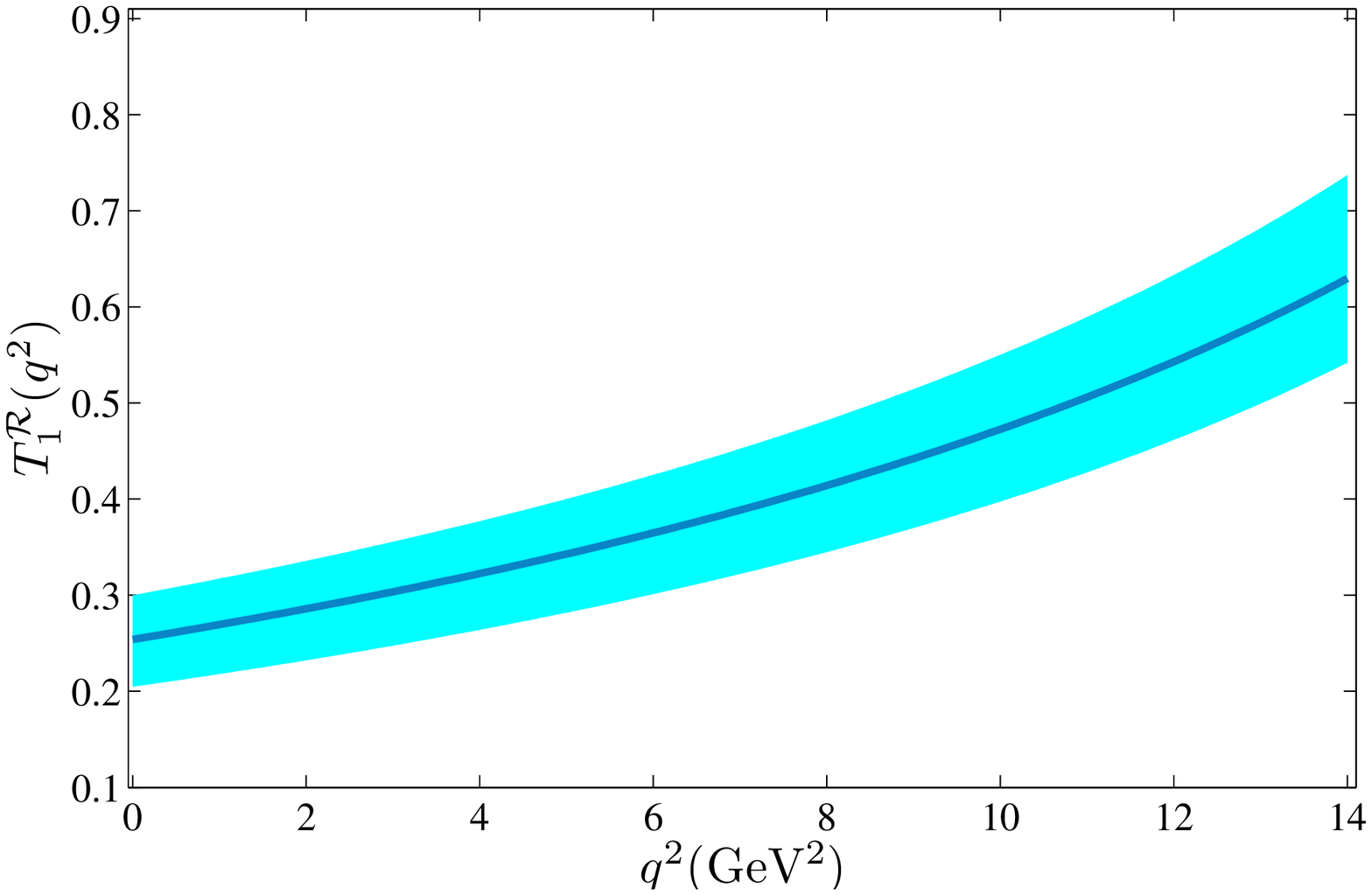}
\includegraphics[width=0.25\textwidth]{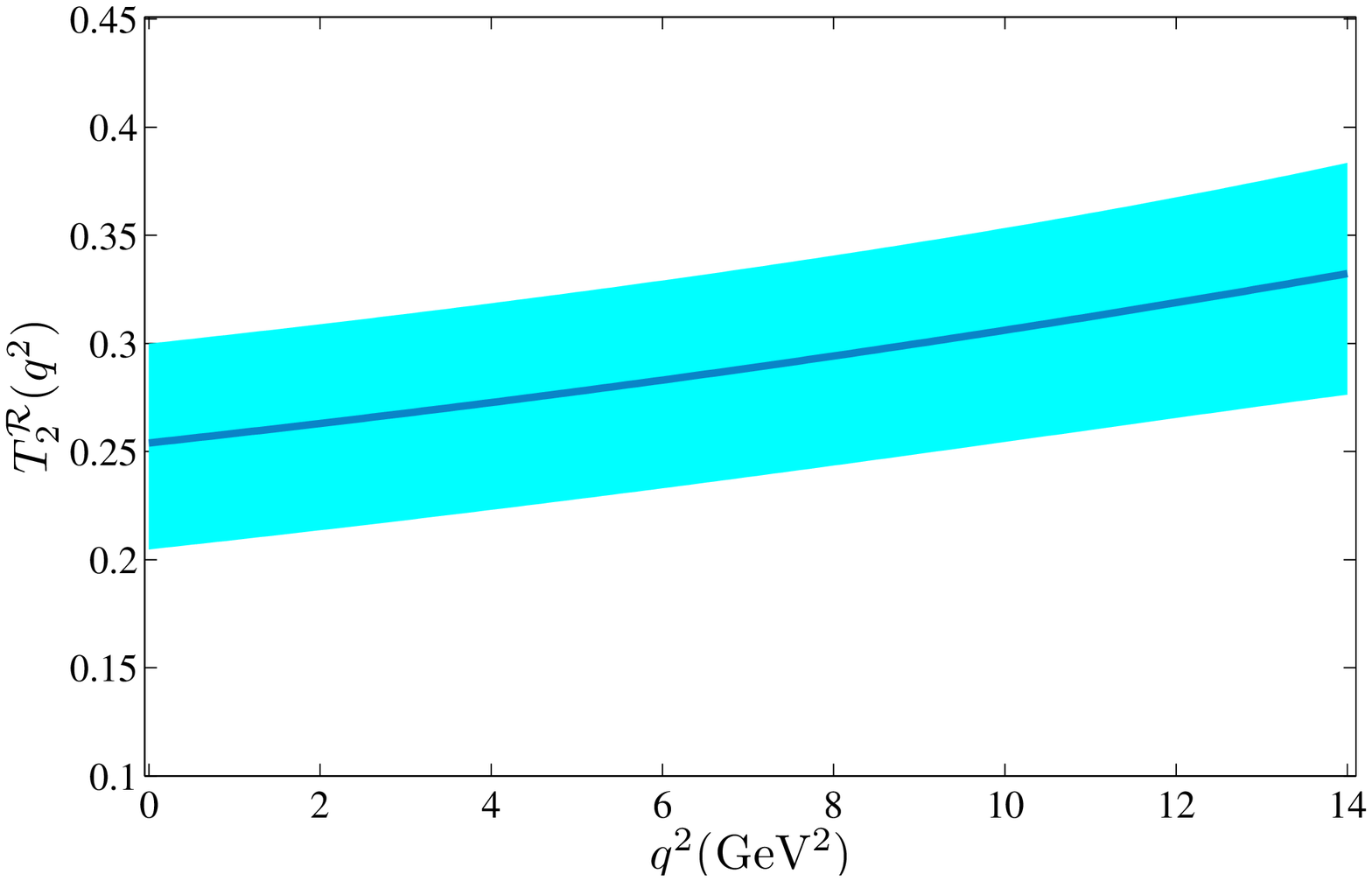}
\includegraphics[width=0.25\textwidth]{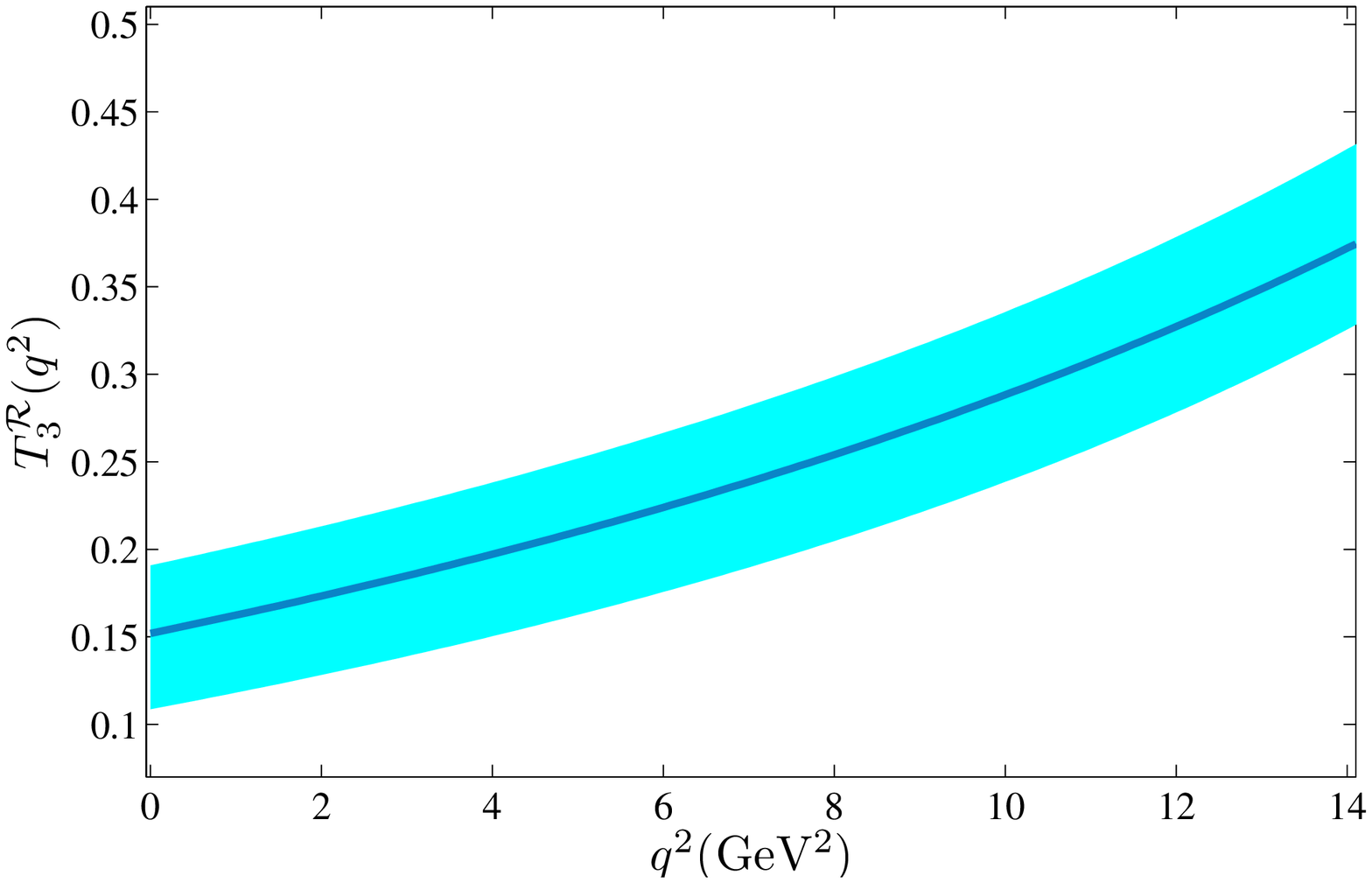}
\caption{The $B\to K^*$ TFFs $T_{1,2,3}(q^2)$ for the right-handed chiral correlator (Lower ones) and the usual correlator (Upper ones), respectively. The solid lines are central values and the shaded bands are their errors.}
\label{TFF:T114T214T314T3W14}
\end{figure*}

\begin{figure*}[htb]
\centering
\includegraphics[width=0.24\textwidth]{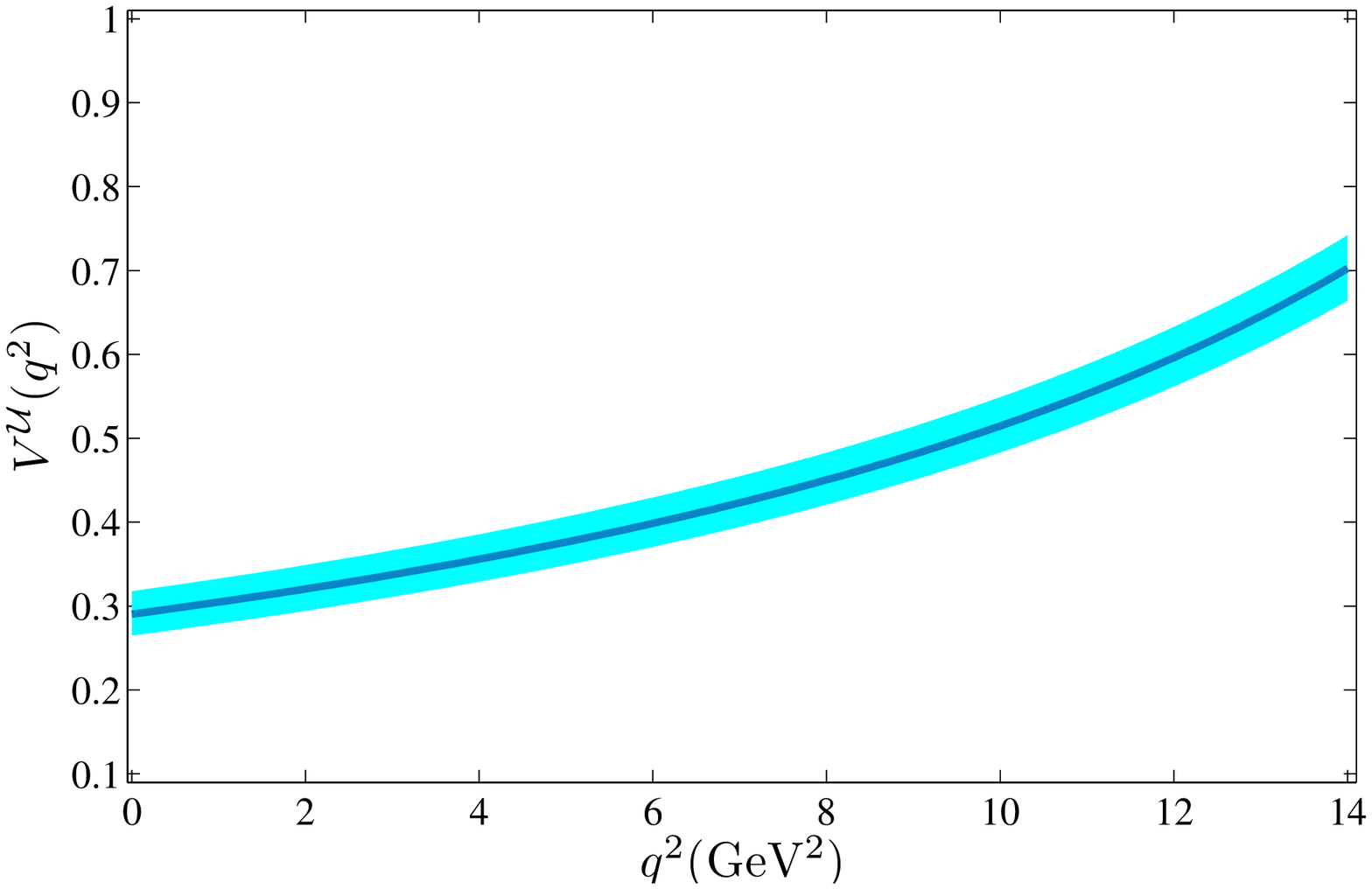}
\includegraphics[width=0.24\textwidth]{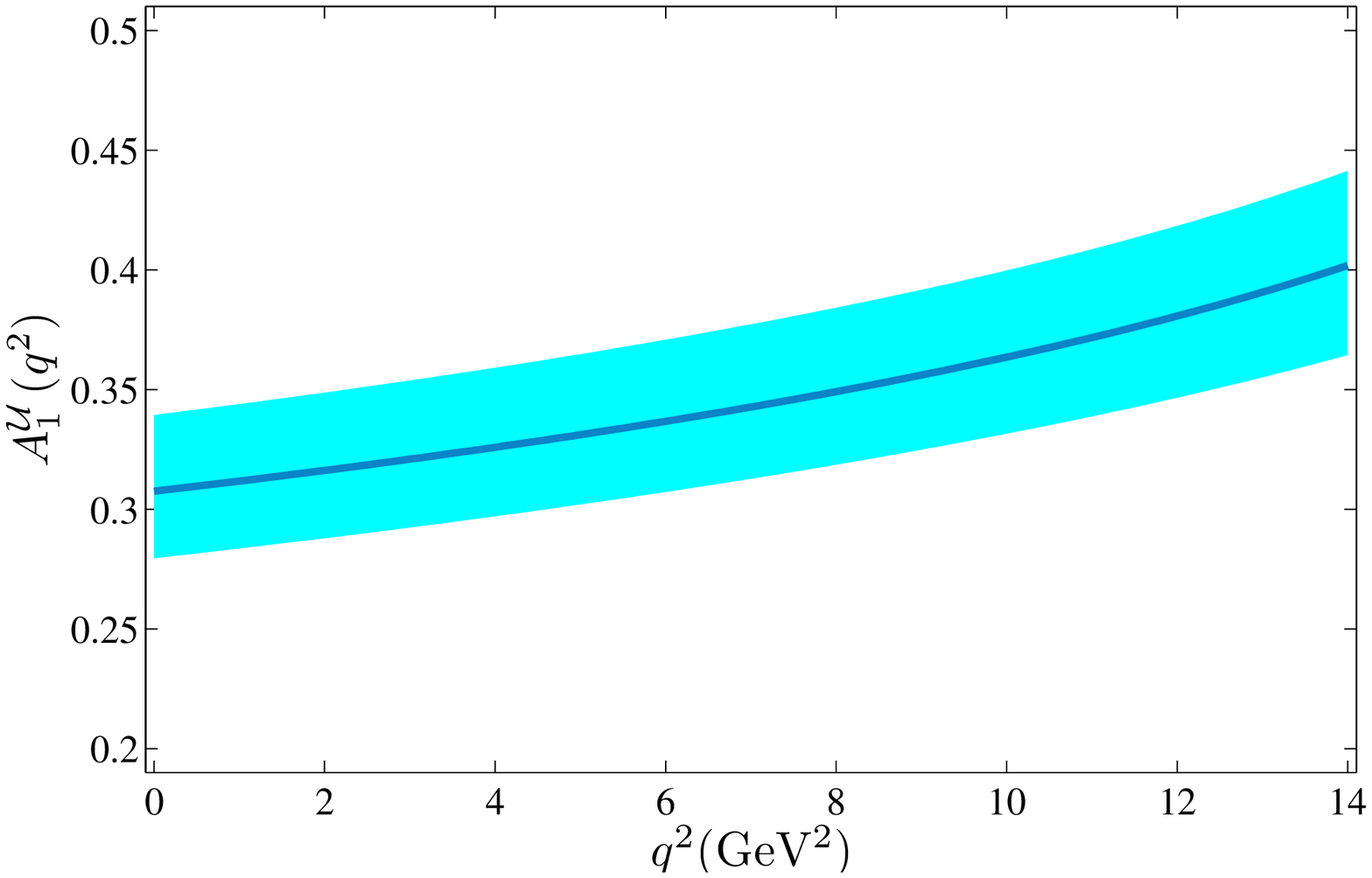}
\includegraphics[width=0.24\textwidth]{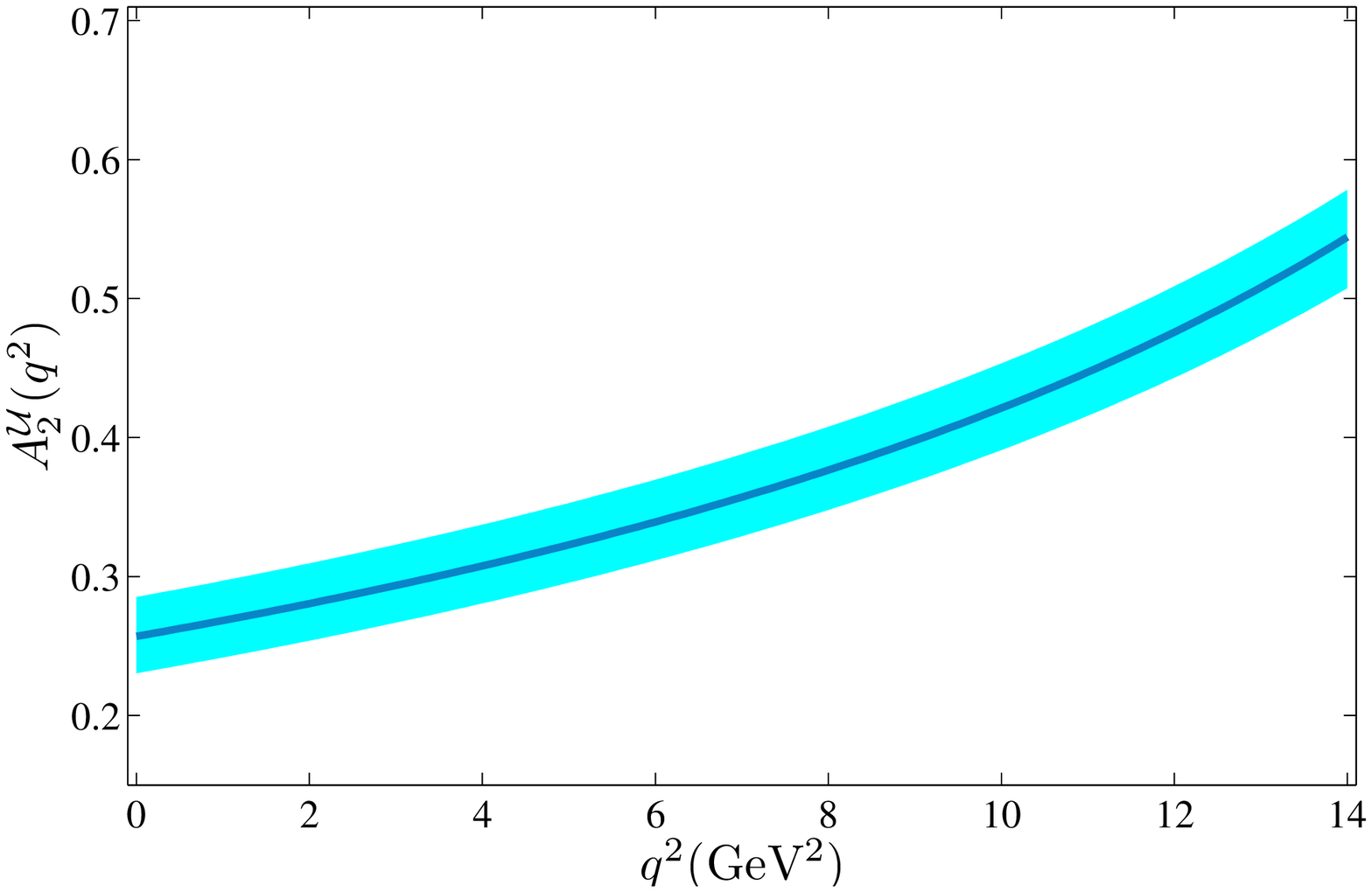}
\includegraphics[width=0.24\textwidth]{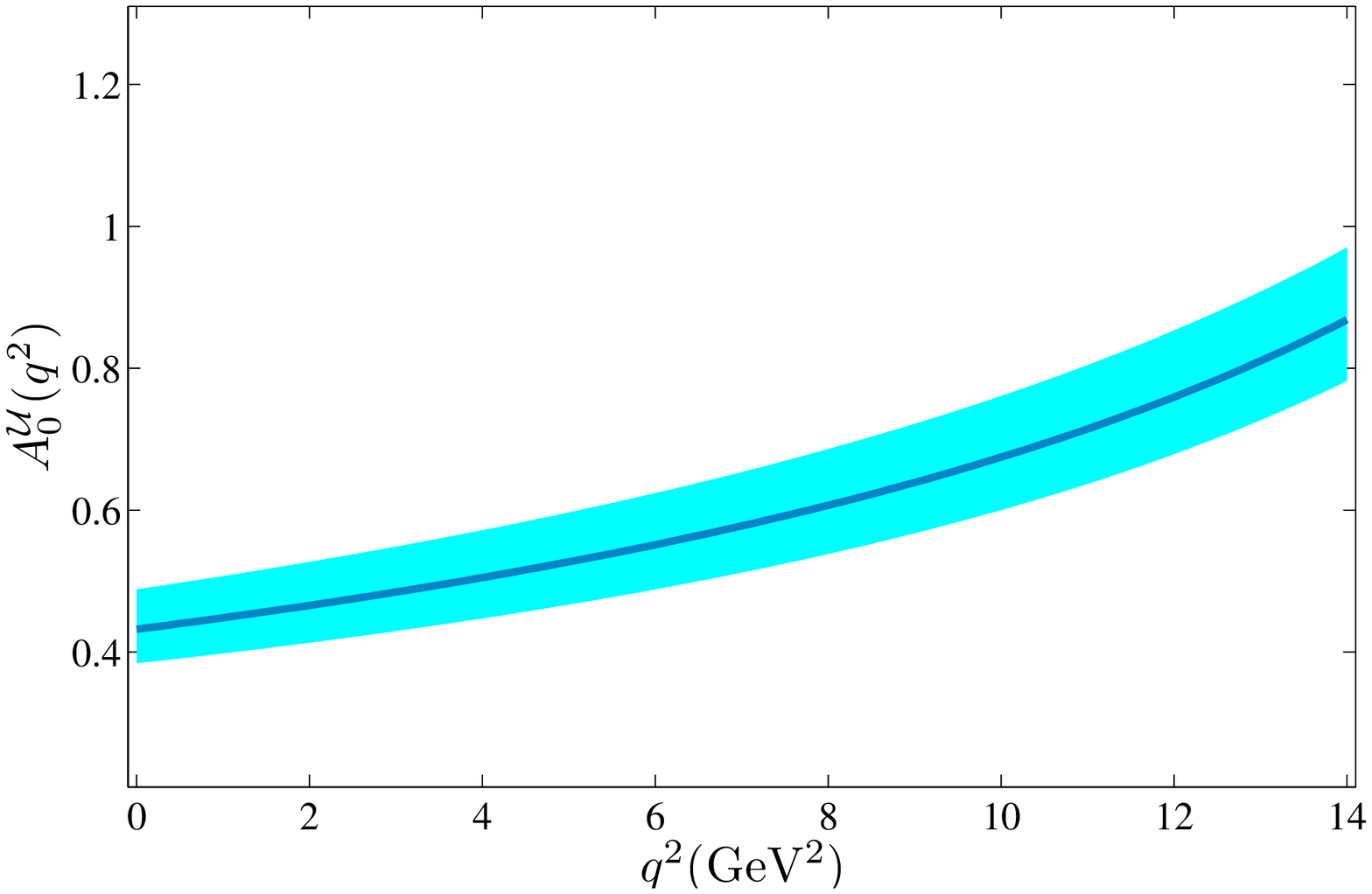}
\includegraphics[width=0.24\textwidth]{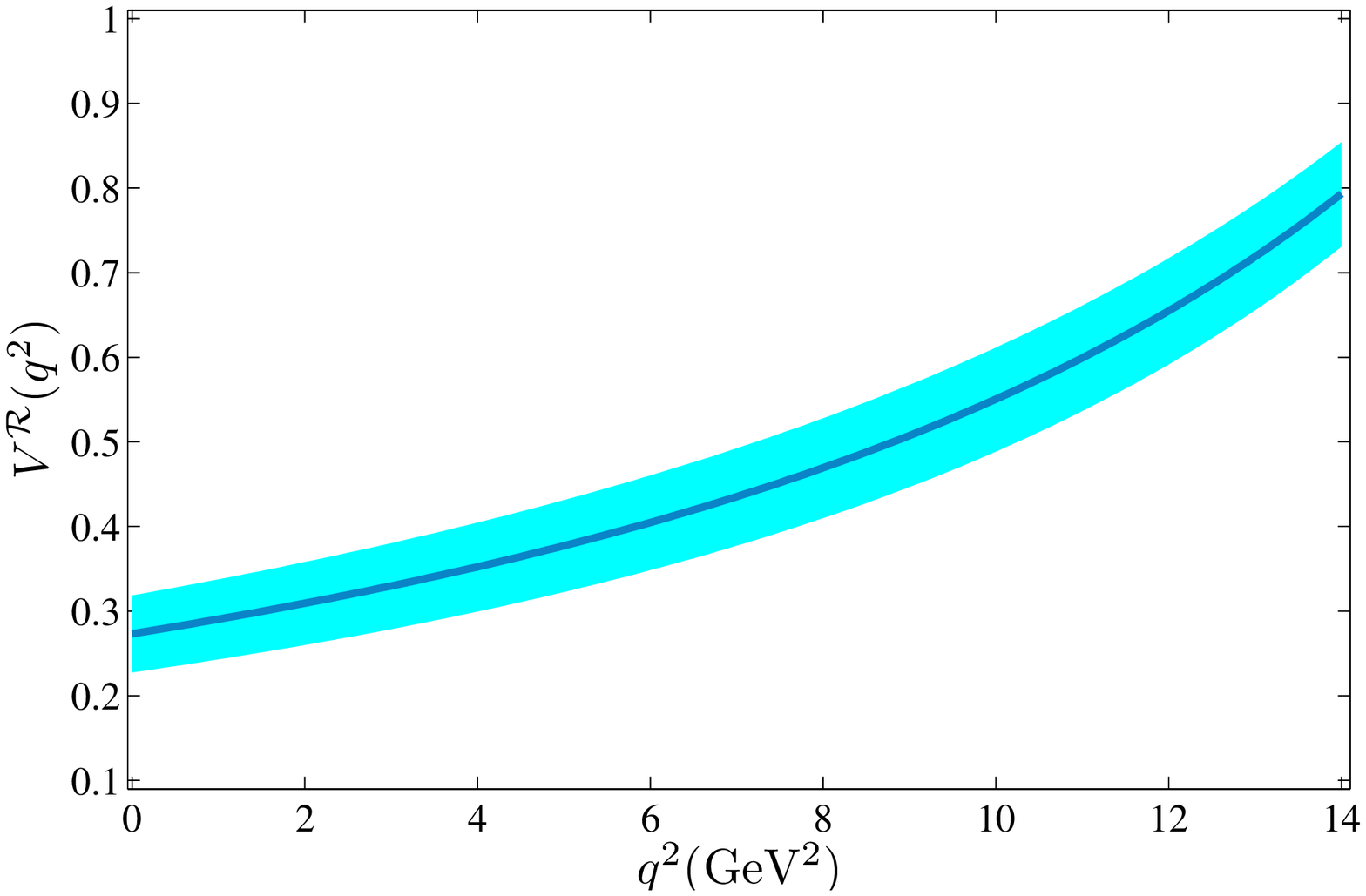}
\includegraphics[width=0.24\textwidth]{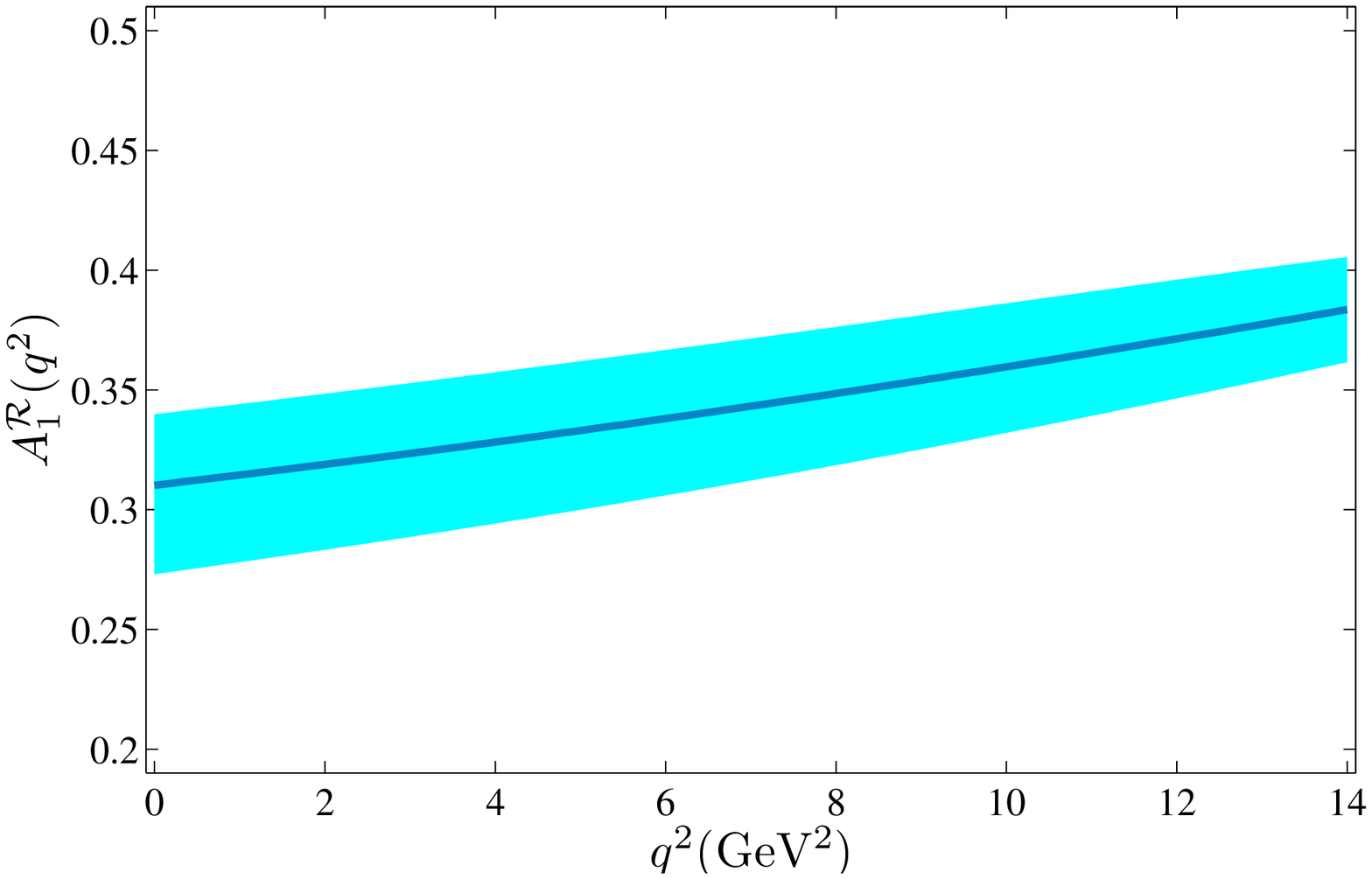}
\includegraphics[width=0.24\textwidth]{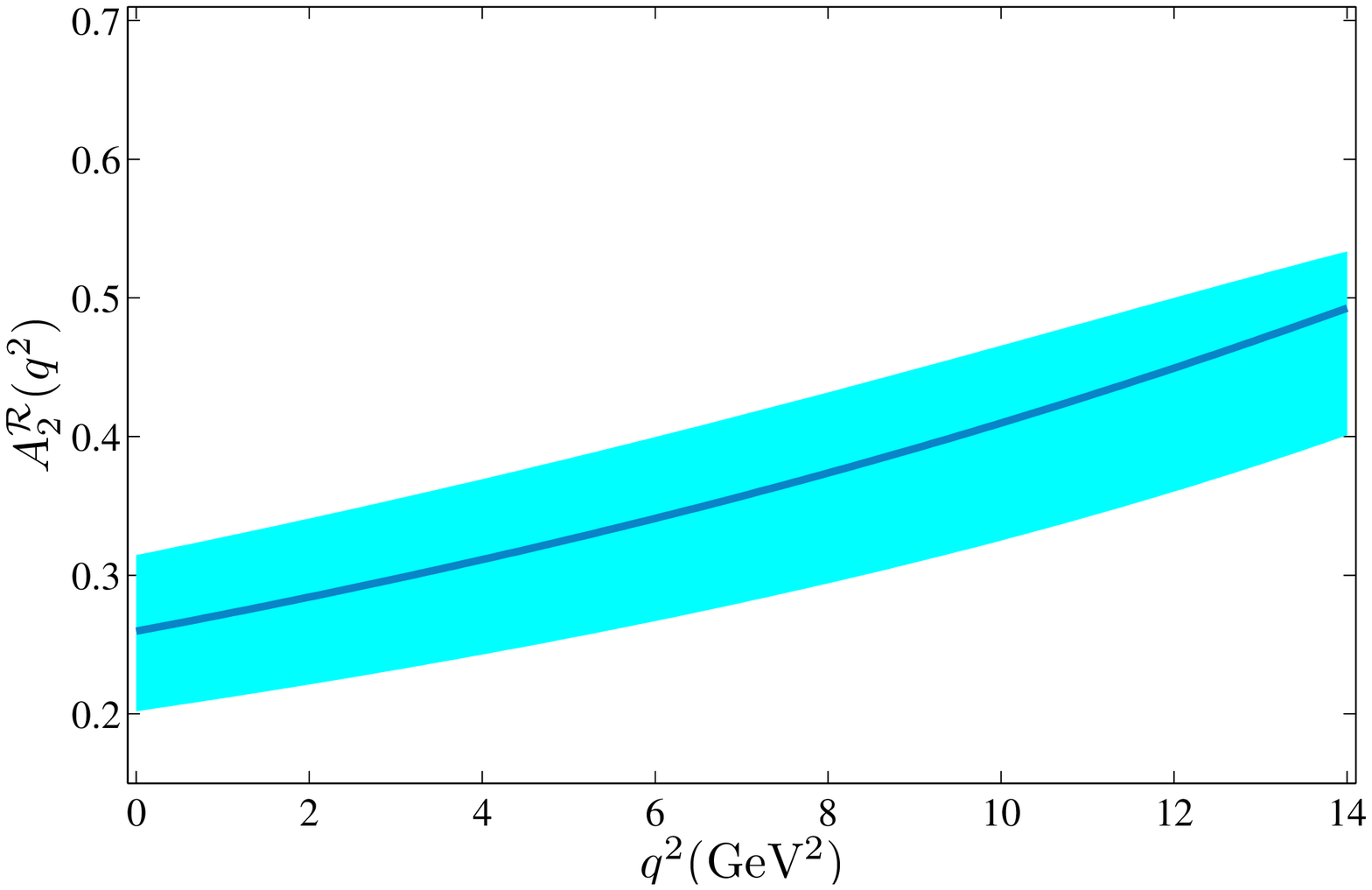}
\includegraphics[width=0.24\textwidth]{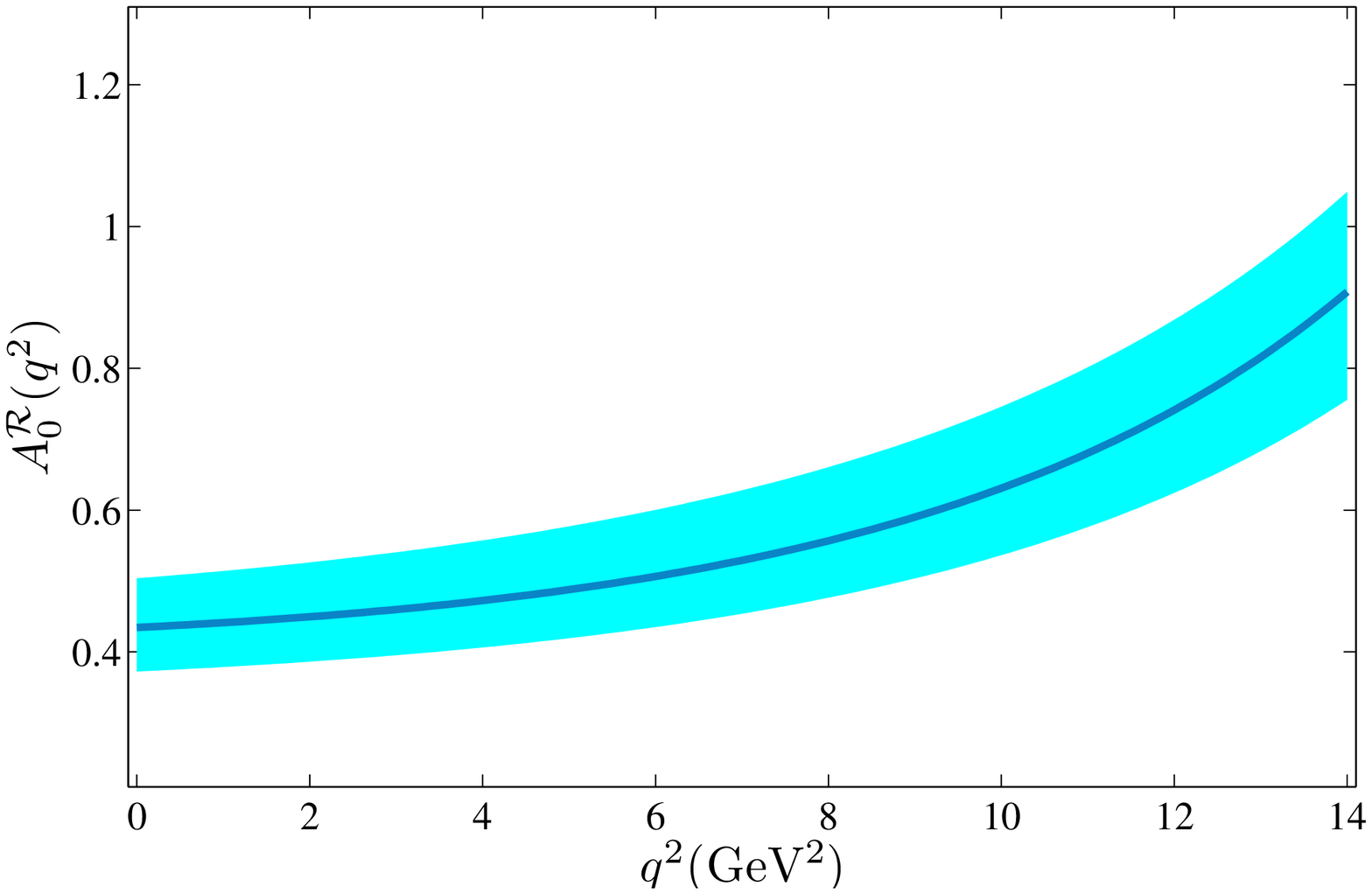}
\caption{The $B\to K^*$ axial-vector and vector TFFs $A_{0,1,2}(q^2)$ and $V(q^2)$ for the right-handed chiral correlator (Lower ones) and the usual correlator (Upper ones), respectively. The solid lines are central values and the shaded bands are their errors. }
\label{TFF:A014A114A214V14}
\end{figure*}

We present the sum rules for the $B\to K^*$ TFFs $T_{1,2,3}(q^2)$, $A_{0,1,2}(q^2)$ and $V(q^2)$ for the right-handed chiral correlator (LCSR-${\cal R}$) and for the usual correlator (LCSR-${\cal U}$) in Figs.(\ref{TFF:T114T214T314T3W14}, \ref{TFF:A014A114A214V14}), in which the solid line stands for its central value and the shaded band is the theoretical error. The error is squared average of errors caused by all the mentioned error sources, e.g. we adopt $\Delta M^2_{{\cal R}/{\cal U}}=\pm0.5{\rm GeV}^2$ and $\Delta s_{0,{\cal R}/{\cal U}}= \pm 0.5{\rm GeV}^2$~\footnote{The TFFs changes very slightly by taking $\Delta M^2_{{\cal R}/{\cal U}}=\pm0.5{\rm GeV}^2$, which is still $\sim3\%$ by setting $\Delta M^2_{{\cal R}/{\cal U}}=\pm1.0{\rm GeV}^2$. Thus our predictions are consistent with the usual flatness criterion for determining the Borel window~\cite{Colangelo:2000dp}.}. Figs.(\ref{TFF:T114T214T314T3W14}, \ref{TFF:A014A114A214V14}) indicate that all the TFFs increase with the increment of $q^2$. We present the LCSRs together with their errors at the large recoil region $q^2 \to 0$ in Table~\ref{F0}. As a comparison, the Ball and Zwicky (BZ) prediction~\cite{Ball:2004rg}, the AdS/QCD prediction~\cite{Ahmady:2014sva}, and the LCSR prediction~\cite{AKhodjamirian:2010} are also presented. Those TFFs are consistent with each other within errors.

\begin{table*}[htb]
\begin{tabular}{ c  c  c c c c c}
\hline
 &~~~$A_1(0)$~~~&~~~$A_2(0)$~~~&~~~$V(0)$~~~&~~$T_1(0)[T_2(0), \widetilde T_3(0)]$~~&~~~$T_3(0)$~~\\
\hline
LCSR-$\cal{R}$                & $0.310^{+0.030}_{-0.037}$       & $0.260^{+0.055}_{-0.058}$     & $0.332^{+0.051}_{-0.051}$    &  $0.254^{+0.046}_{-0.049}$                                &  $0.152^{+0.039}_{-0.043}$    \\
LCSR-$\cal{U}$                & $0.308^{+0.032}_{-0.028}$       & $0.257^{+0.028}_{-0.026}$     & $0.307^{+0.024}_{-0.023}$    &  $0.251^{+0.028}_{-0.024}$                                &  $0.145^{+0.020}_{-0.020}$     \\
LCSR~\cite{AKhodjamirian:2010}          & $0.25^{+0.16}_{-0.10}$       & $0.23^{+0.19}_{-0.10}$     & $0.36^{+0.23}_{-0.12}$    &  $0.31^{+0.18}_{-0.10}$                                &  $0.22^{+0.17}_{-0.10}$     \\
BZ~\cite{Ball:2004rg}         & $0.292 \pm 0.028$       & $0.259 \pm 0.027$     & $0.411 \pm 0.033$    &  $0.333 \pm 0.028$                                & $0.202 \pm 0.018$      \\
AdS~\cite{Ahmady:2014sva} & 0.249       & 0.235     & 0.277    &  0.255                                & 0.155      \\
\hline
\end{tabular}
\caption{The $B\to {K^*}$ TFFs at $q^2=0$. As a comparison, the Ball and Zwicky (BZ) prediction~\cite{Ball:2004rg}, the AdS/QCD prediction~\cite{Ahmady:2014sva}, and the LCSR prediction~\cite{AKhodjamirian:2010} are also presented. $\mu=2.2$ GeV. } \label{F0}
\end{table*}

A smaller Borel parameter indicates a larger $M^2$-dependance due to a weaker convergence over $1/M^2$, and a larger $M^2$-uncertainty could be observed. This explains why a larger $M^2$-uncertainty than our present one is observed in Ref.\cite{AKhodjamirian:2010}, whose Borel parameter is taken as $M^2=1.00 \pm 0.25 {\rm GeV}^2$ by using a rough scaling relation, $M^2\sim 2m_b \tau\sim 1 {\rm GeV}$~\cite{Shifman:1978bx, Shifman:1978by, Pietro Colangelo:2000}, which is much smaller than the $M^2$-values shown by Table~\ref{Borel}.

\begin{table}[htb]
\centering
\begin{tabular}{ c  c  c  c c c }
\hline
& &~~Twist$-2$~~&~~Twist$-3$~~&~~Twist$-4$~~&~~Total~~\\
\hline
&$A_1^{\cal R}$&$1.607$&$-0.029$&$-1.267$&$0.310$\\
&$A_1^{\cal U}$&$0.433$&$0.088$&$-0.214$&$0.308$\\
\hline
&$A_2^{\cal R}$&$0.810$&$-0.634$&$0.084$&$0.260$\\
&$A_2^{\cal U}$&$0.404$&$-0.137$&$-0.010$&$0.257$\\
\hline
&$V^{\cal R}$&$0.359$&$0$&$-0.027$&$0.332$\\
&$V^{\cal U}$&$0.207$&$0.119$&$-0.019$&$0.307$\\
\hline
&$T_{1}^{\cal R}[T_{2}^{\cal R},\widetilde T_{3}^{\cal R}]$&$0.505$&$-0.127$&$-0.124$&$0.254$\\
&$T_{1}^{\cal U}[T_{2}^{\cal U},\widetilde T_{3}^{ \cal U}]$& $0.253$&$0.002$&$-0.004$&$0.251$\\
\hline
&$T_3^{\cal R}$&$0.384$&$-0.275$&$0.043$&0.152\\
&$T_3^{\cal U}$&$0.339$&$-0.199$&$0.005$&0.145\\
\hline
\end{tabular}
\caption{Different twist contributions for the $B\to K^*$ TFFs. The results for LCSR-${\cal R}$ and LCSR-${\cal U}$ are presented. } \label{Twist}
\end{table}

We present the contributions from the $K^*$-meson LCDAs up to twist-4 in Table~\ref{Twist}. For the LCSR-${\cal U}$ of the usual correlator, the relative importance among different twist LCDAs follows the trends, twist-2 $>$ twist-3 $>$ twist-4; For the LCSR-${\cal R}$ of the right-handed chiral correlator, we have, twist-2 $\gg$ twist-3 $\sim$ twist-4. The dominance of the twist-2 term indicates a more convergent twist-expansion could be achieved by using the chiral correlator. In Table~\ref{Twist}, a somewhat larger twist-4 contribution is observed for $A^{{\cal R}/{\cal U}}_1$ and $T^{{\cal R}/{\cal U}}_1$, which comes from the twist-4 LCDA $\psi_{4;{K^*}}^{\bot}$ in the reduced function $H_3=\int_0^u dv \left[\psi_{4;{K^*}}^{\bot}(v)- \phi_{2;{K^*}}^\bot(v)\right]$; because of large suppression from the twist-2 LCDA $\phi_{2;{K^*}}^\bot$, the net contribution of $H_3$ is small, which is about $0.5\%$ of the twist-2 ones. Except for $H_3$, the remaining twist-4 contributions are still about $10\%$ of the twist-2 ones for $A^{{\cal R}/{\cal U}}_{1/2}$, $V^{{\cal R}/{\cal U}}$ and $T^{{\cal R}/{\cal U}}_1$, thus the twist-4 terms are important and should be kept for a sound prediction.

\begin{table}[htb]
\centering
\begin{tabular}{ c  c  c  c c }
\hline
& &~~$\mu=1.2{\rm GeV}$~~&~~$\mu=2.2{\rm GeV}$~~&~~$\mu=3.2{\rm GeV}$~~\\
\hline
&$A_1^{\cal R}$&0.307&0.310&0.310\\
                         &$A_1^{\cal U}$&0.301&0.308&0.308\\\hline
&$A_2^{\cal R}$&0.266&0.260&0.260\\
                         &$A_2^{\cal U}$&0.253&0.257&0.257\\\hline
&$V^{\cal R}$&0.329&0.332&0.332\\
                         &$V^{\cal U}$&0.302&0.307&0.307\\\hline
&$T_{1}^{\cal R}[T_{2}^{\cal R},\widetilde T_{3}^{\cal R}]$&0.254&0.254&0.254\\
                         &$T_{1}^{\cal U}[T_{2}^{\cal U},\widetilde T_{3}^{ \cal U}]$&0.247&0.251&0.252 \\ \hline
&$T_3^{\cal R}$&0.152&0.152&0.153\\
                         &$T_3^{\cal U}$&0.142&0.145&0.145\\
\hline
\end{tabular}
\caption{The factorization scale dependence of the $B \to K^*$ TFFs at large recoil region. $\mu = (2.2 \pm 1.0)$ GeV. }
\label{F0 mu1}
\end{table}

As shown by Table~\ref{F0 mu1}, the factorization scale dependence is small for all the $B\to K^*$ TFFs, e.g. less than $3\%$ by taking $\mu = (2.2 \pm 1.0)$ GeV. Table~\ref{F0 mu1} shows when setting $\mu>2.2$ GeV, TFFs are almost unchanged. This negligible dependence for larger scale value is consistent with the fact that the $K^*$ LCDAs change slightly when running from $2.2$ GeV to higher value via the renormalization group evolution~\cite{Ball:2006nr}.

\subsection{An extrapolation of the TFFs and the correlation coefficient $\rho_{XY}$ for the two LCSRs}

\begin{table*}[htb]
\centering
\begin{tabular}{ c  c  c c c c c c c}
\hline
&~~~~~$A_1$~~~~~&~~~~~$A_2$~~~~~&~~~~~$V$~~~~~&~~~~~$A_0$~~~~~ &~~~~~$T_1$~~~~~&~~~~~$T_2$~~~~~&~~~~~$T_3$~~~~~\\
\hline
 $a_{1,\cal R}^i$       &1.058               &$-0.382$      &$-1.025$        & 1.477       &$-0.900$       &0.730          &$-0.714$    \\
 $a_{2,\cal R}^i$       &0.130            &$-5.008$      &0.318        &14.238       &$-3.330$       &$-0.399$        &$-3.715$    \\
 $\Delta_{\cal R}$      &0.9    &1.0  &0.03 & 1.5      &0.4          &2.8         &2.2   \\
 $a_{1,\cal U}^i$       &1.059               &$-0.275$      &$-0.531$        & $-0.019$        &$-0.136$       &0.719          &$-0.294$    \\
 $a_{2,\cal U}^i$       &1.031           &$-1.339$      &$-0.115$         & $-0.169$        &$-0.708$        &$-0.205$       &$-1.144$    \\
 $\Delta_{\cal U}$     &0.1             &0.2       &0.3         & 1.2      &0.2          &0.3       &0.1   \\
\hline
\end{tabular}
\caption{The fitted parameters $a_{\{1,2\},\{\cal R,\cal U\}}^i$ for the $B \to K^*$ TFFs, where all the LCSR parameters are set to be their central values. $\Delta_{\cal R}$ and $\Delta_{\cal U}$ are the qualities of fits for the right-handed correlator and the usual correlator, respectively. } \label{F0dalta}
\end{table*}

\begin{figure*}[htb]
\centering
\includegraphics[width=0.3\textwidth]{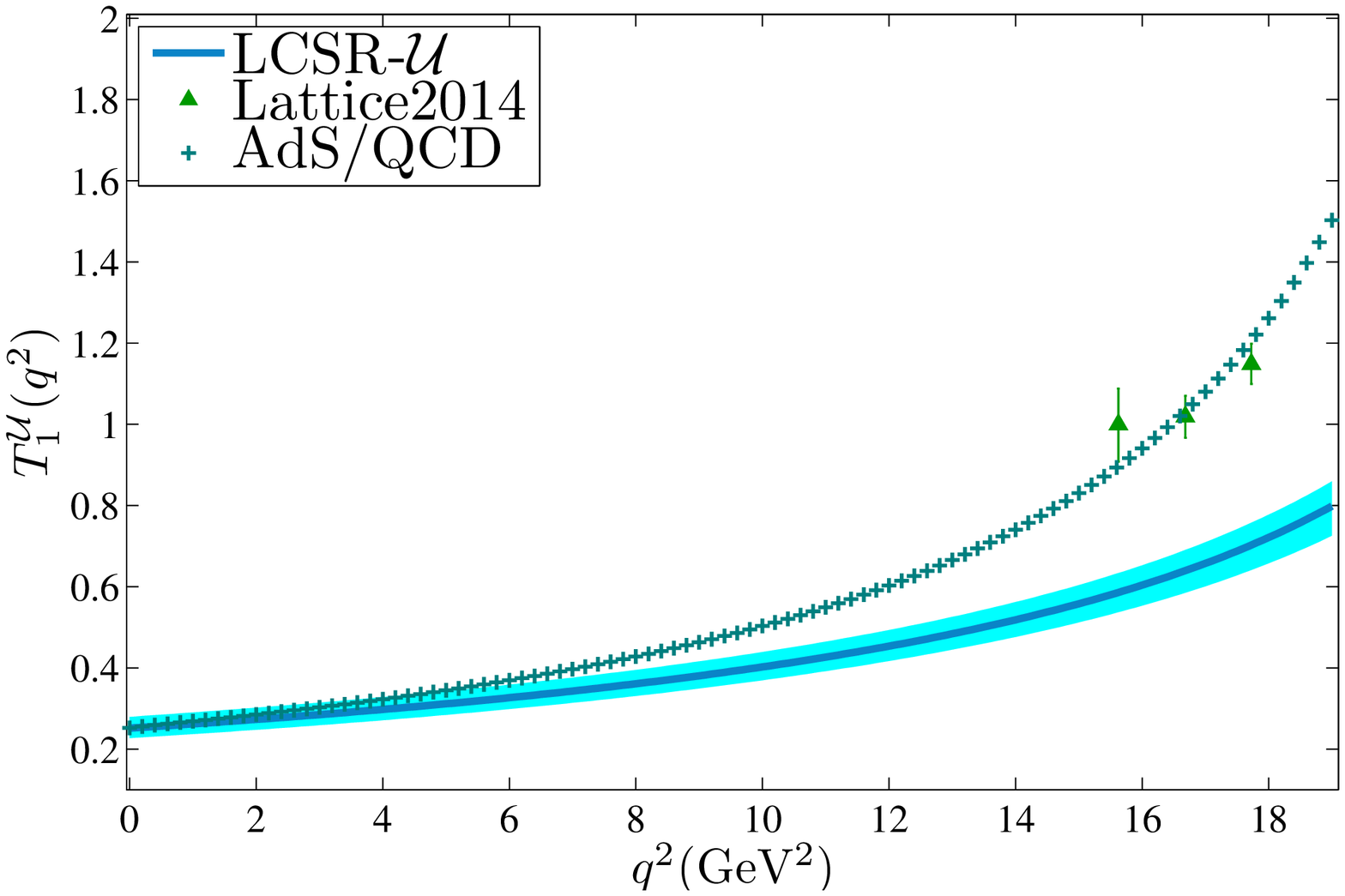}
\includegraphics[width=0.3\textwidth]{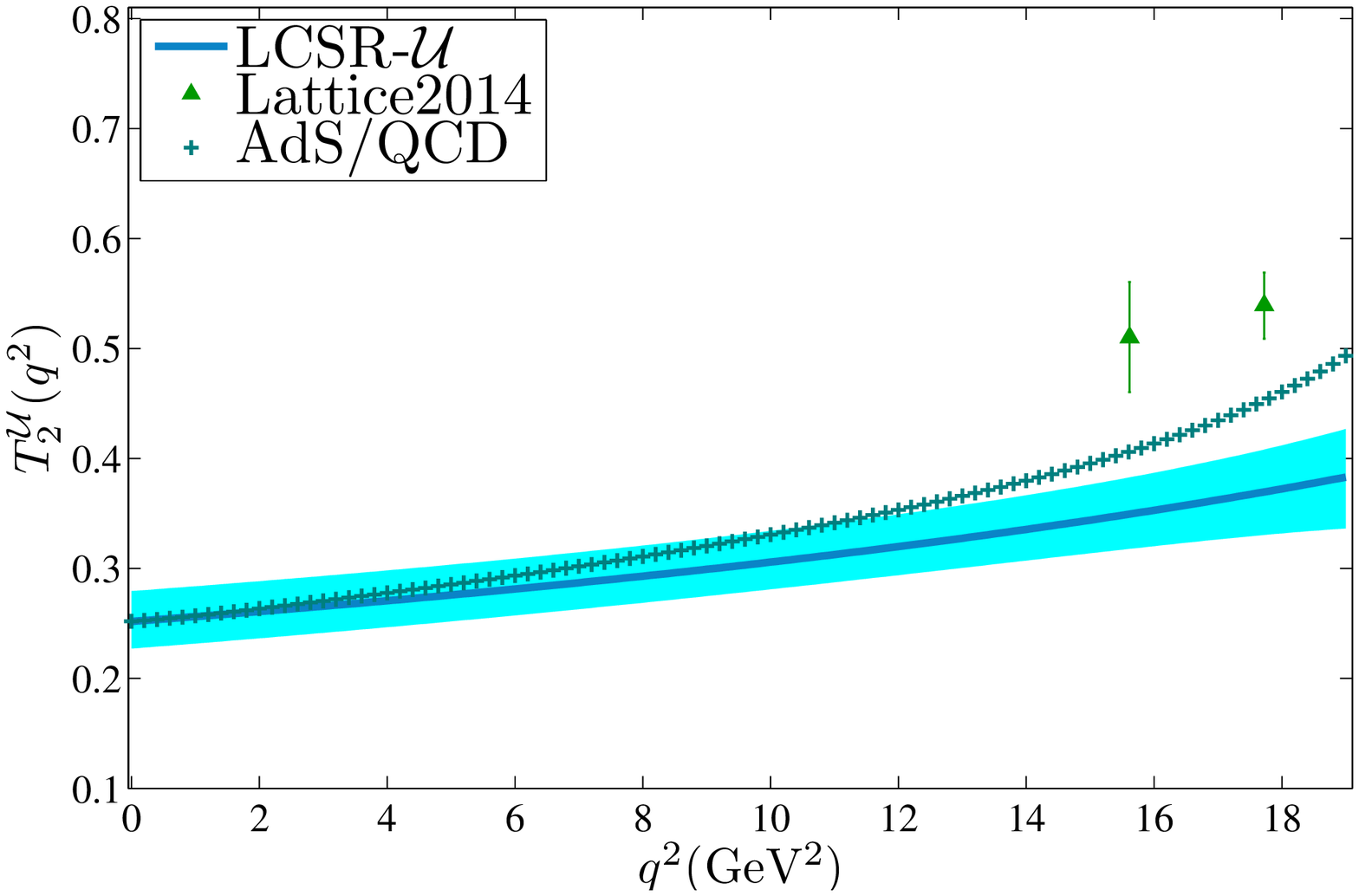}
\includegraphics[width=0.3\textwidth]{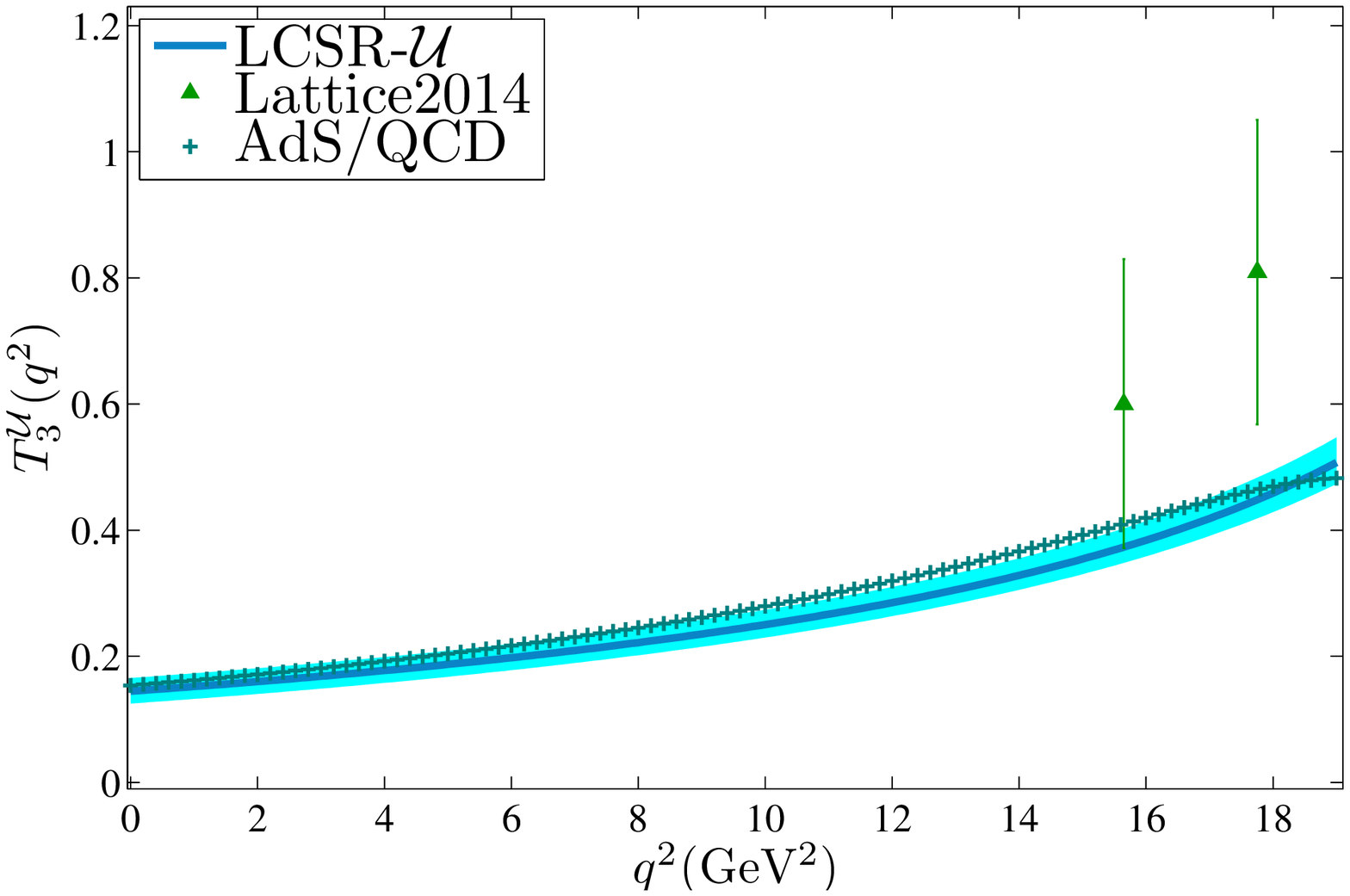}
\includegraphics[width=0.3\textwidth]{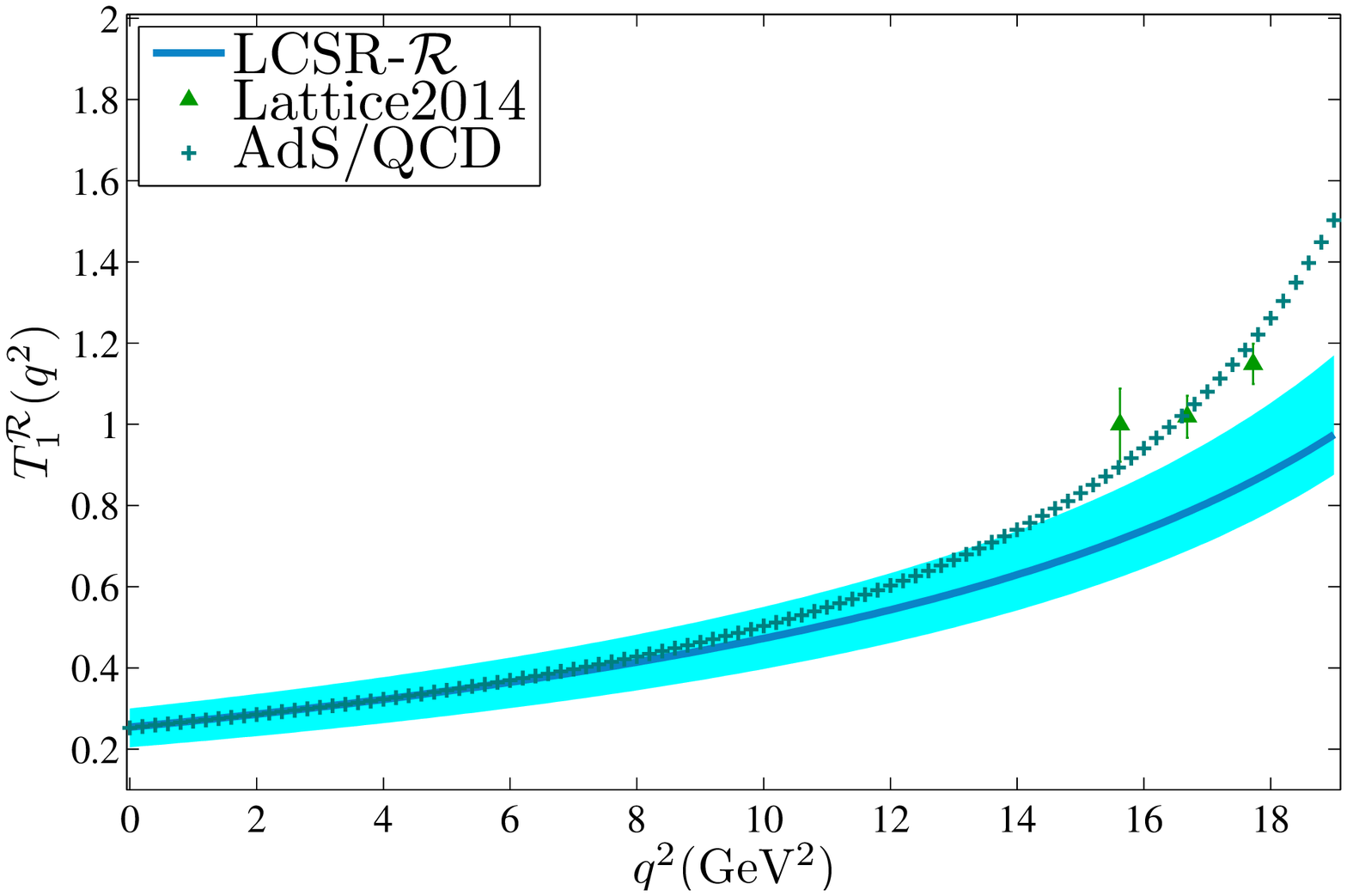}
\includegraphics[width=0.3\textwidth]{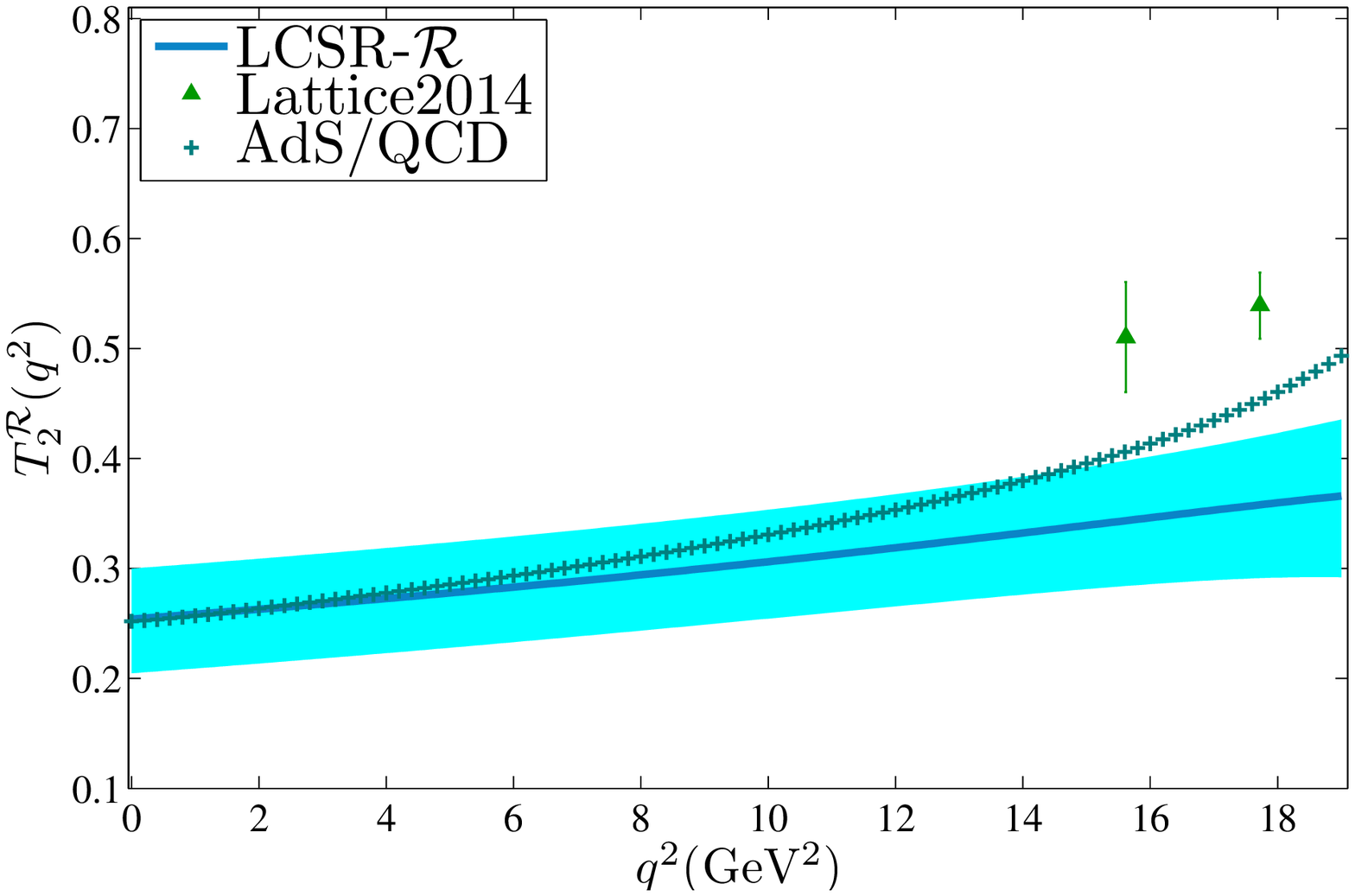}
\includegraphics[width=0.3\textwidth]{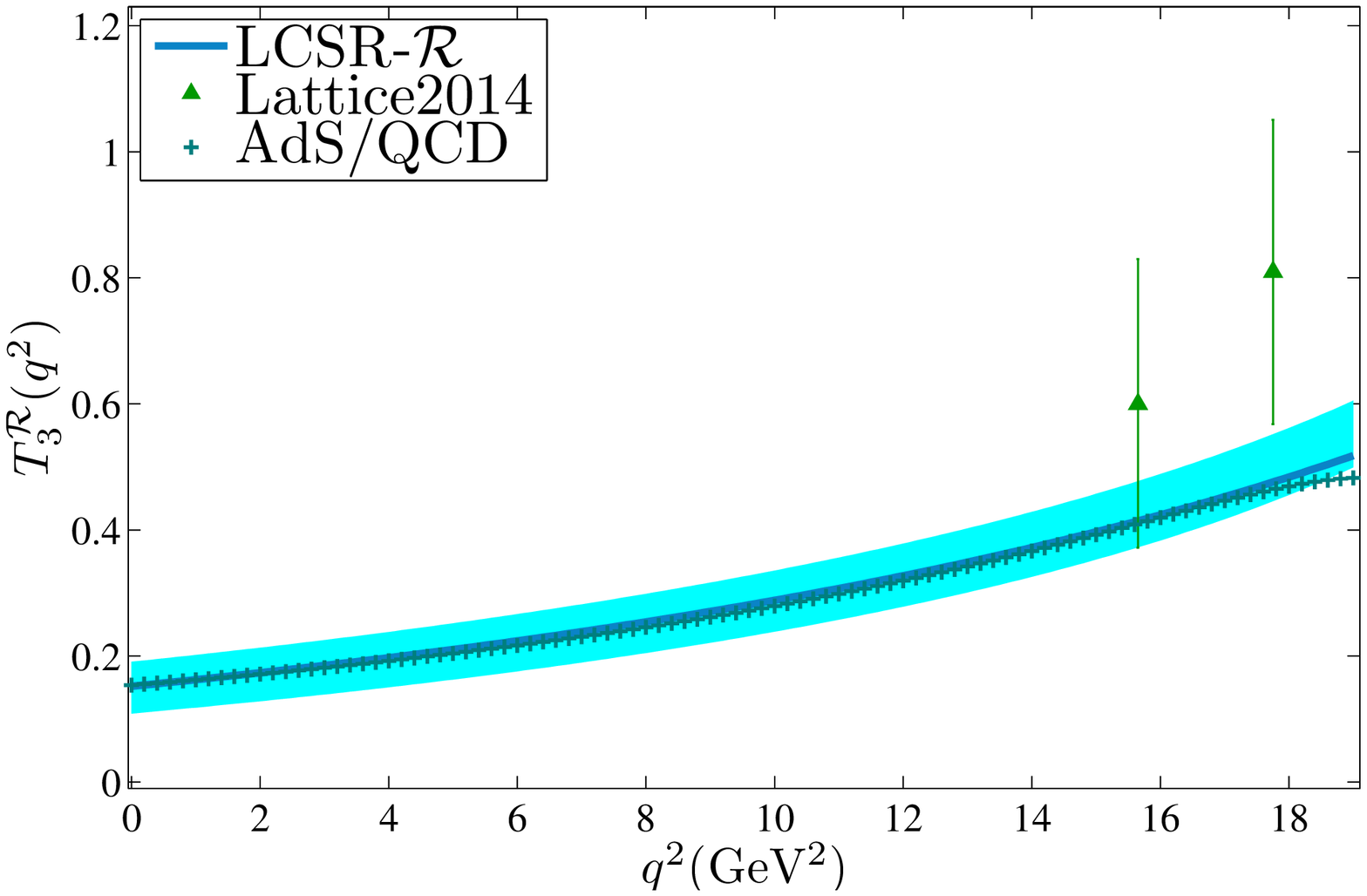}
\caption{The extrapolated $B\to K^*$ tensor TFFs $T_{1,2,3}(q^2)$. The left and right figures stand for LCSR with the usual and right current, respectively. The solid lines are central values and the shaded bands are their errors. As a comparison, the AdS/QCD~\cite{Ahmady:2014sva} and the lattice QCD~\cite{Horgan:2013hoa} and predictions are presented. }\label{TFF:T1T2T3T3W}
\end{figure*}

The LCSRs are valid when the $K^*$-meson energy has large energy in the rest-system of the $B$-meson, $E_{K^*} \>> \Lambda_{\rm QCD}$; using the relation, $q^2=m_B^2-2m_B E_{K^*}$, one usually adopts $0 \leq q^2 \leq 14{\rm GeV}^2$. We adopt the simplified series expansion (SSE) to extrapolate the TFFs to all physically allowable $q^2$-region, i.e. the TFFs $F_i(q^2)$ are expanded as~\cite{Straub:2015ica}
\begin{eqnarray}
F_i(q^2)=\frac1{1-q^2/m_{R,i}^2} \sum_{k=0,1,2} a_k^i \left[ z(q^2) - z(0) \right]^k,   \label{FI}
\end{eqnarray}
where $F_i$ stands for $A_{0,1,2}(q^2)$, $V(q^2)$ and $T_{1,2,3}(q^2)$, respectively. The function
\begin{eqnarray}
z(t)=\frac{\sqrt{t_+-t}-\sqrt{t_+ - t_0}}{\sqrt{t_+ - t} +\sqrt{t_+-t_0}}
\end{eqnarray}
with $t_{\pm}=(m_B \pm m_{K^*})^2$ and $t_0 = t_+ (1 - \sqrt{1 - t_- / t_+})$. The resonance masses $m_{R;i}$ have been given in Ref.\cite{Straub:2015ica}. The coefficients $a_0^i=F_i(0)$, $a_1^i$ and $a_2^i$ are determined such that the quality of fit ($\Delta$) is around several percents. The quality of fit is defined as~\cite{Agashe:2014kda}
\begin{eqnarray}
\Delta = \frac{\sum_t\left|F_i(t)-F_i^{\rm fit}(t)\right|}{\sum_t\left|F_i(t)\right|}\times 100,
\end{eqnarray}
where $t \in \left[0,\frac1{2},\cdots,\frac{27}{2},14 \right]{\rm GeV}^2$. We put the determined parameters $a^i_{1,2}$ in Table~\ref{F0}, in which all the LCSR parameters are set to be their central values.

\begin{figure*}[htb]
\centering
\includegraphics[width=0.24\textwidth]{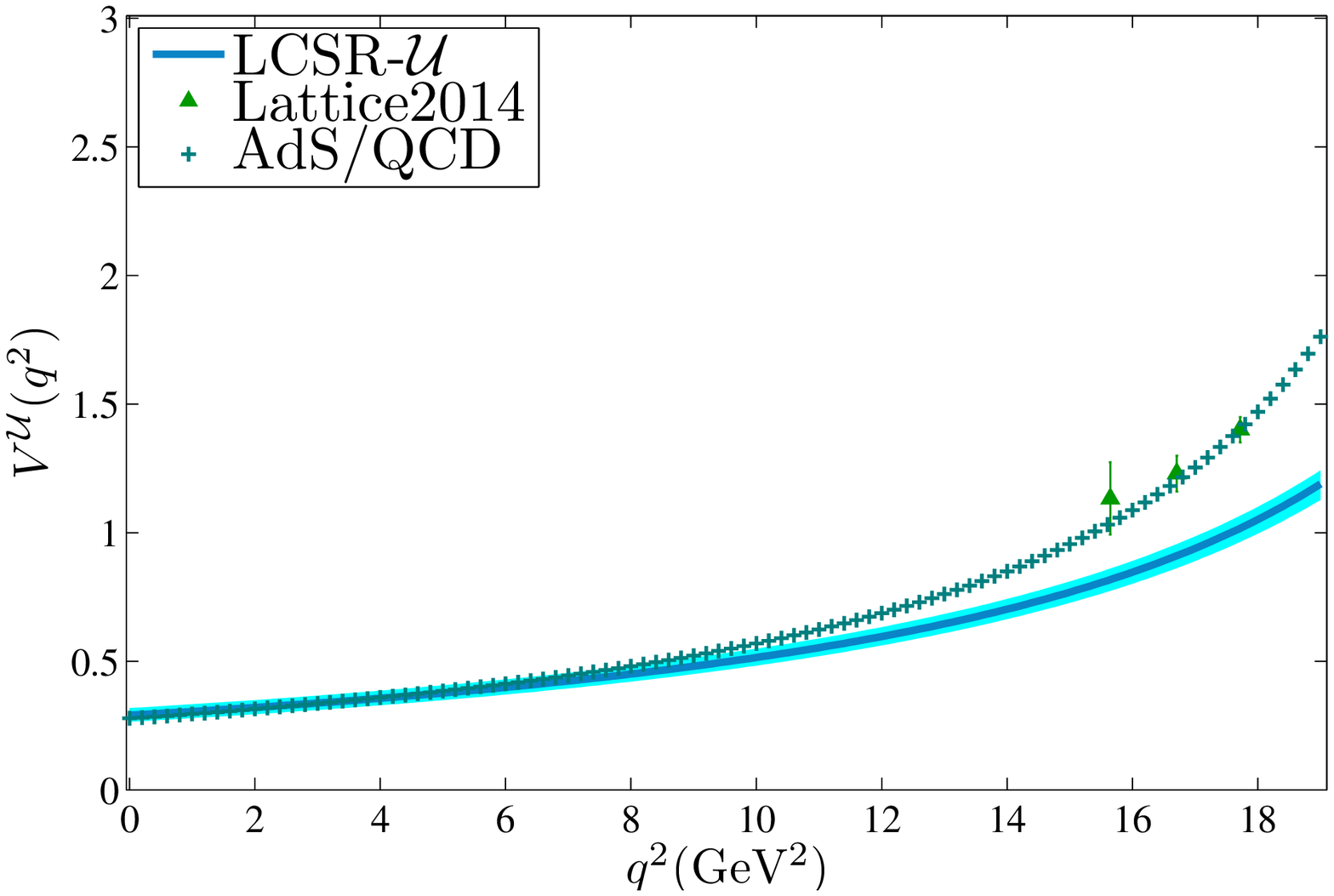}
\includegraphics[width=0.24\textwidth]{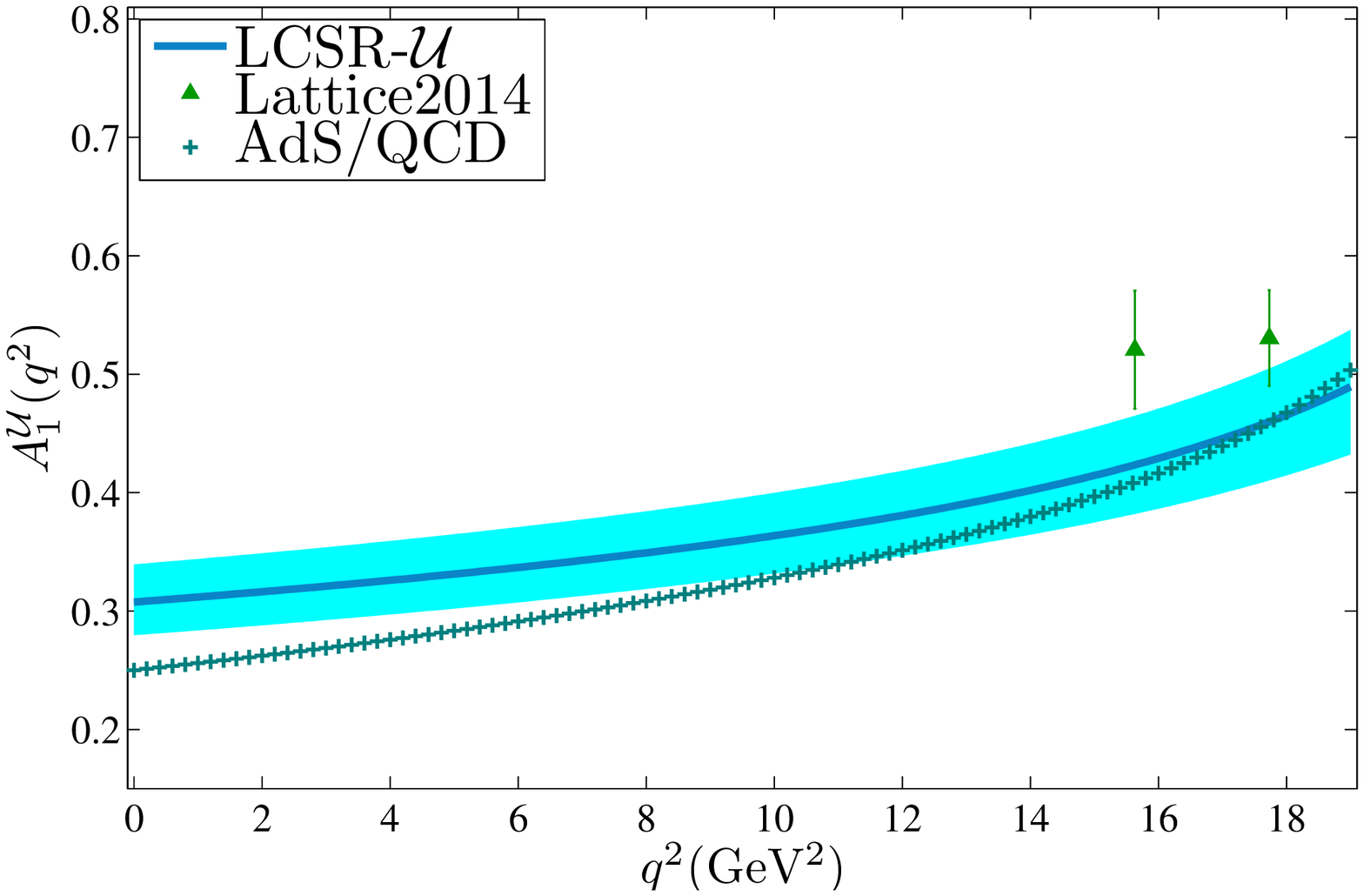}
\includegraphics[width=0.24\textwidth]{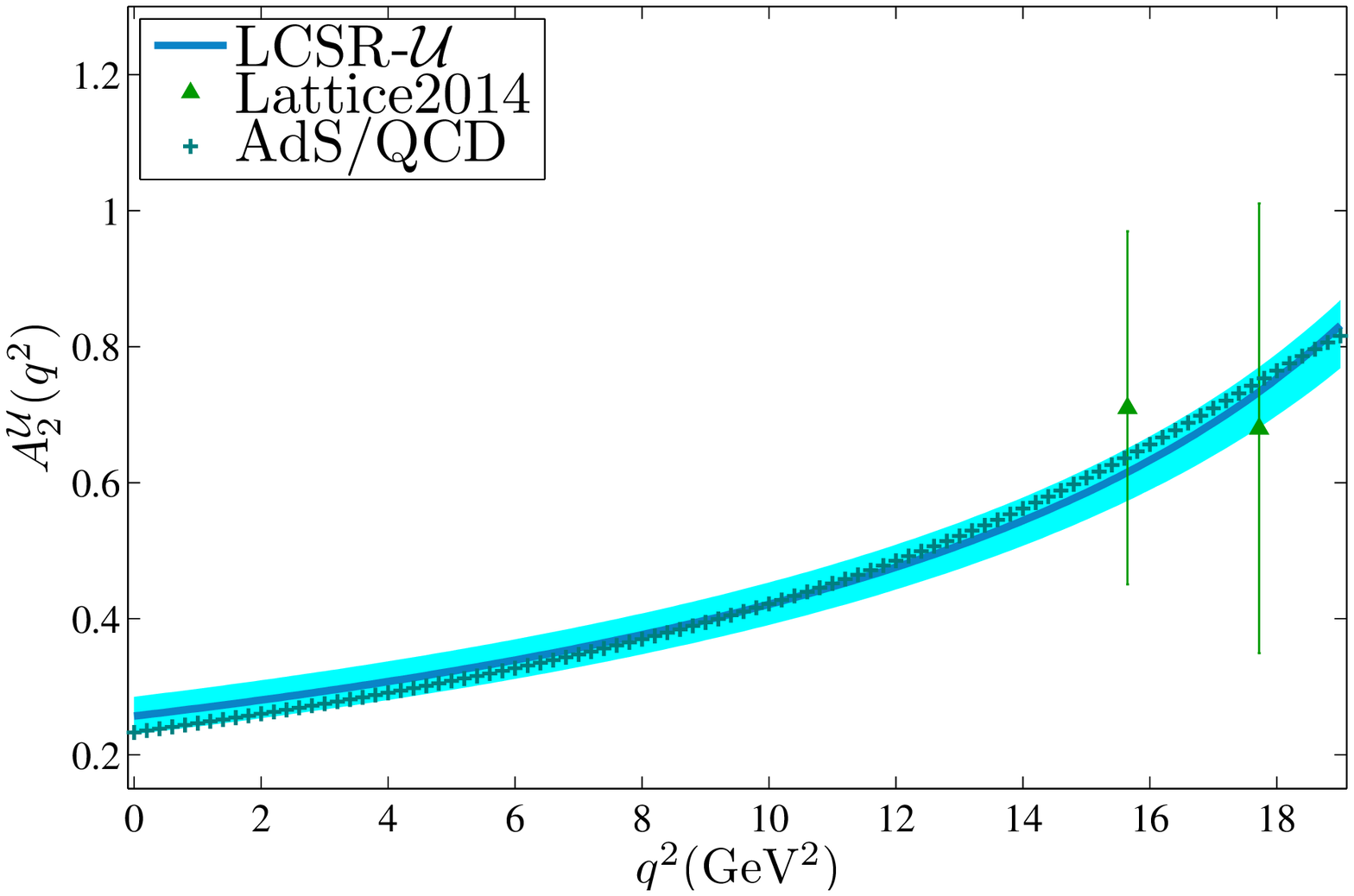}
\includegraphics[width=0.24\textwidth]{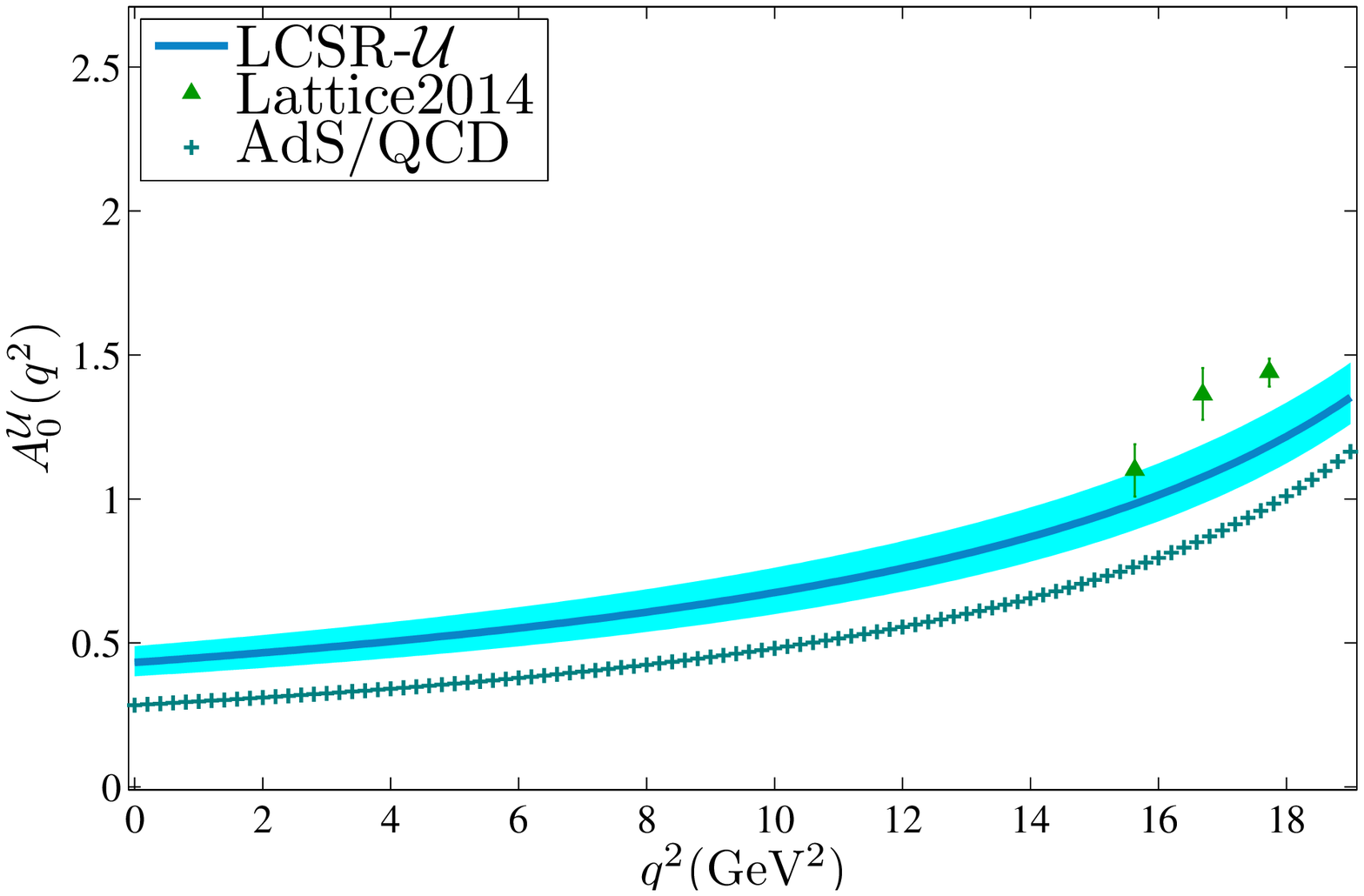}
\includegraphics[width=0.24\textwidth]{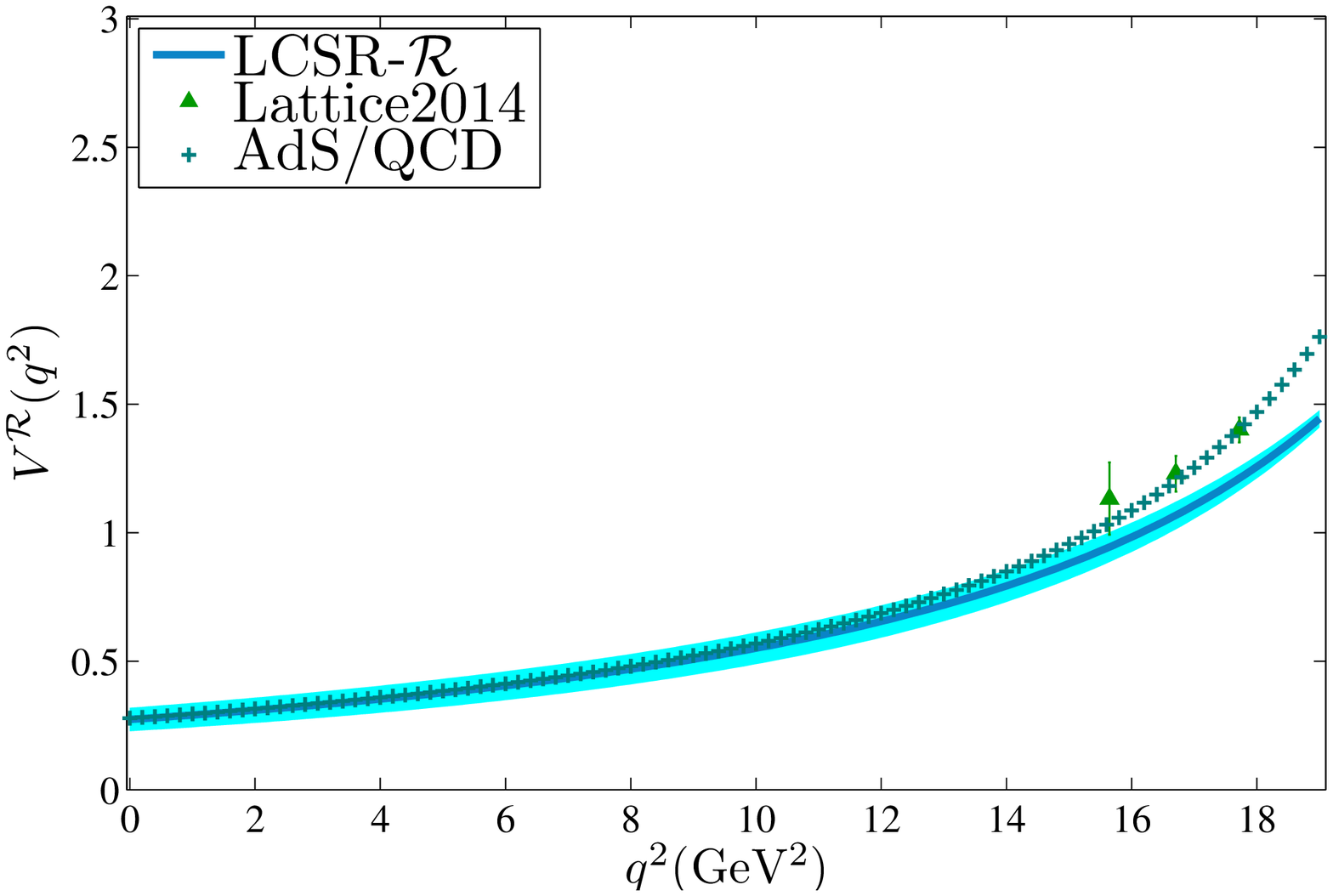}
\includegraphics[width=0.24\textwidth]{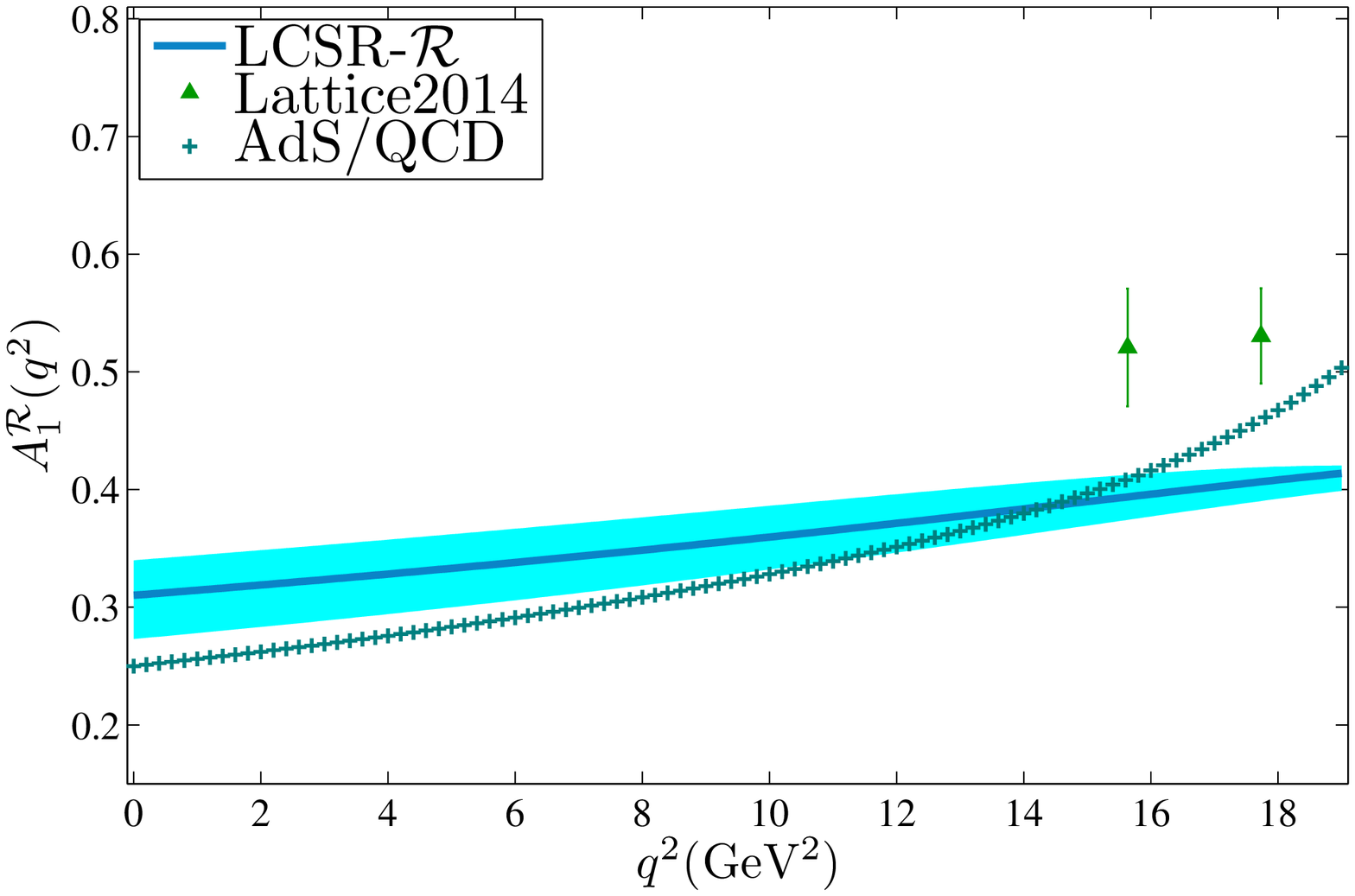}
\includegraphics[width=0.24\textwidth]{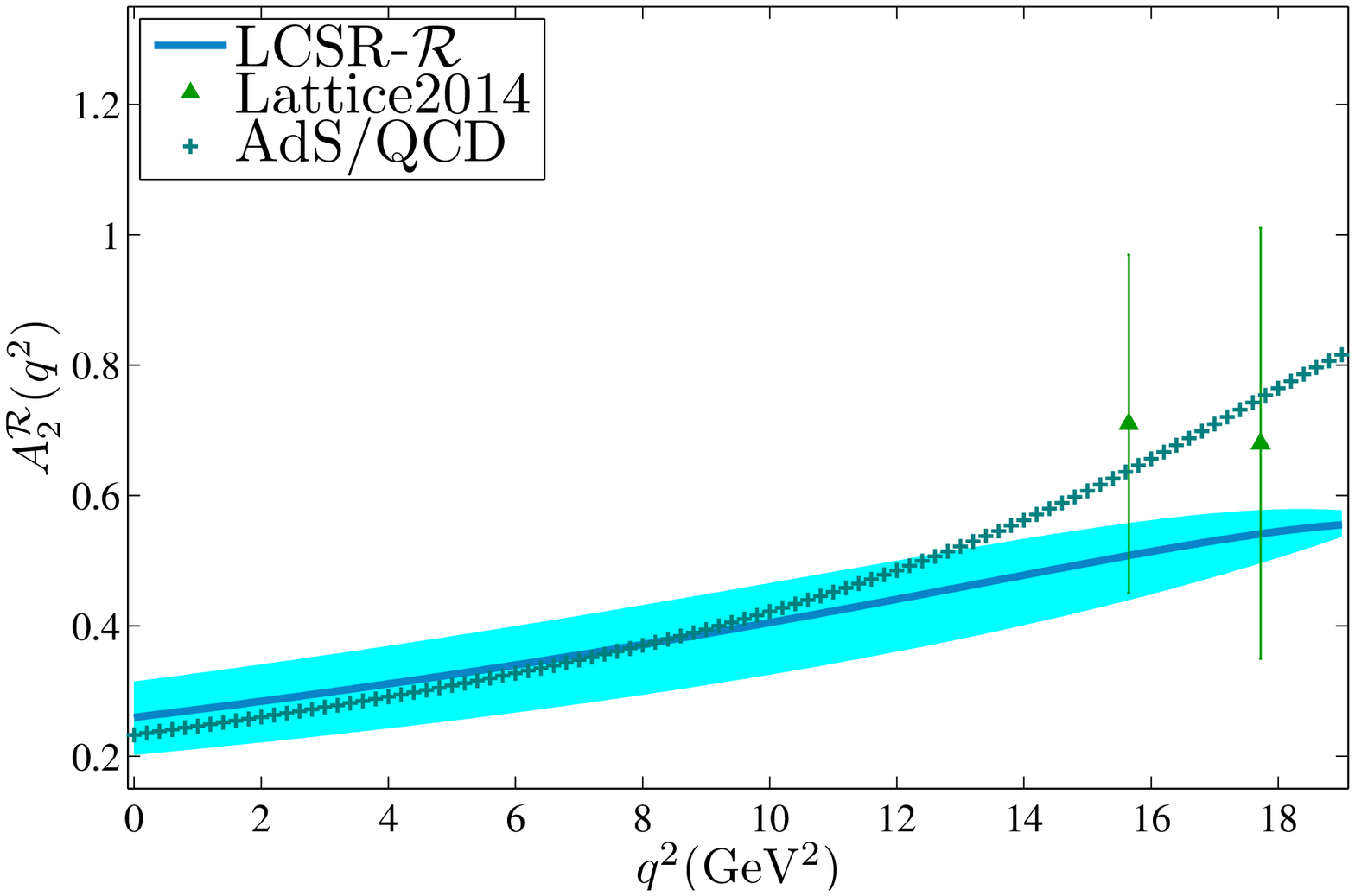}
\includegraphics[width=0.24\textwidth]{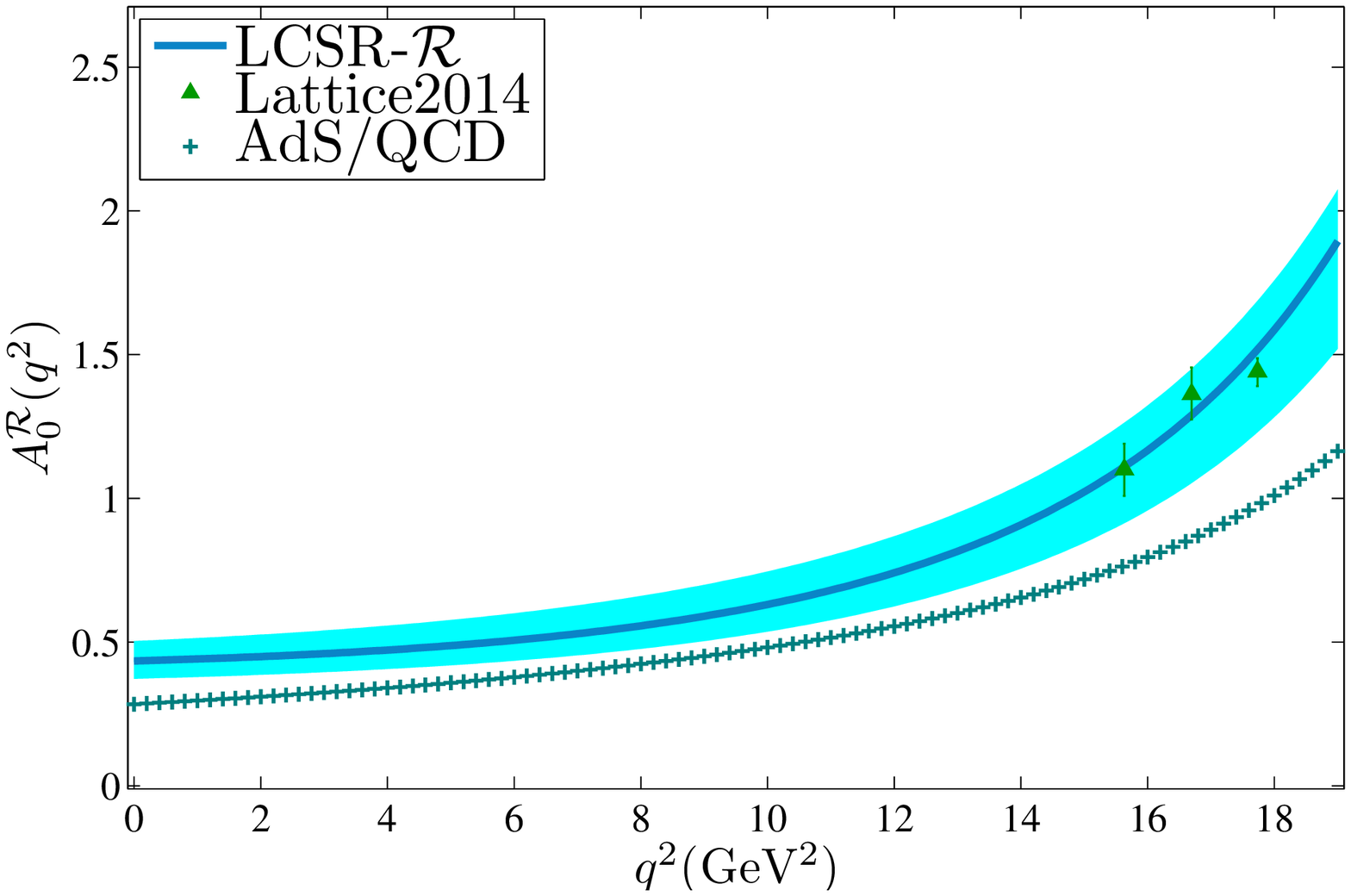}
\caption{The extrapolated $B\to K^*$ axial-vector and vector TFFs $A_{0,1,2}(q^2)$ and $V(q^2)$. The Upper and Lower figures stand for LCSRs with the usual and right-handed correlators, respectively. The solid lines are central values and the shaded bands are their errors. As a comparison, the AdS/QCD~\cite{Ahmady:2014sva} and the lattice QCD~\cite{Horgan:2013hoa} predictions are presented. } \label{TFF:A0A1A2V}
\end{figure*}

\begin{table}[htb]
\begin{tabular}{  c c c  c  c c c c c c }
\hline
      & ~~$A_1$~~ & ~~$A_0$~~ & ~~$T_2$~~ &~~$A_2$~~ & ~~$V$~~  & ~~$T_3$~~& ~~$T_1$~~  \\
\hline
$\rho_{XY}$ &   0.89  &  0.79  &  0.71  &  0.59  & 0.54  &  0.49 & 0.37    \\
\hline
\end{tabular}
\caption{The correlation coefficient $\rho_{\rm XY}$ for the sum rules LCSR-${\cal R}$ and LCSR-${\cal U}$. } \label{correlation parameters}
\end{table}

We present the extrapolated $B\to K^*$ TFFs in Fig.(\ref{TFF:T1T2T3T3W}) and Fig.(\ref{TFF:A0A1A2V}), where the AdS/QCD prediction~\cite{Ahmady:2014sva} and the Lattice QCD prediction~\cite{Horgan:2013hoa} are also given as a comparison. Figs.(\ref{TFF:T1T2T3T3W}, \ref{TFF:A0A1A2V}) show the sum rules of LCSR-${\cal R}$ and LCSR-${\cal U}$ are close in shape. We adopt the correlation coefficient $\rho_{XY}$ to show to what degree those LCSRs are correlated. The correlation coefficient is defined as~\cite{Agashe:2014kda}
\begin{eqnarray}
\rho_{XY}=\frac{{\rm Cov}(X,Y)}{ \sigma_X \sigma_Y}.
\label{rho}
\end{eqnarray}
$X$ and $Y$ stand for the LCSR-${\cal R}$ and LCSR-${\cal U}$ sum rules for the TFFs, respectively. The covariance ${\rm Cov}(X,Y) = E[(X-E(X)) (Y-E(Y))] =E(XY)-E(X)E(Y)$ with $E$ being the expectation value of a function. $\sigma_X$ and $\sigma_Y$ are standard deviations of $X$ and $Y$. The rang of $|\rho_{X,Y}|$ is $0 \sim 1$, a larger $|\rho_{X,Y}|$ indicates a higher consistency among $X$ and $Y$. The correlation coefficients for various TFFs are listed in Table~\ref{correlation parameters}. The magnitudes of the covariance for most of the TFFs are larger than $0.5$, implying those TFFs are consistent with each other, or significantly correlated, even though they are calculated by using different correlators. In the LCSRs, the twist-2, the twist-3, and the twist-4 terms behave differently for different correlators. A larger $\rho_{X}$ shows the net contributions for LCSR-${\cal R}$ and LCSR-${\cal U}$ from various twists are close to each order, not only for their values at the large recoil point $q^2=0$ but also for their ascending trends in whole $q^2$-region.

\subsection{The branching fraction of $B \to K^* \mu^+ \mu^-$}

As an application, we adopt the present TFFs to calculate the branching fraction of the semi-leptonic decay $B \to K^* \mu^+ \mu^-$. We adopt the differential branching fraction derived in Ref.\cite{Aliev:1996hb} as our starting point, where the relations among the coefficients to the TFFs have also been presented.

\begin{figure}[htb]
\centering
\includegraphics[width=0.23\textwidth]{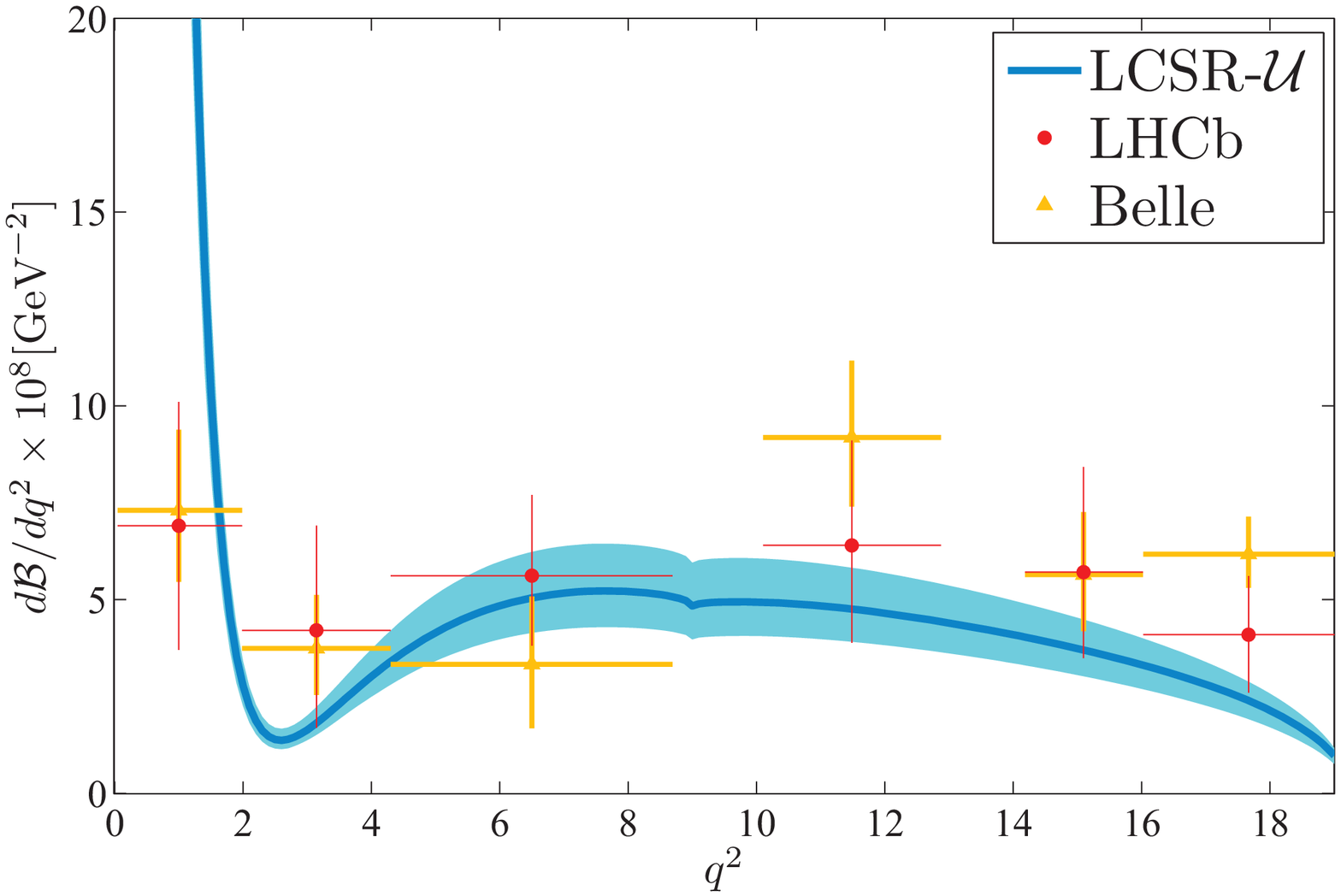}
\includegraphics[width=0.23\textwidth]{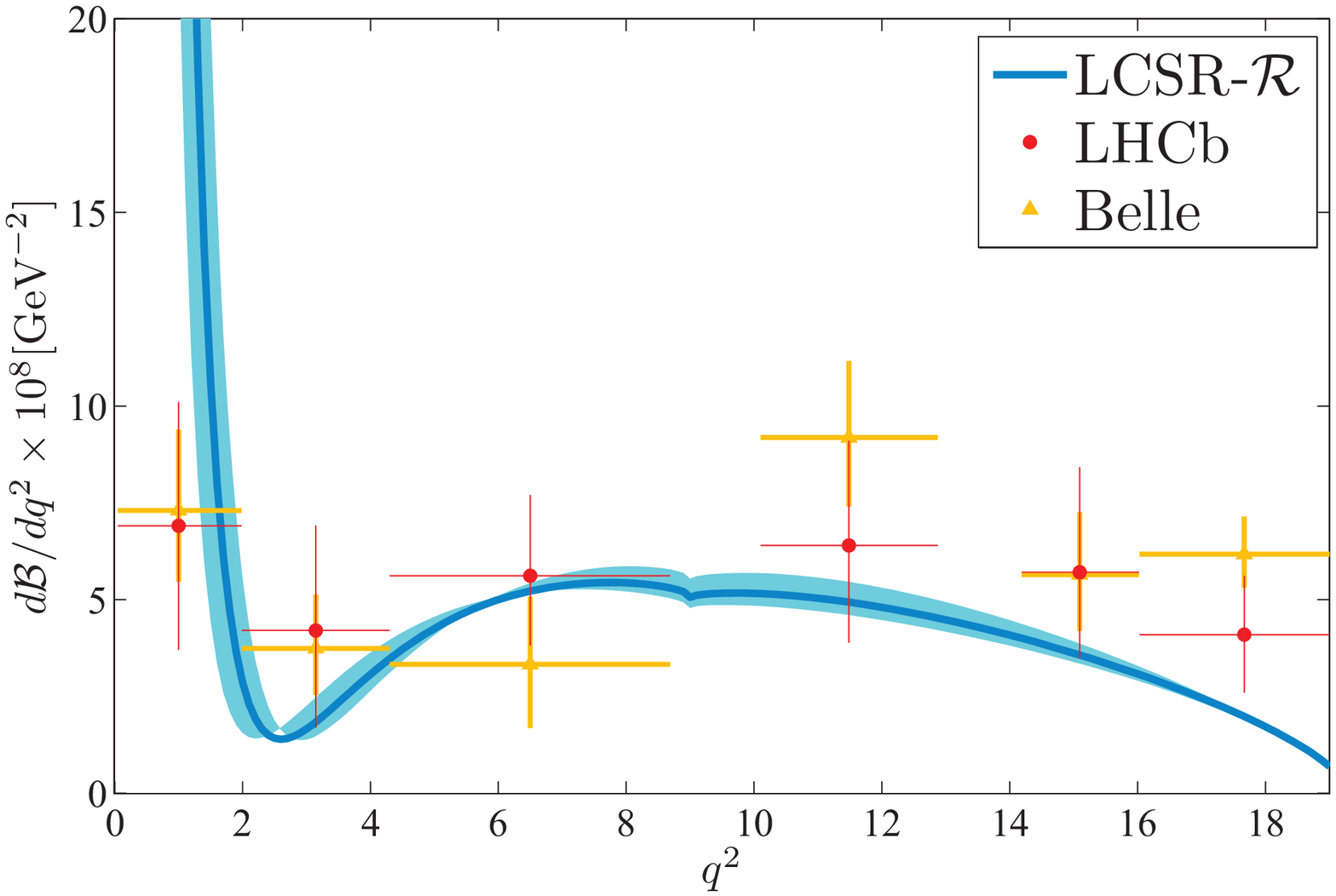}
\includegraphics[width=0.23\textwidth]{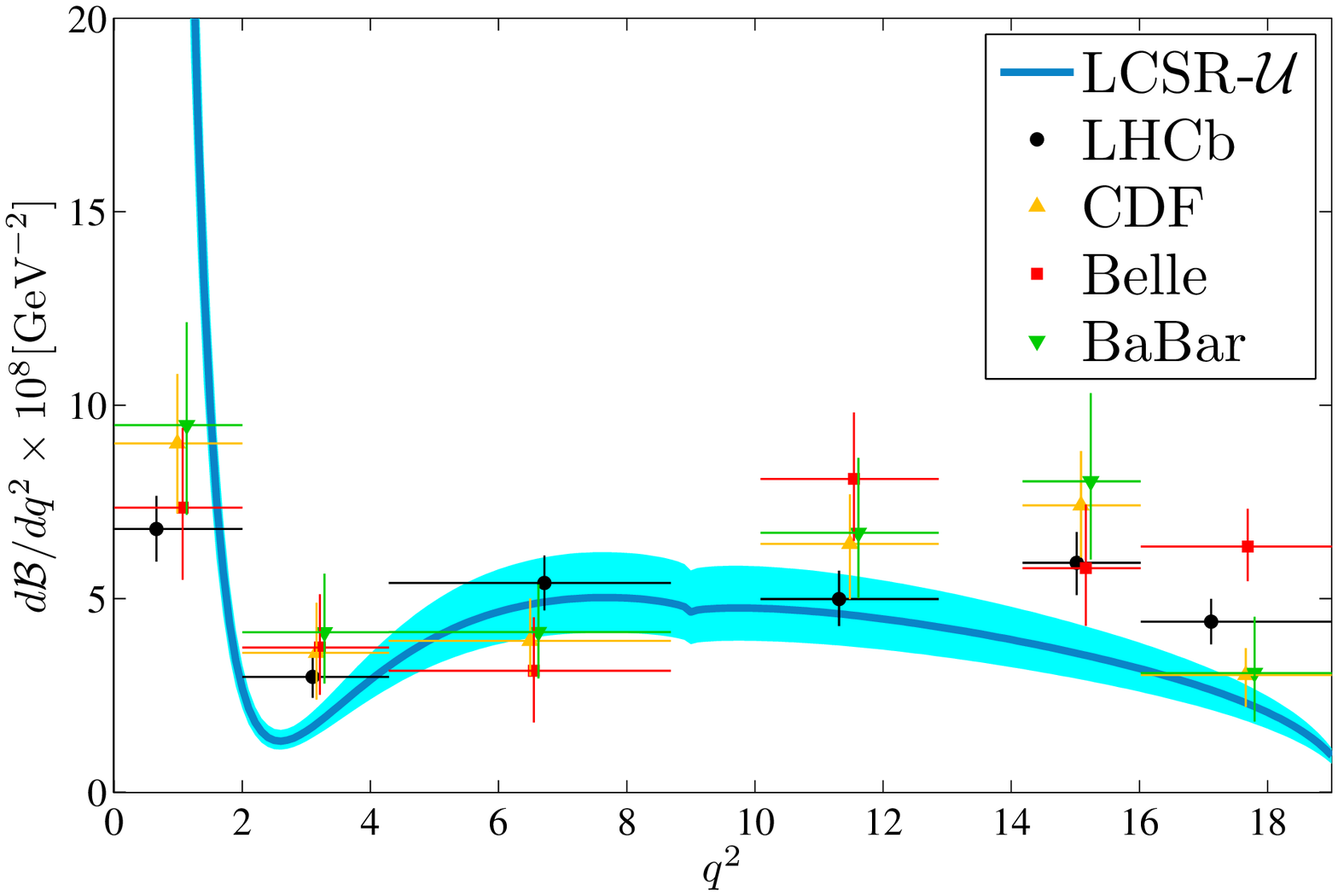}
\includegraphics[width=0.23\textwidth]{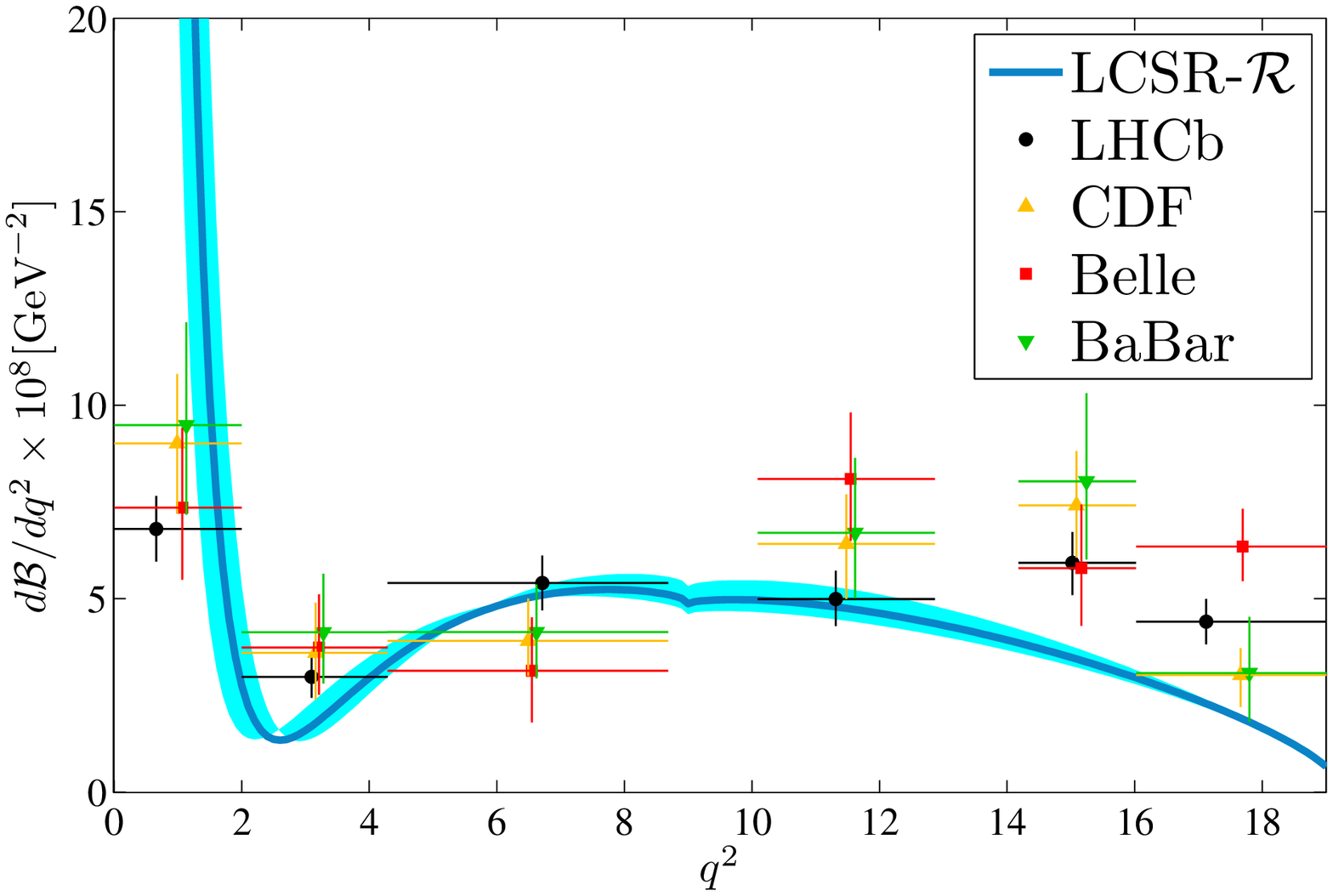}
\caption{Differential branching fraction $d{\cal B}/dq^2$ as a function of $q^2$. The upper figures are for the $B^+$ type and the lower ones are for the $B^0$ type. The solid lines are central values and the shaded bands are their errors. The AdS/QCD prediction~\cite{Ahmady:2014sva}, the Belle data~\cite{Wei:2009zv}, and the LHCb data~\cite{Aaij:2012cq, Aaij:2013qta, LHCb:2012aja} are presented as a comparison. } \label{dB}
\end{figure}

We present the branching fraction $d{\cal B}/{dq^2}$ of the semi-leptonic decay $B \to K^* \mu^+ \mu^-$ in Fig.(\ref{dB}), where the Belle data~\cite{Wei:2009zv} and the LHCb data~\cite{Aaij:2012cq, Aaij:2013qta, LHCb:2012aja} are presented. The branching fractions for $B^+\to K^{*+}\mu^+\mu^-$ ($B^+$-type) and $B^0\to K^{*0}\mu^+\mu^-$ ($B^0$-type) are shown separately. Fig.(\ref{dB}) shows the differential branching fractions from LCSR-${\cal U}$ and LCSR-${\cal R}$ are close in shape, both of which are consistent with the LHCb data. Numerically, we find the correlation coefficient for the branching fractions for the channels $B^+\to K^* \mu^+\mu^-$ and $B^0\to K^{*0}\mu^+\mu^-$ by using the LCSR-${\cal U}$ and the LCSR-${\cal R}$ are the same, both of which have a significant covariance with $\rho_{XY}=0.64$. This is due to the fact that the TFFs $A_1$ and $A_2$ dominate the branching fraction, whose correlation coefficients, as shown by Table~\ref{correlation parameters}, are large.

\section{Summary}\label{Summary}

In the paper, we have studied the $B \to K^*$ TFFs under the LCSR approach by applying two correlators, i.e. the usual one with $j_B^\dag (x)=i m_b \bar b(x) \gamma_5 q(x)$ and the right-handed chiral one with $j_B^\dag (x)=i m_b \bar b(x)(1+ \gamma_5)q(x)$, which lead to different light-cone sum rules for the TFFs, i.e. LCSR-${\cal U}$ and LCSR-${\cal R}$, respectively. The LCSRs for the $B \to K^*$ TFFs are arranged according to the twist structure of the $K^*$-meson LCDA, whose twist-2, twist-3 and twist-4 terms behave quite differently by using different correlators.

The 2-particle and 3-particle LCDAs up to twist-4 accuracy have been kept explicitly in the LCSRs. For the LCSR-${\cal U}$, almost all of the LCDAs come into the contribution, and the relative importance among different twists follows the usual trends, twist-2 $>$ twist-3 $>$ twist-4. For the LCSR-${\cal R}$, only part of the LCDAs are emerged in the TFF, the uncertainty from the unknown high-twist LCDAs are thus greatly suppressed; Moreover, the relative importance among different twists changes to twist-2 $\gg$ twist-3 $\sim$ twist-4. The dominance of the twist-2 term indicates a more convergent twist-expansion could be achieved by using a chiral correlator. Two exceptions for the power counting rule over twists are caused by the twist-4 LCDA $\psi_{4;{K^*}}^{\bot}$; however it contributes to the TFFs via the reduced function $H_3=\int_0^u dv \left[\psi_{4;{K^*}}^{\bot}(v)- \phi_{2;{K^*}}^\bot(v)\right]$, whose net contribution is negligible. Except for $H_3$, the remaining twist-4 contributions are about $10\%$ of the twist-2 ones for the TFFs $A^{{\cal R}/{\cal U}}_{1/2}$, $V$ and $T^{{\cal R}/{\cal U}}_1$, thus the twist-4 terms should be kept for a sound prediction.

We have observed that different LCSRs for the $B\to K^*$ TFFs, i.e. LCSR-${\cal U}$ and LCSR-${\cal R}$, are consistent with each other even though they have been calculated by using different correlators. As shown by Table~\ref{correlation parameters}, large correlation coefficients for most of the TFFs show the net twist-contributions for LCSR-${\cal R}$ and LCSR-${\cal U}$ are close to each order, not only for their values at the large recoil point $q^2=0$ but also for their ascending trends in the whole $q^2$-region. The high correlation of those LCSRs is further confirmed by their application to the branching fraction of the semi-leptonic decay $B \to K^* \mu^+ \mu^-$, i.e. they are significantly correlated with  $\rho_{XY}=0.64$.

The $K^*$-meson LCDAs contribute differently in the LCSRs by using different correlators. The consistency of different LCSRs inversely provide a suitable platform for probing unknown or uncertain LCDAs, i.e. the contributions from those LCDAs to the TFFs can be amplified to a certain degree via a proper choice of correlator, thus amplifying the sensitivity of the TFFs, and hence their related observables, to those LCDAs. \\

{\bf Acknowledgments}: This work was supported in part by Natural Science Foundation of China under Grant No.11625520 and No.11647112.

\appendix

\section{The relations between the LCDAs and the nonlocal matrix elements}

The nonlocal matrix elements in the right-hand-side of the above equation can be reexpressed by the LCDAs of various twists~\cite{Ball:2004rg, Ball:2007zt}, i.e.
\begin{widetext}
\begin{eqnarray}
\langle {K^*}(p,\lambda )|\bar s(x) q_1(0)|0\rangle &=& - \frac{i}{2} f_{K^*}^\bot(e^{*(\lambda )}\cdot x) m_{K^*}^2\int_0^1 du e^{iup\cdot x} \psi_{3;{K^*} }^\parallel(u),\\
\langle {K^*}(p,\lambda )|\bar s (x){\gamma _\beta }{\gamma _5}{q_1}(0)|0\rangle  &=& \frac{1}{4}{\varepsilon^{*(\lambda )}_{\beta}}{m_{{K^*}}}{f_{K^*}^{\parallel}}\int_0^1 d u{e^{iup\cdot x}}\psi_{3;K^*}^\bot(x),\\
\langle {K^*}(p,\lambda )|\overline s(x){\gamma _\beta }{q_1}(0)|0\rangle  &=& {m_{{K^*}}}f_{{K^*}}^\parallel \int_0^1 d u{e^{iu\left (p\cdot x\right)}}\bigg\{ \frac{e^{*(\lambda )} \cdot x}{{p \cdot x}}{p_\beta }\left[ {\phi _{2;{K^*}}^\parallel (u) - \phi _{3;{K^*}}^ \bot (u)} \right] + \frac{e^{*(\lambda )} \cdot x}{{p \cdot x}}{p_\beta }\frac{{m_\rho ^2{x^2}}}{{16}}\phi _{4;{K^*}}^\parallel (u)\nonumber\\
&&+ e^{*(\lambda )}_\beta \phi _{3;{K^*}}^ \bot (u) - \frac{1}{2}{x_\beta }\frac{e^{*(\lambda )} \cdot x}{{{{(p \cdot x)}^2}}}m_{{K^*}}^2\bigg[ {\psi _{4;{K^*}}^\parallel (u) + \phi _{2;{K^*}}^\parallel (u) - 2\phi _{3;{K^*}}^ \bot (u)} \bigg]\bigg\},
\end{eqnarray}
\begin{eqnarray}
\langle {K^*} (p,\lambda)|\bar s(x)\sigma_{\mu \nu}q_1(0)|0\rangle &=& - i f_{K^*}^\bot \int_0^1 du e^{iu(p\cdot x)}\bigg\{(e^{*(\lambda )}_\mu p_\nu - e^{*(\lambda )}_\nu p_\mu)\bigg[\phi_{2;{K^*}}^\bot(u)+ \frac{m_{K^*}^2 x^2}{16} \phi_{4;{K^*}}^\bot(u)\bigg]  \nonumber\\
&& +\left(p_\mu x_\nu - p_\nu x_\mu \right) \frac{e^{*(\lambda )} \cdot x}{(p\cdot x)^2} m_{K^*}^2 \bigg[\phi_{3;{K^*}}^\|(u) - \frac{1}{2} \phi_{2;{K^*}}^\bot(u) - \frac{1}{2} \psi_{4;{K^*}}^\bot(u)\bigg] \nonumber\\
&& +\frac{1}{2} \left( e^{*(\lambda )}_\mu x_\nu  - e^{*(\lambda )}_\nu x_\mu \right) \frac{m_{K^*}^2}{p\cdot x} \bigg[ \psi_{4;{K^*}}^\bot(u)-\phi_{2;{K^*}}^\bot(u) \bigg] \bigg\}. \label{DA1}
\end{eqnarray}
\begin{eqnarray}
\left\langle 0 \right|\overline q (0)g{G_{\mu \nu }}(v x)s( - x)\left| {{K^*}(P,\lambda )} \right\rangle  &=& if_{{K^*}}^ \bot m_{K^*}^2[e_{ \bot \mu }^{(\lambda )}{p_\nu } - e_{ \bot \nu }^{(\lambda )}{p_\mu }]\Psi _{4;{K^*}}^ \bot (v,px)\\
\left\langle 0 \right|\overline q (0)ig{\tilde G _{\mu \nu }}(v x){\gamma _5}s( - x)\left| {{K^*}(P,\lambda )} \right\rangle  &=& if_{{K^*}}^ \bot m_{{K^*}}^2[e_{ \bot \mu }^{(\lambda )}{p_\nu } - e_{ \bot \nu }^{(\lambda )}{p_\mu }]\tilde \Psi  _{4;{K^*}}^ \bot (v,px)\\
\left\langle 0 \right|\overline q (x){\gamma _\mu }{\gamma _5}g{\tilde G _{\alpha \beta }}(vx)s( - x)\left| {{K^*}(P,\lambda )} \right\rangle  &=& {p_\mu }(e_{ \bot \alpha }^{(\lambda )}{p_\beta } - e_{ \bot \beta }^{(\lambda )}{p_\alpha })f_{{K^*}}^{\|}{m_{{K^*}}}\tilde \Phi  _{3;{K^*}}^{\|}(v,px)\nonumber\\
&&+ (p_\alpha g_{\beta \mu }^ \bot  - p_\beta g_{\alpha \mu }^ \bot )\frac{e^{(\lambda )}x}{px} f_{K^*}^{\|}m_{K^*}^3 \tilde \Phi_{4;{K^*}}^{\|}(v,px)\nonumber\\
&& + p_\mu (p_\alpha x_\beta  - p_\beta x_\alpha )\frac{e^{(\lambda )}x}{(p z)^2} f_{K^*}^{\|} m_{{K^*}}^3 \tilde \Psi_{4;{K^*}}^{\|}(v,px)\\
\left\langle 0 \right|\overline q (x)i\gamma _\mu g G_{\alpha \beta }(v x)s( - x)\left| K^*(P,\lambda ) \right\rangle  &=& p_\mu (e_{ \bot \alpha }^{(\lambda )}p_\beta  - e_{ \bot \beta }^{(\lambda )} p_\alpha ) f_{K^*}^{\|} m_{K^*}\Phi _{3;{K^*}}^{\|}(v,px)\nonumber\\
&&+ (p_\alpha g_{\beta \mu }^ \bot  - p_\beta g_{\alpha \mu }^ \bot )\frac{e^{(\lambda )}x}{p x} f_{K^*}^{\|} m_{K^*}^3 \Phi _{4;{K^*}}^{\|}(v,px)\nonumber\\
&&+ p_\mu (p_\alpha x_\beta- p_\beta x_\alpha ) \frac{e^{(\lambda )}x}{(px)^2} f_{K^*}^{\|} m_{K^*}^3 \Psi _{4;{K^*}}^{\|}(v,px)\\
\left\langle 0 \right|\overline q (x)\sigma _{\alpha \beta } g G_{\mu \nu }(v x)s( - x)\left| {K^*}(P,\lambda ) \right\rangle &=& f_{K^*}^ \bot m_{K^*}^2\frac{e^{(\lambda )}\cdot x}{2(p\cdot x)}[p_\alpha p_\mu g_{\beta \nu }^ \bot  - p_\beta p_\mu g_{\alpha \nu }^ \bot  - p_\alpha p_\nu g_{\beta \mu }^ \bot  + p_\beta p_\nu g_{\alpha \mu }^ \bot ]\Phi _{3;{K^*}}^ \bot (v,px)\nonumber\\
&&+ f_{K^*}^ \bot m_{K^*}^2[p_\alpha e_{ \bot \mu }^{(\lambda )}g_{\beta \nu }^ \bot  - p_\beta e_{ \bot \mu }^{(\lambda )}g_{\alpha \nu }^ \bot  - p_\alpha e_{ \bot \nu }^{(\lambda )}g_{\beta \mu }^ \bot  + p_\beta e_{ \bot \nu }^{(\lambda )}g_{\alpha \mu }^ \bot ]\Phi _{4;{K^*}}^{ \bot (1)}(v,px)\nonumber\\
&&+ f_{K^*}^ \bot m_{K^*}^2[p_\mu e_{ \bot \alpha }^{(\lambda )}g_{\beta \nu }^ \bot  - p_\mu e_{ \bot \beta }^{(\lambda )}g_{\alpha \nu }^ \bot  - p_\nu e_{ \bot \alpha }^{(\lambda )}g_{\beta \mu }^ \bot  + p_\nu e_{ \bot \beta }^{(\lambda )}g_{\alpha \mu }^ \bot ]\Phi _{4;{K^*}}^{ \bot (2)}(v,px)\nonumber\\
&&+ \frac{f_{K^*}^ \bot m_{K^*}^2}{p\cdot x}[p_\alpha p_\mu e_{ \bot \beta }^{(\lambda )} x_\nu  - p_\beta p_\mu e_{ \bot \alpha }^{(\lambda )}x_\nu  - p_\alpha p_\nu e_{ \bot \beta }^{(\lambda )}x_\nu + p_\beta p_\nu e_{ \bot \alpha }^{(\lambda )}x_\nu ]\Phi _{4;{K^*}}^{ \bot (3)}(v,px)\nonumber\\
&&+ \frac{f_{K^*}^ \bot m_{K^*}^2}{p\cdot x}[p_\alpha p_\mu e_{ \bot \nu }^{(\lambda )}x_\beta  - p_\beta p_\mu e_{ \bot \nu }^{(\lambda )}x_\alpha  - p_\alpha p_\nu e_{ \bot \mu }^{(\lambda )}x_\beta  + p_\beta p_\nu e_{ \bot \nu }^{(\lambda )}x_\alpha ]\Phi _{4;{K^*}}^{ \bot (4)}(v,px)\nonumber\\
\end{eqnarray}
\end{widetext}
\begin{widetext}
\begin{eqnarray}
\left\langle 0 \right|\overline q (x)i \sigma_{\alpha \beta } g \tilde G_{\mu \nu } (v x)s( - x)\left| {K^*}(P,\lambda ) \right\rangle &=& f_{K^*}^ \bot m_{K^*}^2\frac{e^{(\lambda )}\cdot x}{2(p\cdot x)}[p_\alpha p_\mu g_{\beta \nu }^ \bot  - p_\beta p_\mu g_{\alpha \nu }^ \bot  - p_\alpha p_\nu g_{\beta \mu }^ \bot  + p_\beta p_\nu g_{\alpha \mu }^ \bot ]\Phi _{3;{K^*}}^ \bot (v,px)\nonumber\\
&&+ f_{K^*}^ \bot m_{K^*}^2[p_\alpha e_{ \bot \mu }^{(\lambda )}g_{\beta \nu }^ \bot  - p_\beta e_{ \bot \mu }^{(\lambda )}g_{\alpha \nu }^ \bot  - p_\alpha e_{ \bot \nu }^{(\lambda )}g_{\beta \mu }^ \bot  + p_\beta e_{ \bot \nu }^{(\lambda )}g_{\alpha \mu }^ \bot ] \tilde \Phi _{4;{K^*}}^{ \bot (1)}(v,px)\nonumber\\
&&+ f_{K^*}^ \bot m_{K^*}^2[p_\mu e_{ \bot \alpha }^{(\lambda )}g_{\beta \nu }^ \bot  - p_\mu e_{ \bot \beta }^{(\lambda )}g_{\alpha \nu }^ \bot  - p_\nu e_{ \bot \alpha }^{(\lambda )}g_{\beta \mu }^ \bot  + p_\nu e_{ \bot \beta }^{(\lambda )}g_{\alpha \mu }^ \bot ] \tilde \Phi _{4;{K^*}}^{ \bot (2)}(v,px)\nonumber\\
&&+ \frac{f_{K^*}^ \bot m_{K^*}^2}{p\cdot x}[p_\alpha p_\mu e_{ \bot \beta }^{(\lambda )} x_\nu  - p_\beta p_\mu e_{ \bot \alpha }^{(\lambda )}x_\nu  - p_\alpha p_\nu e_{ \bot \beta }^{(\lambda )}x_\nu + p_\beta p_\nu e_{ \bot \alpha }^{(\lambda )}x_\nu ] \tilde \Phi _{4;{K^*}}^{ \bot (3)}(v,px)\nonumber\\
&&+ \frac{f_{K^*}^ \bot m_{K^*}^2}{p\cdot x}[p_\alpha p_\mu e_{ \bot \nu }^{(\lambda )}x_\beta  - p_\beta p_\mu e_{ \bot \nu }^{(\lambda )}x_\alpha  - p_\alpha p_\nu e_{ \bot \mu }^{(\lambda )}x_\beta  + p_\beta p_\nu e_{ \bot \nu }^{(\lambda )}x_\alpha ] \tilde \Phi _{4;{K^*}}^{ \bot (4)}(v,px)\nonumber \\
\end{eqnarray}
\end{widetext}
Here $f_{K^*}^\bot$ and $f_{K^*}^\|$ are ${K^*}$-meson decay constants, which are defined as $\left\langle {{K^*}(P,\lambda )} \right|\bar s (0){\gamma _\mu }q(0)\left| 0 \right\rangle  = f_{{K^*}}^\parallel {m_{{K^*}}}e^{*(\lambda )}_\mu$ and $\left\langle {{K^*}(P,\lambda )} \right|\bar s (0){\sigma _{\mu \nu }}q(0)\left| 0 \right\rangle = if_{{K^*}}^ \bot (e^{*(\lambda )}_\mu {p_\nu } - e^{*(\lambda )}_\nu {p_\mu })$.

\section{LCSRs for the $B \to K^*$ TFFs by using the right-handed chiral correlator}

We list the LCSRs for the $B \to K^*$ TFFs by using the right-handed chiral correlator in the following
\begin{widetext}
\begin{eqnarray}
A_1^{\mathcal R}(q^2) &=&\frac{m_b m_{K^*}^2 f_{K^*}^\bot}{f_B m_B^2(m_B+m_{K^*})} \bigg\{ \int_0^1\frac{du}{u}e^{\left( {m_{B}^2 - s(u)} \right) / M^2} \bigg\{\frac{\cal C}{u m_{K^*} ^2}\Theta(c(u,s_0))\phi_{2;{K^*}}^\bot(u,\mu) +\Theta(c(u,s_0))(u)  \nonumber\\
&& \times \psi_{3;{K^*}}^\|-\frac{1}{4}\bigg[ \frac{m_b^2{\cal C}}{u^3M^4} \widetilde{\widetilde\Theta}(c(u,s_0)) + \frac{{\cal C}-2m_b^2}{u^2M^2}\widetilde\Theta(c(u,s_0))- \frac{1}{u}\Theta(c(u,s_0))\bigg]\phi_{4;{K^*}}^\bot(u)- 2\bigg[\frac{\cal C}{u^2M^2}\nonumber\\
&&\times \widetilde\Theta(c(u,s_0)) -\frac{1}{u}\Theta(c(u,s_0))\bigg]  I_L(u)-\bigg[\frac{2m_b^2}{uM^2}\widetilde \Theta(c(u,s_0)) + \Theta(c(u,s_0)) \bigg] H_3(u)\bigg\}+\int_0^1 dv \int_0^1 du \nonumber\\
&& \times \int_0^1 d {\mathcal D} e^{\left( {m_{B}^2 - s(u)} \right) / M^2} \frac{f_{K^*}^\bot m_b m_{K^*}^2} {12 f_B m_B^2( m_B + m_{K^*})} \frac{\widetilde\Theta(c(u,s_0))}{u^2 M^2}\bigg[\widetilde {\Psi} _{4;{K^*}}^\bot (\underline \alpha ) - 12\bigg(4v \Phi _{4;{K^*}}^{\bot(2)} (\underline \alpha )-2v \Psi _{4;{K^*}}^\bot (\underline \alpha )\nonumber\\
&& +2\Phi _{4;{K^*}}^{\bot(1)} (\underline \alpha )-2\Phi _{4;{K^*}}^{\bot(2)} (\underline \alpha )+\Psi _{4;{K^*}}^\bot (\underline \alpha )\bigg)\bigg] (m_B^2-m_{K^*}^2+2 u m_{K^*}^2 ),
\label{R A1}
\\
A_2^{\mathcal R}(q^2) &=&\frac{m_b(m_B + m_{K^*} )m_{K^*}^2 f_{K^*}^\bot }{f_B m_B^2} \bigg\{ \int_0^1 \frac{du}{u} e^{\left( {m_{B}^2 - s(u)} \right) / M^2} \bigg\{\frac{1}{m_{K^*} ^2}\Theta(c(u,s_0))\phi_{2;{K^*}}^\bot(u,\mu )- \frac{1}{M^2}\nonumber \\
&& \times \widetilde\Theta (c(u,s_0))\psi_{3;{K^*}}^{\|}(u)- \frac{1}{4}\bigg[\frac{m_b^2}{u^2M^4} \widetilde{\widetilde\Theta}(c(u,s_0)) + \frac{1}{uM^2} \widetilde\Theta(c(u,s_0))]\phi_{4;{K^*}}^\bot(u)+ 2\bigg[\frac{{\cal C} - 2m_b^2}{u^2 M^4}\nonumber\\
&&\times \widetilde{\widetilde\Theta}(c(u,s_0)) - \frac{1}{uM^2}\widetilde\Theta (c(u,s_0))]I_L(u)- \frac{1}{M^2}\widetilde \Theta (c(u,s_0))H_3(u)\bigg\}+\int_0^1 dv \int_0^1 du \int_0^1 d {\mathcal D}  \nonumber\\
&&\times e^{\left( {m_{B}^2 - s(u)} \right) / M^2} \frac{f_{K^*}^\bot m_b m_{K^*}^2} {12 f_B m_B^2} \frac{ m_B + m_{K^*}}{u^2 M^2}\widetilde\Theta(c(u,s_0))\bigg[\widetilde {\Psi} _{4;{K^*}}^\bot (\underline \alpha ) + 12\bigg((4v-2) \Phi _{4;{K^*}}^{\bot(1)} (\underline \alpha )\nonumber\\
&&+2v \Psi _{4;{K^*}}^\bot (\underline \alpha )+2\Phi _{4;{K^*}}^{\bot(2)} (\underline \alpha )-\Psi _{4;{K^*}}^\bot (\underline \alpha )\bigg)\bigg],
\label{R A2}
\end{eqnarray}
\begin{eqnarray}
A_3^{\mathcal R}(q^2) &-& A_0^{\mathcal R}(q^2) =\frac{m_b m_{K^*} f_{K^*}^\bot q^2}{2f_B m_B^2} \bigg\{ \int_0^1 \frac{du}{u} e^{\left( {m_{B}^2 - s(u)} \right) / M^2} \bigg\{-\frac{1}{m_{K^*}^2}\Theta (c(u,s_0))\phi_{2;K^*}^\bot(u)- \frac{2-u}{uM^2}\nonumber\\
&&\times \widetilde \Theta (c(u,s_0))\psi_{3;K^*}^\| (u,\mu)+ \frac{1}{4}\bigg[\frac{m_b^2}{u^2M^4} \widetilde{\widetilde\Theta}(c(u,s_0))+ \frac{1}{uM^2}\widetilde\Theta(c(u,s_0))\bigg]\phi_{4;K^*}^\bot(u)+\bigg[(4-2u)
\nonumber\\
&&\times \bigg(\frac{\cal C}{u^3 M^4}\widetilde{\widetilde\Theta}(c(u,s_0))- \frac{2}{u^2 M^2} \widetilde\Theta(c(u,s_0))\bigg)+ \bigg(\frac{4m_b^2}{u^2 M^4}\widetilde{\widetilde\Theta}(c(u,s_0)) + \frac{2}{uM^2} \widetilde\Theta(c(u,s_0))\bigg)\bigg]I_L(u)
\nonumber\\
&&- \frac{2-u}{uM^2}\widetilde\Theta(c(u,s_0))H_3(u)\bigg\}-\int_0^1 dv \int_0^1 du \int_0^1 d {\mathcal D}  e^{\left( {m_{B}^2 - s(u)} \right) / M^2} \frac{ m_b q^2 f_{K^*}^\bot} {24 f_B m_B^2} \frac{\widetilde\Theta(c(u,s_0))}{u^2 M^2} \bigg[ \widetilde {\Psi} _{4;{K^*}}^\bot (\underline \alpha )\nonumber\\
&& + 12\bigg((4v-2) \Phi _{4;{K^*}}^{\bot(1)} (\underline \alpha )+2v \Psi _{4;{K^*}}^\bot (\underline \alpha )+2\Phi _{4;{K^*}}^{\bot(2)} (\underline \alpha )-\Psi _{4;{K^*}}^\bot (\underline \alpha )\bigg)\bigg],
\label{R A30}\\
T_1^{\mathcal R}(q^2) &=& \frac{m_b^2 m_{K^*}^2 f_{K^*}^\bot }{m_B^2f_B}\int_0^1  \frac{du}{u} e^{\left( {m_{B}^2 - s(u)} \right) / M^2} \bigg\{ \frac{1}{m_{K^*}^2}\Theta (c(u,s_0))\phi _{2;K^*}^\bot (u) - \frac{m_b^2}{4u^2M^4} \widetilde{\widetilde \Theta} (c(u,s_0)) \phi_{4;K^*}^\bot(u)\nonumber\\
&&-\frac{2}{uM^2}\widetilde\Theta(c(u,s_0))I_L(u) - \frac{1}{M^2} \widetilde\Theta(c(u,s_0)) H_3(u) \bigg\}+\int_0^1 dv \int_0^1 du \int_0^1 d {\mathcal D} e^{\left( {m_{B}^2 - s(u)} \right) / M^2} \frac{\widetilde\Theta(c(u,s_0))}{u^2 M^2} \nonumber\\
&&\times \frac{f_{K^*}^\bot  m_{K^*}^2 m_b} {12 f_B m_B^2}\bigg[\widetilde {\Psi} _{4;{K^*}}^\bot (\underline \alpha )-12\bigg(2\Phi _{4;{K^*}}^{\bot(1)} (\underline \alpha ) - 2 \Phi _{4;{K^*}}^{\bot(2)} (\underline \alpha )+\Psi _{4;{K^*}}^\bot (\underline \alpha )\bigg)\bigg],
\label{R T1}
\end{eqnarray}
\begin{eqnarray}
T_2^{\mathcal R}(q^2) &=&\frac{m_b^2f_{K^*}^\bot m_{K^*}^2}{m_B^2 f_B} \int_0^1 \frac{du}{u}e^{\left( {m_{B}^2 - s(u)} \right) / M^2} \bigg\{ \frac{1 - {\mathcal H}}{m_{K^*}^2}
\Theta (c(u,s_0)) \phi_{2;K^*}^\bot(u) - \frac{m_b^2} {4u^2M^4}
(1 - {\mathcal H}) \widetilde{\widetilde\Theta}(c(u,s_0)) \nonumber\\
&&\times \phi_{4;K^*}^\bot(u)- \frac{2(1 - {\mathcal H})}{uM^2}\widetilde\Theta(c(u,s_0))I_L(u) - \frac{1}{M^2}\bigg[ 1+ \bigg( \frac{2}{u}-1 \bigg){\mathcal H} \bigg]\widetilde\Theta (c(u,s_0)) H_3(u) \bigg\}+\int_0^1 dv\nonumber\\
&& \int_0^1 du \int_0^1 d {\mathcal D} e^{\left( {m_{B}^2 - s(u)} \right) / M^2} \frac{f_{K^*}^\bot  m_{K^*}^2 m_b} {12 f_B m_B^2} \frac{\widetilde\Theta(c(u,s_0))}{u^2 M^2}\bigg[\widetilde {\Psi} _{4;{K^*}}^\bot (\underline \alpha )-12\bigg(2\Phi _{4;{K^*}}^{\bot(1)} (\underline \alpha ) - 2 \Phi _{4;{K^*}}^{\bot(2)} (\underline \alpha )\nonumber\\
&&+\Psi _{4;{K^*}}^\bot (\underline \alpha )\bigg)\bigg],
\label{R T2}\\
T_3^{\mathcal R}(q^2) &=& \frac{m_b^2f_{K^*}^\bot m_{K^*}^2}{m_B^2f_B} \int_0^1\frac{du}{u} e^{\left( {m_{B}^2 - s(u)} \right) / M^2} \bigg\{\frac{1}{m_{K^*}^2} \Theta(c(u,s_0))\phi_{2;K^*}^\bot(u) - \frac{m_b^2}{4u^2M^4}\widetilde{\widetilde\Theta}(c(u,s_0))\phi_{4;K^*}^\bot(u)\nonumber\\
&&- \bigg[\frac{2}{uM^2}\widetilde\Theta(c(u,s_0)) + \frac{4}{u^2 M^4}\widetilde{\widetilde \Theta}(c(u,s_0))(m_B^2 - m_{K^*}^2)\bigg] I_L(u) + \bigg[ \frac{2}{uM^2} - \frac{1}{M^2} \bigg] \widetilde\Theta(c(u,s_0))H_3(u) \bigg\} \nonumber\\
&&+\int_0^1 dv\int_0^1 du \int_0^1 d {\mathcal D}  e^{\left( {m_{B}^2 - s(u)} \right) / M^2} \frac{f_{K^*}^\bot  m_{K^*}^2 m_b} {12 f_B m_B^2 u^2 M^2}\widetilde\Theta(c(u,s_0))\bigg[\widetilde {\Psi} _{4;{K^*}}^\bot (\underline \alpha )-12\bigg(2\Phi _{4;{K^*}}^{\bot(1)} (\underline \alpha )\nonumber\\
&& - 2 \Phi _{4;{K^*}}^{\bot(2)} (\underline \alpha )+\Psi _{4;{K^*}}^\bot (\underline \alpha )\bigg)\bigg],
\label{R T3} \\
V^{\mathcal R}(q^2) &=&\frac{m_b(m_B + m_{K^*})f_{K^*}^\bot}{f_B m_B^2}\int_0^1 \frac{du}{u} e^{\left( {m_{B}^2 - s(u)} \right) / M^2} \bigg\{ \Theta(c(u,s_0)) \phi_{2;{K^*}}^\bot(u,\mu) - \bigg[\frac{m_b^2}{u^2M^4}\widetilde {\widetilde\Theta}(c(u,s_0))
\nonumber\\
&&+\frac{1}{uM^2}\widetilde\Theta(c(u,s_0))\bigg]\frac{m_{K^*}^2}{4}\phi_{4;{K^*}}^\bot (u)\bigg\}+\int_0^1 dv \int_0^1 du \int_0^1 d {\mathcal D}  e^{\left( {m_{B}^2 - s(u)} \right) / M^2} \frac{f_{K^*}^\bot m_{K^*}^2} {6u^2 M^2}\widetilde\Theta(c(u,s_0))\nonumber\\
&&\times \bigg[(2v-1)\widetilde {\Psi} _{4;{K^*}}^\bot (\underline \alpha ) + 12\bigg(\Psi _{4;{K^*}}^\bot (\underline \alpha )-2(v-1)(\Phi _{4;{K^*}}^{\bot(1)} (\underline \alpha )-\Phi _{4;{K^*}}^{\bot(2)} (\underline \alpha ))\bigg)\bigg].
\label{R V}
\end{eqnarray}
\end{widetext}
To compare with previous LCSRs given by Ref.\cite{Fu:2014uea}, in the above formulas, we keep all the three-particle twist-4 terms in the LCSRs.

\end{document}